\documentclass[11pt]{article}
\pdfoutput=1

\usepackage[utf8x]{inputenc} 
\usepackage{jheppub}
\usepackage{amsmath}
\usepackage{amssymb}
\usepackage{dcolumn}
\usepackage{bm}
\usepackage{multirow}
\usepackage{blkarray}     
\usepackage{slashed}                             
\usepackage{hyperref,graphicx}           
\usepackage{url}
\usepackage{caption}
\usepackage{subcaption}
\usepackage{multirow}
\usepackage{units}
\usepackage{xcolor}
\usepackage{graphicx}
\usepackage{tikz}
\usepackage{textcomp}
\usepackage{tabularx}
\graphicspath{{./Figure/}}

\begin{document}

\title{Testing Lepton Flavor Universality at Future $Z$ Factories}
\author[a]{Tin Seng Manfred Ho,}
\emailAdd{tsmho@connect.ust.hk}
\author[a]{Xu-Hui Jiang,}
\emailAdd{xjiangaj@connect.ust.hk}
\author[a]{Tsz Hong Kwok,}
\emailAdd{thkwokae@connect.ust.hk}
\author[b,c]{Lingfeng Li,}
\emailAdd{lingfeng\_li@brown.edu}
\author[a,b]{Tao Liu}
\emailAdd{taoliu@ust.hk}
\affiliation[a]{Department of Physics, The Hong Kong University of Science and Technology, Hong Kong S.A.R., P.R.China}
\affiliation[b]{Jockey Club Institute for Advanced Study, The Hong Kong University of Science and Technology, Hong Kong S.A.R., P.R.China}
\affiliation[c]{Department of Physics, Brown University, Providence, RI, 02912, USA}
\abstract{Department of Physics, Brown University, Providence, RI, 02912, USA}
\abstract{As one of the hypothetical principles in the Standard Model (SM), lepton flavor universality (LFU) should be tested with a precision as high as possible such that the physics violating this principle can be fully examined. The run of $Z$ factory at a future $e^+e^-$ collider such as CEPC or FCC-$ee$ provides a great opportunity to perform this task because of the large statistics and high reconstruction efficiencies for $b$-hadrons at $Z$ pole. In this paper, we present a systematic study on the LFU test in the future $Z$ factories. The goal is three-fold. Firstly, we study the sensitivities of measuring the LFU-violating observables of $b\to c \tau \nu$, {\it i.e.}, $R_{J/\psi}$, $R_{D_s}$,  $R_{D_s^\ast}$ and  $R_{\Lambda_c}$, where $\tau$ decays muonically. For this purpose, we develop the strategies for event reconstruction, based on the track information significantly. Secondly, we explore the sensitivity robustness against detector performance and its potential improvement with the message of event shape or beyond the $b$-hadron decays. A picture is drawn on the variation of analysis sensitivities with the detector tracking resolution and soft photon detectability, and the impact of Fox-Wolfram moments is studied on the measurement of relevant flavor events. Finally, we interpret the projected sensitivities in the SM effective field theory, by combining the LFU tests of $b\to c \tau \nu$ and the measurements of $b\to s  \tau^+\tau^-$ and $b\to s  \bar{\nu} \nu$. We show that the limits on the LFU-violating energy scale can be pushed up to $\sim \mathcal{O} (10)$~TeV for $\lesssim \mathcal O(1)$ Wilson coefficients at Tera-$Z$.}

\maketitle
\flushbottom
\section{Introduction}
\label{sec:intro}

Lepton flavor universality (LFU), as one of the hypothetical principles in the Standard Model (SM), requires the leptons of all three generations to couple to gauge bosons universally. Any deviation from the LFU would be an unambiguously signal for physics beyond the SM. So, the LFU should be tested with a precision as high as possible such that the relevant physics can be fully explored. 

Given its significance in particle physics, the LFU has been tested in various experiments. One class of such tests involves the $b\to c\tau\nu$ transitions mediated by flavor changing charged current (FCCC). The relevant observables are usually defined as 
\begin{equation}
R_{H_c}\equiv \frac{{\rm Br}(H_b\to H_c\tau\nu)}{{\rm Br}(H_b\to H_c\ell\nu)}~,
\end{equation}
where $H_b$ and $H_c$ refer to exclusive $b$- and $c$-hadron states~\footnote{Throughout this paper, we take a notation implicitly including the relevant charge-conjugation mode.}. 
Since systematic errors from hadron physics tend to be canceled for the observables defined in such a way, any noteworthy deviation from the SM predictions in statistics may indicate the existence of LFU-violating new physics. In Tab.~\ref{tab:FCCC}, we have summarized SM prediction and experimental measurement for a set of $R_{H_c}$ observables. Notably, some anomalies in relation to $R_{H_c}$ were reported in the last years. Addressing these anomalies further strengthens the necessity and significance of performing dedicated and more complete LFU measurements.   

\begin{table}[h!]
\centering
\begin{tabular}{ccccc}
\hline
 & $H_b$ & $H_c$ & SM Prediction~\footnotemark & Experimental Average\\
\hline
$R_D$ &  $B^0$, $B^\pm$ & $D^0$, $D^\pm$ & $0.307$~\cite{Sakaki:2013bfa, Hu:2019bdf} & $0.340\pm 0.030$~\cite{Amhis:2019ckw} \\
$R_{D^\ast}$ & $B^0$, $B^\pm$ & $D^{\ast 0}$, $D^{\ast \pm}$& $0.253$~\cite{Sakaki:2013bfa, Hu:2019bdf} & $0.295\pm 0.014$~\cite{Amhis:2019ckw}\\
$R_{J/\psi}$ & $B_c$ & $J/\psi$ & $0.289$~\cite{Wang:2012lrc, Watanabe:2017mip, Asadi:2019xrc} & $0.71\pm 0.17\pm 0.18$~\cite{Aaij:2017tyk}\\
$R_{D_s}$ & $B_s$ & $D_s$ & $0.393$~\cite{Fan:2013kqa, Zhang:2022opp, Hu:2019bdf, Faustov:2012mt, Monahan:2017uby, Dutta:2018jxz, Soni:2021fky} & N/A\\
$R_{D_s^\ast}$ & $B_s$ & $D_s^\ast$ & $0.303$~\cite{Fan:2013kqa, Hu:2019bdf, Faustov:2012mt, Soni:2021fky} & N/A\\
$R_{\Lambda_c}$  & $\Lambda_b$ &  $\Lambda_c$ & $0.334$~\cite{Shivashankara:2015cta, Gutsche:2015mxa, Detmold:2015aaa, Dutta:2015ueb, Datta:2017aue} & $0.242\pm 0.076$~\cite{LHCb:2022piu}\\
\hline
\end{tabular}
\caption{SM prediction and experimental measurement for $R_{H_c}$ observables.}\label{tab:FCCC}
\end{table}
~\footnotetext{The calculation of $R_{J/\psi}$, $R_{D_s}$, $R_{D_s^\ast}$ and $R_{\Lambda_c}$ and the relevant references are shown in App.~\ref{app:calc}. The results listed in Tab.~\ref{tab:FCCC} are slightly different from those in the literatures, due to the update of form factor values or the variation of parameter setup.}

Future $Z$ factories, namely the $Z$-pole runs of next-generation $e^+e^-$ colliders~\cite{CEPCStudyGroup:2018ghi,Abada:2019zxq,Fujii:2019zll}, would provide a great opportunity for performing this task. Their advantages are generic, manifested as relatively high production rate and reconstruction efficiency of heavy flavored hadrons.

Consider first the expected $b$-hadron yields in Belle II, LHCb and two representative future $Z$ factories (see Tab.~\ref{tab:Bnum}). At Tera-$Z$, the statistics of $B^0/\bar{B^0}$ and $B^\pm$ are $\sim 1.2\times 10^{11}$, about twice as those in Belle II~\cite{Kou:2018nap}. However, for the heavier $B_s$/$\bar{B}_s$, the difference in statistics between the Tera-$Z$ and Belle II increases to nearly two orders of magnitude. The future $Z$ factories are thus especially suitable for studying flavor physics involving such heavy $b$-hadrons. Unlike Belle II and $Z$ factories, LHCb produces $b$-hadrons mainly through parton-level QCD processes. However, although the expected yields can be even larger at LHCb~\cite{Bediaga:2018lhg}, the event reconstruction efficiency is significantly limited by its noisy data environment.

\begin{table}
\centering
\begin{tabular}{cccccc}
\hline 
   & Belle~II & LHCb & Tera-$Z$ & $10 \times$Tera-$Z$   \\ 
\hline 
$B^0$, $\bar{B}^0$ & $5.3\times 10^{10}$ & $ 6\times 10^{13}$  & $1.2 \times 10^{11}$ & $1.2 \times 10^{12}$\\
$B^\pm$ & $5.6\times 10^{10}$ & $ 6\times 10^{13}$  & $1.2 \times 10^{11}$ & $1.2 \times 10^{12}$ \\
$B_s$, $\bar{B}_s$ & $5.7 \times 10^{8}$ & $ 2\times 10^{13}$  & $3.1\times 10^{10}$ & $3.1\times 10^{11}$ \\
$B_c^\pm$ & - & $ 4 \times 10^{11}$  & $1.8\times 10^8$ & $1.8\times 10^9$ \\
$\Lambda_b$, $\bar{\Lambda}_b$ & - & $ 2\times 10^{13}$  & $2.5\times 10^{10}$ & $2.5\times 10^{11}$ \\
\hline
\end{tabular}
\caption{Expected $b$-hadron yields in Belle~II, LHCb and the Tera-$Z$, $10 \times$Tera-$Z$ factories~\cite{Wang:2022nrm}. There is no statistics on the $B_c^\pm$ and $\Lambda_b/\bar{\Lambda}_b$ productions at Belle~II because of the limitation of energy threshold.  \label{tab:Bnum} }
\end{table}

The boosted kinematics of $b$-hadrons at $Z$-pole and the relatively clean environment for their production represent another set of advantages for the future $Z$ factories to measure the $b\to c\tau\nu$ transitions. The $b$ hadrons produced at $Z$-pole tend to be energetic. This feature weakens the multiple scattering of charged particles such as the ones from the $\tau$-lepton and $c$-hadron decays inside the tracker, improving their energy/momentum~\cite{Berger:2016vak} and motion direction~\cite{CEPCStudyGroup:2018ghi,Abada:2019zxq} resolutions. Moreover, the boosted particles tend to displace more before decay, which may further reduce the uncertainties of reconstructing their decay vertexes. Several recent studies~\cite{Descotes-Genon:2022qce,Descotes-Genon:2022gcp,Li:2022tov,Li:2022tlo,Aleksan:2021gii,Aleksan:2021fbx,Kamenik:2017ghi,Monteil:2021ith,Chrzaszcz:2021nuk,Qin:2017aju,Li:2018cod,Calibbi:2021pyh,Dam:2018rfz,Yu:2020bxh,Zheng:2020emi,Li:2020bvr,Amhis:2021cfy} have illustrated the potential of the future $Z$ factories in measuring the $\tau$-related physics.
Separately, the clean data environment can benefit the measurement of missing energy, a crucial observable for reconstructing the $b\to c\tau\nu$ events. With relatively few particles in final states, negligible pile-up effect and fixed $\sqrt{s}$ value, the measurement of missing energy is expected to be significantly improved at $Z$-pole~\cite{Li:2022tov}. In this paper, we will focus on the four representative measurements of $R_{H_c}$ listed in Tab.~\ref{tab:FCCC}: $R_{J/\psi}$, $R_{D_s}$, $R_{D_s^{\ast}}$, and $R_{\Lambda_c}$. Currently, the experimental constraints on these observables are either weak or unavailable.

From a broader perspective, the LFU can be tested also in the $b\to s\ell_3^{+}\ell_3^{-}$ transitions mediated by flavor changing neutral current (FCNC). Here $\ell_3^{\pm}$ denotes the charged leptons of all three generations. Different from the FCCC, the FCNC in the SM are loop-suppressed, with the leading contributions arising from electroweak (EW) penguin and box diagrams. So the width of the FCNC-mediated $b$-hadron decays is typically smaller than that of the FCCC-mediated ones by a factor $\sim\mathcal{O}(\alpha^2/16\pi^2)$.  
This fact has motivated the introduction of the LFU-violating observable 
\begin{equation}
R_{H_s}\equiv \frac{{\rm BR}(H_b\to H_s \mu^+ \mu^-)}{{\rm BR}(H_b\to H_s e^+ e^-)} 
\label{eq:RHsdefinition}
\end{equation}
which involves the first two generations of leptons only, where $H_b$ and $H_s$ stand for the exclusive $b$ and $s$ hadronic states. Interestingly, anomalies were reported in the LHCb measurements of $R_{K^{(*)}}$~\cite{Aaij:2017vbb}, where $H_b = B$ and $H_s = K^{(*)}$. If LFU is respected, $R_K$ and $R_{K^\ast}$ shall be close to one. However, the measurements indicate that $R_{K}$ and $R_{K^\ast}$ are both lower than this prediction~\cite{Aaij:2017vbb}, with a significance $\sim 2-3 \sigma$. 

The test of LFU firmly calls for the extension of FCNC measurements from $R_{H_s}$ to the $b\to s\tau^+\tau^-$ transitions since there is no known first principle that forbids large FCNC amplitudes in the third lepton generation. Such a measurement will benefit our understanding of the $R_{K^{(*)}}$ anomalies also. Moreover, some models addressing these anomalies predict an enhancement of the $b\to s \tau^+\tau^-$ transitions, such as the singlet-triplet model~\cite{Crivellin:2017zlb,Crivellin:2019dwb}. The measurements of the $b\to s\tau^+\tau^-$ transitions are highly challenging, given the complexity of reconstructing multiple-$\tau$ events. None of the $b\to s\tau^+\tau^-$ channels have been experimentally observed so far. However, the future $Z$ factories could perform the $b\to s\tau^+\tau^-$ measurements, as explored at detector level recently~\cite{Li:2020bvr}, with a precision sufficient for probing the SM predictions.

Besides $b\to s\ell_3^{+}\ell_3^{-}$, another class of FCNC measurements relevant to the LFU test involves the $b\to s\nu\bar{\nu}$ transitions. These measurements cannot be applied to probe the LFU violation directly since neutrino flavor is untagged at colliders. However, the inclusive signal rate contributed by the neutrinos of all three flavors is still relevant, which can yield an overall constraint on the possible LFU-violating couplings. Notably, neutrinos do not couple with gluons or photons directly. The $b\to s\nu\bar{\nu}$ processes receive weak radiative corrections only and thus enjoy a lower theoretical uncertainty for their SM predictions. Currently, the upper limits for the $b\to s\nu\bar{\nu}$ branching ratios are $\sim\mathcal{O}(10^{-4}  -10^{-5})$, not far from their SM predictions~\cite{ParticleDataGroup:2020ssz}.

Each of these FCCC and FCNC measurements provides an independent test of LFU in experiments. Any deviation in data from their SM predictions could be a hint or indication of the violation of this principle. Theoretically, the LFU-violating physics could either yield a signal correlating these observables or leave an imprint in only a subset of these measurements. For example, the $SU(2)$ gauge symmetry may relate the  $b\to s\nu\bar{\nu}$ amplitudes with those of $b\to c\tau \nu$ or $b\to s\tau \tau$ or both of them. More discussions about these issues can be found in Sec.~\ref{sec:EFT}. So the LFU measurements should be performed with a coverage broad enough and a precision as high as possible. The future $Z$ factories allow us to extend the existing measurements to the more challenging ones, which suffer from either a small production rate in Belle II or low reconstruction efficiency at LHCb, of heavy flavored hadrons. To demonstrate the potential capability of these machines in testing the LFU, in the paper, we will take a sensitivity interpretation in the SM Effective Field Theory (SMEFT), where the SM gauge symmetries are respected. We will focus on a subset of 6D operators which encode the LFU violation arising from the third generation only to converge the discussions. Especially, considering the possible hierarchy between the measurement scale and the new physics scale, the effects of renormalization of the relevant Wilson coefficients will be taken into account. 

This paper is organized as follows. In Sec.~\ref{sec:simulaton}, we introduce general strategies for our simulations and analyses. The analysis of measuring $R_{J/\psi}$ is taken in Sec.~\ref{sec:RJpsi}, while the ones for measuring $R_{D_s^{(\ast)}}$ and $R_{\Lambda_c}$ are performed in Sec.~\ref{sec:RDs} and~\ref{sec:RLambdac}, respectively. The sensitivity robustness against detector resolution and potential improvements from event shape for these analyses are then explored in Sec.~\ref{sec:Results}. We present the SMEFT interpretations for the projected sensitivities at the future $Z$ factories in Sec.~\ref{sec:EFT}, and finally conclude in Sec.~\ref{sec:conclusion}.

\section{Strategy for Event Simulation}
\label{sec:simulaton}

We use Pythia8~\cite{Sjostrand:2007gs} to simulate both signal and background events for the $R_{H_c}$ measurements. In each of them, two signal modes are involved, namely $H_b \to H_c \tau^+ \nu_\tau$ and $H_b \to H_c \mu^+ \nu_\mu$. The signal events of these two modes contribute as the mutual backgrounds also in their respective measurements. Signal samples are generated via the $Z\to b\bar{b}$ production at the $Z$-pole, forcing the $b$-hadrons (together with  $H_c$) to decay into the relevant states exclusively. These events are then reweighted according to the $d\Gamma/d q^2$ differential cross section obtained in App.~\ref{app:calc} to reproduce the correct kinematic distributions. Background samples are generated via the $Z\to b\bar{b}$ process also.

The detector effects are simulated using Delphes 3~\cite{deFavereau:2013fsa}. Given that the relative impact on the results is of percent level and hence tiny between the ILD~\cite{Chen:2017yel} and  IDEA concepts~\cite{Antonello:2020tzq}, we take the former detector profile as our benchmark in the analyses below. Notably, some features, such as particle identification (ID) efficiency and impact-parameter resolution for tracks, are not hardcoded in these profiles. As these features may play a crucial role in our analysis, we simulate them with a set of benchmark values and discuss the potential impacts of their variance in Sec.~\ref{ssec:trackingnoise}.

One such feature is muon ID. Our study relies on muon tagging significantly. The four  $R_{H_c}$ analyses are either based on the three-muon system or requesting at least one tagged muon. However, due to the comparable mass of $\pi^\pm$ with muons and their large multiplicity in hadronic final states~\cite{Yu:2021pxc}, the $\pi^\pm$ could be misidentified as muons and yield visible negative impact for the $B_c$ reconstruction. So we will consider this effect in our analysis. Concretely, we assume the muon mis-ID probability $\epsilon_{\mu\pi}$ to be $1\%$~\cite{Yu:2021pxc}, an optimal value which is expected to achieve at FCC-$ee$~\cite{Abada:2019lih} and CEPC~\cite{CEPCStudyGroup:2018ghi} by the time of their operation. As for the ID for charged hadrons ($e.g.$, $\pi/K$, $K/p$, and $\pi/p$ mis-ID), it is less relevant for reconstructing the $H_c$ resonances. So we will not simulate its effects directly. 
At last, to simulate the effects of finite spatial resolution, we smear the decay vertex of particles by turning on independent and isotropic Gaussian noise in the tracker. 
Such smearing is also applied to the impact parameter of the muon tracks, which arise from (semi-)leptonic hadron and $\tau$ decays. We set the overall noise level to be $10$~$\mu$m, a typical tracker resolution suggested in~\cite{CEPCStudyGroup:2018ghi,Abada:2019lih}.

The background analysis is highly involved for the $R_{H_c}$ measurements. Because of the complexity of the $b$-hadron decay chains, it is not realistic to make an exhaustive list of the backgrounds. But it is beneficial to understand the general background sources and their characteristics first. Motivated by this, we classify these backgrounds into five categories: inclusive, cascade, combinatoric, muon mis-ID, and fake-$H_c$-resonance backgrounds.

\paragraph{Inclusive backgrounds}
We refer to $H_b\to H_c\tau(\mu)\nu +X$ as ``inclusive backgrounds''. Here $H_b$ decays semi-leptonically. $X$ arises from either resonant $H_c^{\ast}$ decay or non-resonant contribution. In the simulation, any non-signal $b$-hadron events, if containing the $H_c+\mu$ produced via semileptonic $b$-hadron decays at the truth level, will be recognized as inclusive backgrounds.

\paragraph{Cascade backgrounds}  
We refer to $H_b\to H_c\tau(\mu)\nu +X$ as ``cascade backgrounds''. Here $H_b$ decays hadronically. 
In the simulation, any non-signal $b$-hadron events, if containing the $H_c+\mu$ produced not via semileptonic $b$-hadron decay at truth level, will be recognized as the cascade backgrounds.

\paragraph{Combinatoric backgrounds}
We refer to $H_c\tau(\mu)\nu +X$ as ``combinatoric backgrounds''. Here $H_c$ and $\tau(\mu)$ do not share a parent particle at the truth level. In the simulation, any reconstructed $b$-hadron events, if containing the $H_c+\mu$ but not identified as the inclusive and cascade backgrounds, will be recognized as the combinatoric backgrounds.

\paragraph{Muon mis-ID backgrounds}
We refer to $H_c \mu_\pi +X$ as ``muon mis-ID backgrounds''. Here $\mu_\pi$ denotes the muon misidentified from pion.
In the simulation, any $H_c \pi +X$ events will be recognized as the mis-ID background, weighted by the mis-ID probability $\epsilon_{\mu\pi}=1\%$ as mentioned above.

\paragraph{Fake $H_c$ backgrounds}
We refer to $H_{c,F} \mu +X$ as ``fake $H_c$ backgrounds''. Here $H_{c,F}$ denotes the fake $H_c$ resonance, with the latter decaying as: $J/\psi\to\mu^+\mu^-$, $D_s^- \to K^+ K^- \pi^-$, or $\Lambda_c^- \to \bar{p} K^+ \pi^- $ in this study. 
These backgrounds represent the chance that the remnants for reconstructing $H_c$ are not from $H_c$ decays at the truth level. 
In the analysis, they appear as a continuous distribution of the reconstructed $m_{H_c}$. 
A good width resolution of resonance is thus essential for suppressing these backgrounds. In practice, the resonance width is determined by the resolution of the tracking system, given $\Gamma_{H_c}\lesssim \mathcal{O}(\rm keV) \ll \Delta_{\rm track}$, where $\Delta_{\rm track}$ denotes the tracker smearing effect.  
We can estimate the level of these backgrounds from the relevant LHCb studies~\cite{Aaij:2017tyk,Aaij:2020hsi,Aaij:2017svr}. As summarized in Tab.~\ref{tab:fakeresonance}, the rations of the $H_c$ events and the continuous backgrounds in the resonant bin for the reconstructed $m_{H_c}$ are at most a few percent. The reconstructed resonance widths are expected to be further improved at the future $Z$ factories~\cite{Aaij:2017tyk,Aaij:2020hsi,Aaij:2017svr}. Furthermore, the fake $H_c$ background sizes can easily be extrapolated by sideband $m_{H_c}$ distributions. So the effect of this type of background can be safely neglected in $R_{H_c}$ precision projections.  

\begin{table}
\begin{center}
\begin{tabular}{cccccc}
\hline
$H_c$ & fake $H_c$ ratio & $H_c$ width$_{\rm Ref.}$ & $H_c$ width$_{Z~{\rm factory}}$ & Estimated\\
\hline
$J/\psi(\to \mu^+\mu^-)$ & $4.5\%$ & $9.1$~MeV~\cite{Aaij:2017tyk} & $8.3$~MeV &  $\lesssim 2.3\%$ \\
$D_s^-(\to \phi\pi^-)$ & $3.8\%$ & $7.6$~MeV~\cite{Aaij:2020hsi} & $6.1$~MeV & $\lesssim 3.8\%$\\
$\Lambda_c^-(\to \bar{p}K^+\pi^-)$ & $1.5\%$& $5.5$~MeV~\cite{Aaij:2017svr} & $4.5$~MeV & $\lesssim 0.3\%$\\
\hline
\end{tabular}
\end{center}
\caption{Estimation of the fake $H_c$ backgrounds. The first column represents the estimated yield ratio of the fake $H_c$ background over the real $H_c $ resonance from the reference studies. The second and third columns are the reconstructed $H_c$ resonance standard deviation values of the reference and our study, respectively. The last one is the estimated yield ratio of the fake $H_c$ over the real $H_c$ resonance contributing to our studies. }
\label{tab:fakeresonance}
\end{table}

\section{Measurement of $R_{J/\psi}$}
\label{sec:RJpsi}

\subsection{Method}
\label{subsec:RJpsi}

\begin{figure}[]
	\centering
	\includegraphics[width=10cm]{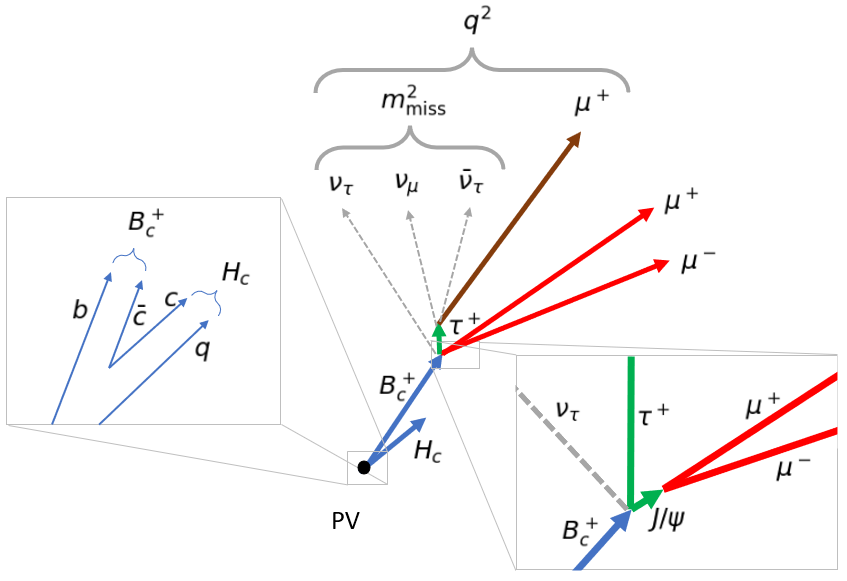}
	\caption{Schematic of the $B_c^+\to J/\psi \tau^+ \nu_\tau$ process. Since $J/\psi$ is short-lived, its decay vertex can serve as a good approximation of the $B_c^+$ decay vertex. Additionally, the $c$ quark paired produced with the $B_c^+$ is hadronized to another $c$-hadron ($H_c$), which tends to move along with the $B_c^+$.  } 
	\label{fig:cartoons2}
\end{figure}

\begin{figure}[h!]
	\centering
	\includegraphics[width=0.95\textwidth]{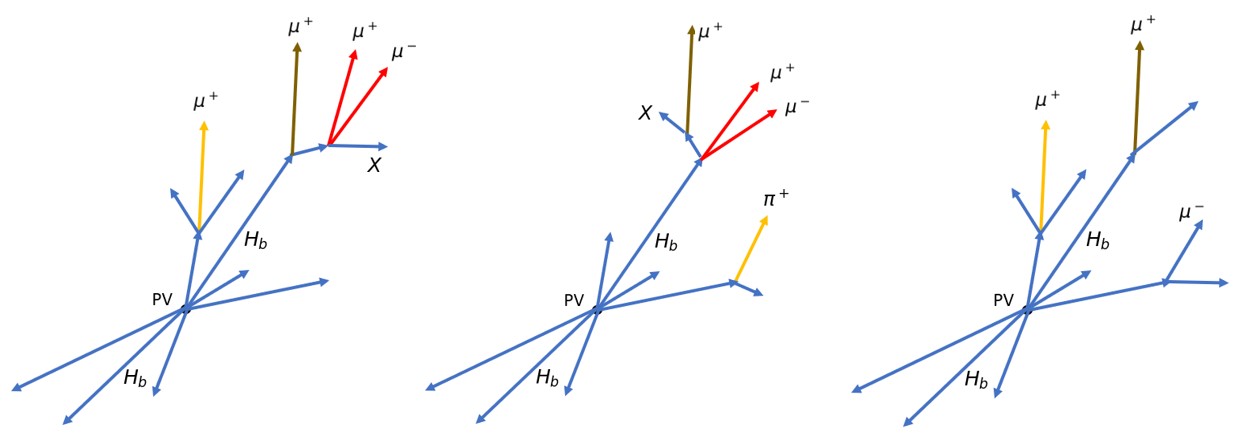}
	\caption{Schematics of the universal backgrounds in the $R_{J/\psi}$ measurement.  \textbf{Left:} The typical topology for the inclusive backgrounds and the combinatoric backgrounds, where $B_c^+$ is reconstructed combining muons produced by the $J/\psi$ (red), and the unpaired muon from semi-leptonic $H_b$ decay (brown) or irrelevant particle decay (orange), respectively. \textbf{Middle:} The typical topology for the cascade backgrounds and the Mis-ID backgrounds, where $B_c^+$ is reconstructed combining the muons decayed from $J/\psi$ (red), and the unpaired muon from intermediate hadron decay (brown) and pion misidentification (orange), respectively. \textbf{Right:} The typical topology for the fake $H_c$ backgrounds, where the muons which do not share a parent particle (brown and orange) are used to reconstruct $J/\psi$.}
	\label{fig:cartoons1}
	
\end{figure}

To measure $R_{J/\psi}$, we consider the exclusive $B_c^+$ decays, $i.e.$, $B_c^+ \to J/\psi(\to\mu^+ \mu^-)\mu^+\nu_\mu$ and $B_c ^+\to J/\psi(\to\mu^+ \mu^-)\tau^+(\to \mu^+\nu_\mu\bar{\nu}_\tau)\nu_\tau$, as the signals. Both signal modes contain $3\mu$ in their final states. The schematic of the $B_c^+\to J/\psi \tau^+ \nu_\tau$ process is shown in Fig.~\ref{fig:cartoons2}. The same decay modes have been considered in the $R_{J/\psi}$ measurement at LHCb also~\cite{Aaij:2017tyk}. We also show the schematics of several topologies for the universal backgrounds in Fig.~\ref{fig:cartoons1}. Below are a set of cuts applied to preselect such events. 
\begin{itemize}

\item {\bf The $3\mu$ selection.}  The events with exactly three muon tracks ($p_T>0.1$~GeV), and at least two of them sharing the same vertex, are selected. 

\item {\bf The $J/\psi$ selection.}  Two of the three muons need to be oppositely charged. Their momentum satisfies $|\vec{p}|>2.5$~GeV. The leading transverse momentum must be $>0.75$~GeV, while their total $p_T$ must be $> 1$~GeV. These two muons form a common vertex, with its distance to the primary vertex (PV) $> 0.1$~mm. Besides, these two muons must have an invariant mass with $|m_{\mu^+\mu^-}-m_{J/\psi}|<27.5$~MeV for them to be considered as the $J/\psi$ decay products. 

\item {\bf The $B_c^+$ selection.} We divide the space into signal and tag hemispheres with a plane perpendicular to the displacement of the reconstructed $J/\psi$. The $J/\psi$ vertex appears in the signal hemisphere. The unpaired third muon ($\mu_3$) appears in the signal hemisphere also and has $p_T>0.375$~GeV and $|\vec{p}|>1.5$~GeV.  The $3\mu$ system needs to have an invariant mass smaller than $m_{B_c^+}$. 
\end{itemize}
The Tera-$Z$ yields for the preselected signals and the backgrounds are summarized in Tab.~\ref{tab:Jpsi_yields}. The requirement of narrow $J/\psi$ and $B_c^+$ reconstruction excludes most of the backgrounds except the inclusive ones, as expected.

\begin{table}[h!]
\begin{center}
\fontsize{9pt}{10.8pt}\selectfont 
\begin{tabular}{cccccc}	
	\hline
	Channel
	& Events at Tera-$Z$ 
	& $N(3\mu)$
	& $N(J/\psi)$ 
	& $N(B_c^+)$ 
	& Total eff.\\
	\hline
	
	$B_c^+\to J/\psi \tau^+\nu_\tau$
	& $9.83\times 10^3$
	& $6.53\times 10^3$
	& $3.83\times 10^3$
	& $3.08\times 10^3$
	& $31.34\%$\\
	
	$B_c^+\to J/\psi \mu^+\nu_\mu$
	& $2.39\times 10^5$
	& $1.63\times 10^5$
	& $9.66\times 10^4$
	& $8.40\times 10^4$
	& $35.13\%$\\
	
	Inclusive bkg.
	& $1.27\times 10^4$
	& $8.20\times 10^3$
	& $5.29\times 10^3$
	& $3.90\times 10^3$
	& $30.63\%$\\

	Cascade bkg.
	& $1.81\times 10^4$
	& $4.89\times 10^3$
	& $3.32\times 10^3$
	& $1.84\times 10^3$
	& $10.15\%$\\
	
	Combinatoric bkg.
	& $4.64\times 10^7$
	& $3.93\times 10^7$
	& $2.66\times 10^7$
	& $7.78\times 10^4$
	& $0.17\%$\\

	Mis-ID bkg.
	& $\epsilon_{\mu\pi}\times 1.45\times 10^9$
	& $\epsilon_{\mu\pi}\times 1.03\times 10^9$
	& $\epsilon_{\mu\pi}\times 6.96\times 10^8$
	& $\epsilon_{\mu\pi}\times 1.10\times 10^8$
	& $7.61\%$\\
	
	\hline
\end{tabular}
\caption{Tera-$Z$ yields for the preselected signals and the backgrounds in the $R_{J/\psi}$ measurement. The preselection criteria are defined in the text.} \label{tab:Jpsi_yields}
\end{center}
\end{table}

\begin{figure}[h!]
	\centering
	\includegraphics[width=7.2cm]{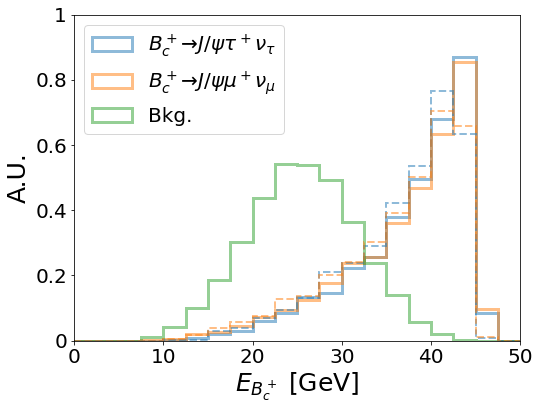}	
	\includegraphics[width=7cm]{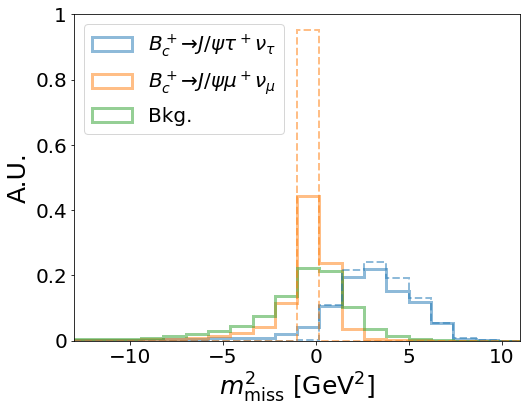}   
	\includegraphics[width=7cm]{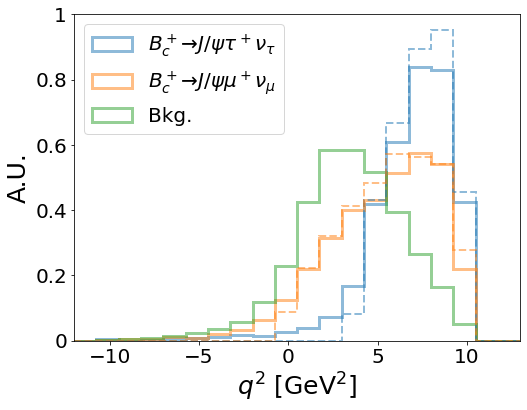}  
	\includegraphics[width=7cm]{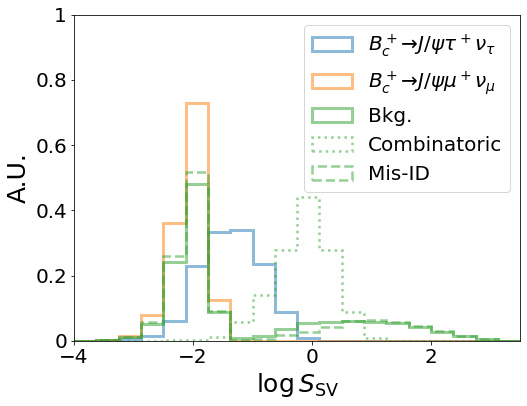}	
	\caption{Distributions of the reconstructed $E_{B_c^+}$, $q^2$, $m_{\text{miss}}^2$ and $\log{S_{\rm SV}}$ in the $R_{J/\psi}$ measurement. The solid and dashed lines represent the simulated and truth-level messages respectively. }
	\label{fig:JpsipB}
\end{figure}

The preselected events are then subjected to the $B_c^+$ reconstruction. Such a task is highly involved since the signal events contain at least one neutrino. For reconstructing the four-momentum of $B_c^+$  ($p_{B_c^+}$), thus we will take several approximations. Firstly, as $J/\psi$ decays promptly, we will use the $J/\psi$ decay vertex to approximate the $B_c^+$ decay vertex and define its displacement from the PV as the $\vec{p}_{B_c^+}$ direction. Secondly, we calculate the total energy of the particles inside the signal hemisphere $E_{\rm sig}$ with the relation
\begin{equation}
E_{\rm sig} = \frac{m_{\rm tag}^2 + m_Z^2-m_{\rm sig}^2}{2m_Z}~,
\label{eq:reconstruction}
\end{equation}
where $m_{\rm sig}$ and $m_{\rm tag}$ are the invariant masses of visible particles in the signal and tag hemispheres respectively. This relation is generated by applying the energy- and momentum-conservation conditions to the two-body decay of a $Z$ boson at rest~\cite{Li:2022tov}. No missing particles are involved in this case. To calculate $E_{\rm sig}$, we have mimicked these two bodies with the collection of particles in the signal and tag hemispheres, and replaced their invariant masses with $m_{\rm sig}$ and $m_{\rm tag}$. Clearly, this relation becomes exact  only if no neutrinos have been produced. With this calculation, the $B_c^+$ energy $E_{B_c^+}$ is reconstructed as 
\begin{equation}
E_{B_c^+}=E_{\rm sig}-\sum_{i\in \text{sig-hem}} E_{i}+E_{J/\psi}+E_{\mu_3}~,
\end{equation}
where the index $i$ goes over all visible particles inside the signal hemisphere. With the direction message of $\vec{p}_{B_c^+}$ and the value of $E_{B_c^+}$, the four-momentum $p_{B_c^+}$ can be completely determined using the $B_c^+$ on-shell condition. We show the distributions of the reconstructed $E_{B_c^+}$ for the $B_c^+ \to J/\psi \tau^+ \nu_\tau$ and $B_c^+ \to J/\psi \mu^+ \nu_\mu$ signals and their common backgrounds in Fig.~\ref{fig:JpsipB}. A sharp edge at $m_Z/2$ can be seen for the signal distributions where the $Z\to b\bar{b}$ events tend to be hadronized into two $b$-hadrons only.

With the reconstructed four-momentum of $B_c^+$, we are able to define two Lorentz-invariant observables:
\begin{equation}
q^2 \equiv (p_{B_c^+} - p_{J/\psi})^2~, \quad m_{\rm miss}^2 \equiv (p_{B_c^+} - p_{J/\psi} - p_{\text{unpaired}~\mu})^2~. 
\end{equation}
These two observables are visualized in Fig.~\ref{fig:cartoons2}. For the SM events, they measure the mass of off-shell $W$ boson and produced neutrinos, respectively. Similar observables can be defined for the other $R_{H_c}$ measurements. As $q^2 $ and $m_{\rm miss}^2$ receive contributions from more neutrinos for the signal events of $B_c^+\to J/\psi\tau^+\nu_\tau$, compared to the ones of $B_c^+\to J/\psi\mu^+\nu_\mu$, their values and variances tend to be bigger in the former case. This feature is important since the signal events of these two modes can serve as the backgrounds mutually in their measurements. Finally we have the reconstruction errors of $q^2$ and $m_{\rm miss}^2$:  1.88(1.80)~GeV$^2$ and 1.90(1.61)~GeV$^2$. Here the numbers outside and inside the brackets are for the $\tau$- and $\mu$-modes, respectively. Other than the reconstructed $B_c^+$ kinematics, the signal events of the $\tau$- and $\mu$-modes can be further separated using the message on $\tau$ lepton displacement. The lifetime of $\tau$ lepton is relatively long. It may travel a detectable distance before its decays to other particles. The minimal distance ($S_{\rm SV}$, in the unit of mm) between the $\mu_3$ track and the secondary vertex (SV) ($i.e.$, the $B_c^+$ decay vertex) thus can be applied to discriminate the signal events of $B_c ^+\to J/\psi\tau^+ \nu_\tau$ from the $B_c ^+\to J/\psi\mu^+ \nu_\mu$ ones. We demonstrate these features in Fig.~\ref{fig:JpsipB}.

\begin{figure}[h]
	\centering
	\includegraphics[width=7.5 cm]{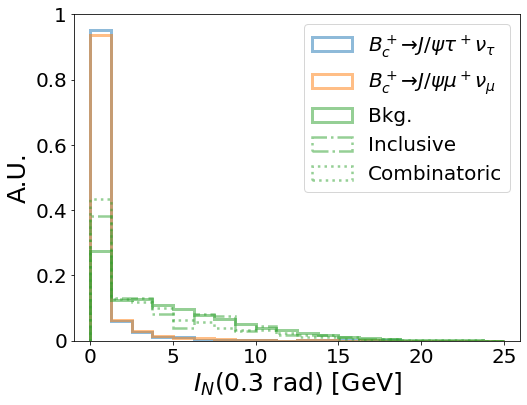} \hspace{-0.6 cm}	
	\caption{ Distributions of $I_N(0.3$~rad$)$ in the $R_{J/\psi}$ measurement. }
	\label{fig:Jpsiisolation}
\end{figure}

The observables introduced above can also separate the signals of different modes from the universal backgrounds to various extents. To further suppress these backgrounds, we may use the message on the signal $b$-hadron ($B_c^+$ here) isolation. Different from the reconstructed background events, the signal $B_c^+$ mesons tend to be isolated. Thus we can introduce the isolation observables $I_N(\Omega)$ and $I_T(\Omega)$ to facilitate the selection of the signal events. Here $I$ is the total energy of some specific particles within a cone around the reconstructed momentum of $B_c^+$. $\Omega$ denotes the angular size of this cone. $N$ represents neutral particles such as neutral hadrons ($I_H$) and photons ($I_\gamma$), while $T$ represents tracks which can be either from the PV ($I_{T,{\rm PV}}$) or away from the PV ($I_{T,{\rm dis}}$). This feature is demonstrated in Fig.~\ref{fig:Jpsiisolation} with $I_N(0.3$~rad$)$.

To optimize the sensitivity of measuring $R_{J/\psi}$, we apply the tool of Boosted Decision Tree (BDT) in this analysis and the subsequent ones for the $R_{D_s^{(*)}}$ and $R_{\Lambda_c}$ measurements. We include more observables on the track impact parameter other than the ones discussed above and some observables used in~\cite{LHCb:2020cyw} as the BDT discriminators. The BDT classifier is trained in a three-class mode to address its two signal patterns. The full list of the discriminators is summarized below:
\begin{itemize}
\itemsep-0.5em 
\item Kinematics of the three-muon system:
	\begin{itemize}
	\itemsep-0.25em 
	\item Invariant mass $m_{3\mu}$
	\item Energy and momentum of the reconstructed $J/\psi$ and the unpaired muon $\mu_3$: $E_{J/\psi}$, $|\vec{p}_{J/\psi}|$, $E_{\mu_3}$, $|\vec{p}_{\mu_3}|$
	\end{itemize} 

\item Observables of the reconstructed $B_c^+$:
	\begin{itemize}
	\itemsep-0.25em 
	\item Energy and momentum of the reconstructed $B_c^+$: $E_{B_c^+}$, $|\vec{p}_{B_c^+}|$
	\item Lorentz-invariant observables: $m_{\rm miss}^2$, $q^2$
	\end{itemize}

\item Vertex information:
	\begin{itemize}
	\itemsep-0.25em 
	\item Minimal distance between the $B_c^+$ (or $J/\psi$) decay vertex and the $\mu_3$ track ($S_{\rm SV}$)
	\item Minimal distance between the $\mu_3$ track and its closest track
	\item Minimal distance between the reconstructed $J/\psi$ trajectory and its closest track
	\item Distance between the $J/\psi$ decay vertex and the PV
	\end{itemize}

\item Isolation observables of $B_c^+$: 
	\begin{itemize}
	\itemsep-0.25em 
	\item Neutral particles: $I_N(0.3  \ {\rm rad})$, $I_N(0.6 \ {\rm rad})$
	\item Neutral hadrons: $I_{H}(0.3 \ {\rm rad})$, $I_{H}(0.6 \ {\rm rad})$
	\item Photons: $I_\gamma(0.3 \ {\rm rad})$, $I_\gamma(0.6 \ {\rm rad})$
	\item Charged particles: $I_T(0.3 \ {\rm rad})$, $I_T(0.6 \ {\rm rad})$
	\item Tracks from the PV: $I_{T,\rm PV}(0.3 \ {\rm rad})$, $I_{T,\rm PV}(0.6 \ {\rm rad})$ 
	\item Tracks not from the PV: $I_{T,\rm dis}(0.3 \ {\rm rad})$, $I_{T, \rm dis}(0.6 \ {\rm rad})$ 
	\end{itemize} 
	
\item Impact parameter of the tracks in the signal hemisphere: 
	\begin{itemize}
	\itemsep-0.25em 
	\item Maximum and sum of transverse impact parameters
	\item Maximum and sum of longitudinal impact parameters
	\end{itemize} 
	
\item Some other discriminators~\cite{LHCb:2020cyw}:
	\begin{itemize}
	\itemsep-0.25em 
	\item $J/\psi\mu^+$ momentum transverse to the $B_c^+$ moving direction: $p_{\perp}(J/\psi\mu^+)$	
	\item Corrected mass: $m_{\rm corr} = \sqrt{m^2(J/\psi\mu^+)+p_\perp^2(J/\psi\mu^+)} + p_\perp(J/\psi\mu^+)$
	\end{itemize} 
\end{itemize}

\subsection{Results}

\begin{figure}[h!]
\centering
	\includegraphics[width=7.5 cm]{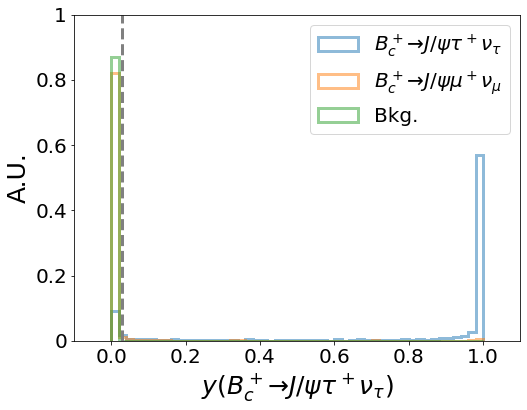}	
		\includegraphics[width=7.5 cm]{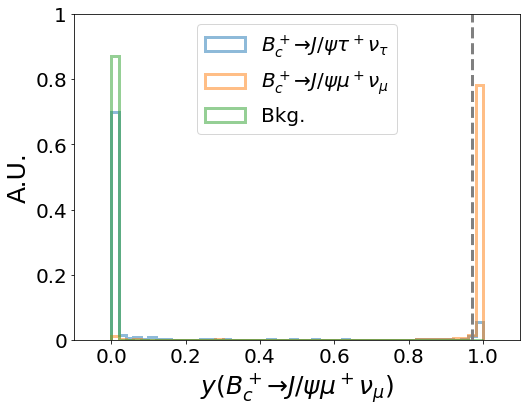}	
\caption{Distributions of BDT response in favor of $B_c^+\to J/\psi \tau^+ \nu_\tau$ ($y_{J/\psi}^\tau$) and $B_c^+\to J/\psi \mu^+ \nu_\mu$ ($y_{J/\psi}^\mu$) in the $R_{J/\psi}$ measurement. The vertical dashed lines represent optimal thresholds for sensitivity analysis. }
\label{fig:Jpsi_BDT_Response}
\end{figure}

In Fig.~\ref{fig:Jpsi_BDT_Response}, we show the distributions of BDT response in favor of $B_c^+\to J/\psi \tau^+ \nu_\tau$ and $B_c^+\to J/\psi \mu^+ \nu_\mu$ in the $R_{J/\psi}$ measurement. The two classes of signal events also serve as the mutual backgrounds of their measurements. Unless otherwise specified, in this paper the BDT thresholds are always defined to be the ones maximizing the statistical analysis sensitivity. We summarize the event counts in the relevant signal regions in Tab.~\ref{tab:Jpsi_Final_Yield} and the expected precisions of measuring $R_{J/\psi}$ at Tera-$Z$ and $10\times$Tera-$Z$ in Tab.~\ref{tab:Jpsi_BDT_yields} accordingly. Essentially, the precisions of measuring $R_{J/\psi}$ are limited by the relatively low counts of the $B_c^+\to J/\psi \tau^+ \nu_\tau$ events. Signal events are recognized to be of high- or low-$q^2$ by comparing their reconstructed $q^2$ with the 7.15~GeV$^2$ reference value~\cite{Aaij:2017tyk}. As shown in Tab.~\ref{tab:Jpsi_BDT_yields}, in the high $q^2$ region where a larger $R_{J/\psi}$ has been predicted, a better precision can be achieved compared to the low $q^2$ region. At last, we point out that the relatively high $S/B$ ratios in all scenarios ensure the robustness of the sensitivity analysis of measuring $R_{J/\psi}$ against the potential systematic uncertainties.

\begin{table}[h!]
\begin{center}
\begin{tabular}{c|c|c}
\hline
	& $y_{J/\psi}^\tau \geq0.03  ~ \cap ~  y_{J/\psi}^\mu<0.97$ & $y_{J/\psi}^\tau<0.03   ~ \cap ~ y_{J/\psi}^\mu \geq 0.97$ \\
	\hline
	$B_c^+\to J/\psi\tau^+\nu_\tau$ &
	$2.68\times 10^3$ &
	$2.14\times 10^2$
	\\
	$B_c^+\to J/\psi\mu^+\nu_\mu$ &	
	$4.30\times 10^3$  &
	$7.62\times 10^4$
	\\
	Inclusive bkg. & 
	$3.17\times 10^2$ &
	$4.08\times 10^2$
	\\
	Cascade bkg. &
	$6.21\times 10^2$ &
	$8.87\times 10^1$
	\\
	Combinatoric bkg. &
	$2.04\times 10^3$ &
	$2.66\times 10^2$
	\\
	Mis-ID bkg. &
	$\epsilon_{\mu\pi}\times 2.09\times 10^5$ &
	$\epsilon_{\mu\pi}\times 3.30\times 10^4$\\
	\hline

\end{tabular}
\caption{Event counts in the signal regions of $B_c^+\to J/\psi \tau^+ \nu_\tau$ and $B_c^+\to J/\psi \mu^+ \nu_\mu$ for the $R_{J/\psi}$ measurement at Tera-$Z$. 
 \label{tab:Jpsi_Final_Yield}}

\end{center}
\end{table}

\begin{table}[h!]
\begin{center}
\begin{tabular}{c|cc|cc|c}	
	\hline
	\hline
	\multirow{2}{*}{$q^2$ range} &
	\multicolumn{2}{c|}{$B_c^+\to J/\psi\tau^+\nu_\tau$} &
	\multicolumn{2}{c|}{$B_c^+\to J/\psi\mu^+\nu_\mu$} &
	$R_{J/\psi}$\\
	 & 
	Rel. precision & $S/B$ & 
	Rel. precision & $S/B$ & 
	Rel. precision \\ 
	\hline
	\hline
	\multirow{2}{*}{$q^2<7.15$~GeV$^2$}
	& $8.19\times 10^{-2}$ & \multirow{2}{*}{$0.18$} 
	& $5.18\times 10^{-3}$ & \multirow{2}{*}{$48.80$} & $8.20\times 10^{-2}$\\
	& $(2.59\times 10^{-2})$ &  
	& $(1.64\times 10^{-3})$ & 
	& $(2.59\times 10^{-2})$
	\\
	\hline
	\multirow{2	}{*}{$q^2 \geq 7.15$~GeV$^2$}
	& $4.56\times 10^{-2}$ & \multirow{2}{*}{$0.47$} 
	& $6.93\times 10^{-3}$ & \multirow{2}{*}{$96.27$} & $4.61\times 10^{-2}$\\
	& $(1.44\times 10^{-2})$ &  
	& $(2.19\times 10^{-3})$ & 
	& $(1.46\times 10^{-2})$
	\\
	\hline	
	\multirow{2}{*}{Full $q^2$}
	& $4.23\times 10^{-2}$ & \multirow{2}{*}{$0.29$} 
	& $4.15\times 10^{-3}$ & \multirow{2}{*}{$58.31$} & $4.25\times 10^{-2}$\\
	& $(1.34\times 10^{-2})$ &  
	& $(1.31\times 10^{-3})$ & 
	& $(1.35\times 10^{-2})$
	\\
	\hline		
\end{tabular}
\caption{Expected BDT (relative) precisions of measuring $R_{J/\psi}$ at Tera-$Z$ ($10\times$Tera-$Z$). 
 \label{tab:Jpsi_BDT_yields}}
\end{center}
\end{table}

\section{Measurement of $R_{D_s^{(\ast)}}$}
\label{sec:RDs}

\subsection{Method}

\begin{figure}[]
	\centering
	\includegraphics[width=10cm]{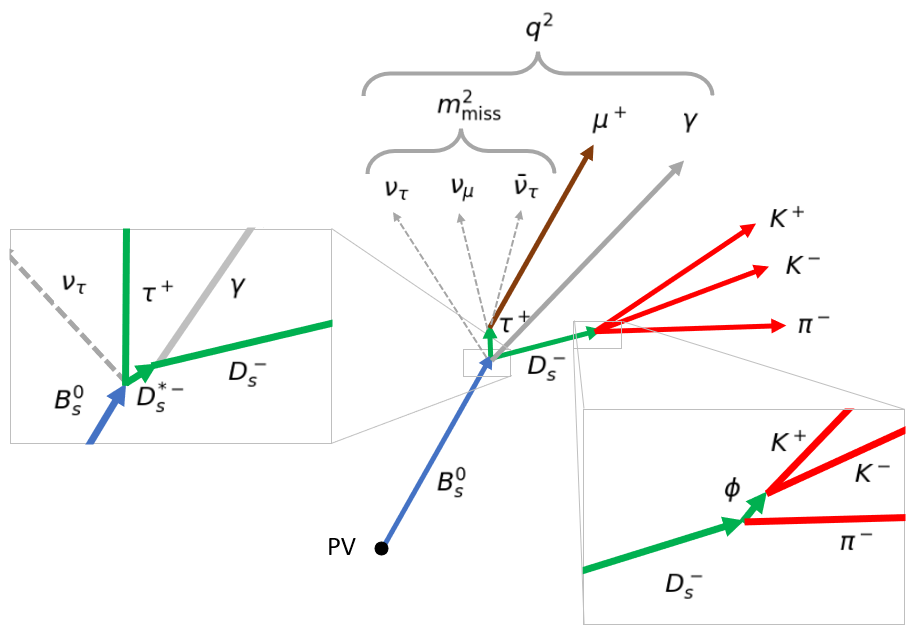}
	\caption{Schematic of the $B_s^0\to D_s^{*-}\tau^+\nu$ process. $D_s^{*-}$ decays to $D_s^-$ with extra photon. Compared to that of $J/\psi$ in the $R_{J/\psi}$ measurement, the lifetime of $D_s^{-}$ here is longer. }
	\label{fig:cartoons3}
\end{figure}

To measure $R_{D_s^{(*)}}$, we consider the exclusive $B_s^0$ decays, $i.e.$, $B_s^0\to D_s^-\mu^+\nu_\mu$ and $B_s^0\to D_s^-\tau^+\nu_\tau$ with $D_s^- \to \phi(\to K^+ K^-)\pi^-$, as the signals. All signal modes contain $K^+K^- \pi^-\mu^+$ in their final states. The schematic of the $B_s^0\to D_s^{*-}\tau^+\nu_\tau$ process is shown in Fig.~\ref{fig:cartoons3}. Below are a set of cuts applied to preselect such events. 

\begin{itemize}

\item {\bf The $K^+K^- \pi^-\mu^+$ selection.} 
The events with two oppositely charged kaon tracks, one charged pion track sharing a secondary vertex, and exactly one muon track with a charge opposite to the identified pion track are selected. All tracks need to have $p_T>0.1$~GeV.

\item {\bf The $D_s^-$ selection.} The two kaons should satisfy $|m_{K^+K^-}-m_\phi|<12$~MeV, with the displacement of their vertex from the PV being greater than 0.5~mm. Moreover, we require the reconstructed $K^+K^-\pi^-$ system to have $|m_{K^+K^-\pi^-}-m_{D_s}|<25$~MeV. The $D_s$ trajectory is inferred from the system's momentum $p_{K^+K^-\pi^-}$ and its vertex. The minimum distance between the reconstructed $D_s$ trajectory and any other tracks (except the muon one) needs to be  $>0.02$~mm. 

\item {\bf The $B_s^0$ selection.} We divide the space into signal and tag hemispheres with a plane perpendicular to the displacement of the reconstructed $D_s^-$. The $D_s^-$ vertex appears in the signal hemisphere. The muon track must appear in the signal hemisphere, having $p_T>1.2$~GeV and a minimal distance greater than 0.02~mm from all tracks except the reconstructed $D_s$ trajectory. The $K^+K^-\pi^-\mu^+$ system needs to have an invariant mass smaller than $m_{B_s}$.

\end{itemize}
The Tera-$Z$ yields for the preselected signals and the backgrounds are summarized in Tab.~\ref{tab:Ds_yields}. The requirement of narrow $D_s^-$ and $B_s^0$ resonances excludes most of the backgrounds except the inclusive ones, as expected.

\begin{figure}[h!]
	\centering
	\includegraphics[width=7.15cm]{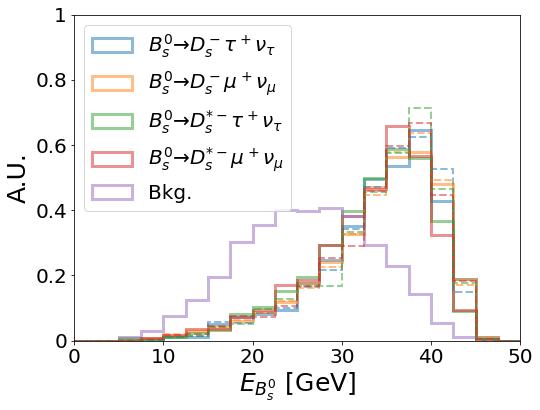}
	\includegraphics[width=6.8cm]{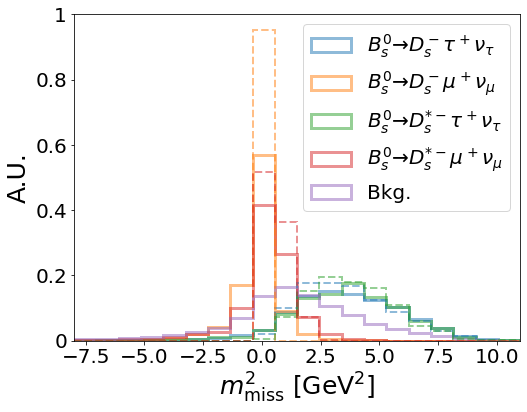}
	\includegraphics[width=7cm]{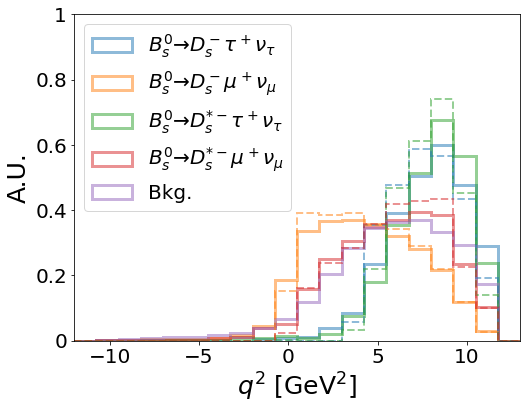}
	\includegraphics[width=7.15cm]{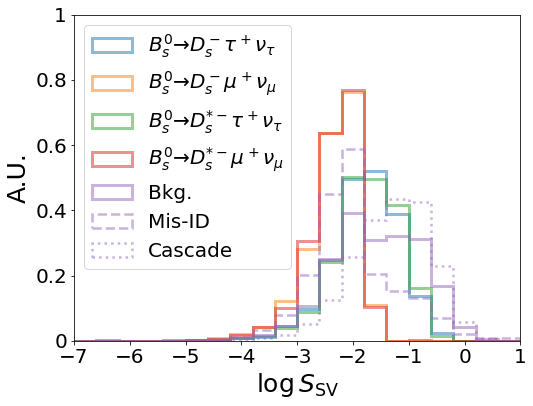}
	\caption{Distributions of the reconstructed $E_{B_s^0}$, $q^2$, $m_{\text{miss}}^2$ and $\log{S_{\rm SV}}$ in the $R_{D_s^{(*)}}$ measurement. The solid and dashed lines represent the simulated and truth-level values, respectively.}	
	\label{fig:DspB}
\end{figure}

The $B_s^0$ four-momentum can be reconstructed using the method introduced in Subsec.~\ref{subsec:RJpsi}. However, the $D_s^-$ decay vertex does not approximate the $B_s^0$ one well, as shown in Fig.~\ref{fig:cartoons3}, due to its macroscopic $D_s^-$ decay length. 
So we determine the $B_s^0$ decay vertex instead as the point on the $D_s^-$ track closest to the muon track. Here the $D_s^-$ tack is deduced from its decay vertex and momentum. Then, the $B_s^0$ four-momentum gets reconstructed by combining its displacement from the PV, total energy~\footnote{As a universal treatment, the energy of the $D_s^{*}$ photon has not been included in the $E_{B_s^0}$ reconstruction. But one can do so for a more dedicated analysis of $R_{D^*}$ to improve the reconstruction quality of $E_{B_s^0}$.} (see Fig.~\ref{fig:DspB} for its distribution) 
\begin{equation}
E_{B_s^0}=E_{\rm sig}-\sum_{i\in \text{sig-hem}} E_{i}+E_{D_s^-}+E_{\mu}~,
\end{equation}
and $B_s^0$ on-shell condition.

As done for the $R_{J/\psi}$ measurement, we introduce the kinematic variables $q^2$, $m_{\text{miss}}^2$ and the minimal distance between the $\mu$ track and the secondary vertex $S_{\rm SV}$ to distinguish the signal events of the $\tau$- and $\mu$-modes. Their distributions are shown in Fig.~\ref{fig:DspB}. The events of the $\tau$-modes tend to have larger $q^2$, $m_{\text{miss}}^2$ and $S_{\rm SV}$, compared to those of the $\mu$-modes. 
Notably, the reconstruction errors of $q^2$ (1.49(1.25)~GeV$^2$ for $R_{D_s}$ and 1.54(1.34)~GeV$^2$ for $R_{D_s^*}$) and $m_{\text{miss}}^2$ (1.46(1.12)~GeV$^2$ for $R_{D_s}$ and 1.46(1.23)~GeV$^2$ for $R_{D_s^*}$) in this analysis are smaller than those of the $R_{J/\psi}$ measurement; and the peaks for the $\log S_{\rm SV}$ distributions here are also shifted slightly to the left of those in the latter case. This is because the $B_s^0$ lifetime is about three times as long as $B_c^+$. A larger displacement from the PV can reduce the uncertainty in determining the $b$-hadron momentum direction.

\begin{figure}[h!]
	\centering	\includegraphics[width=7.5cm]{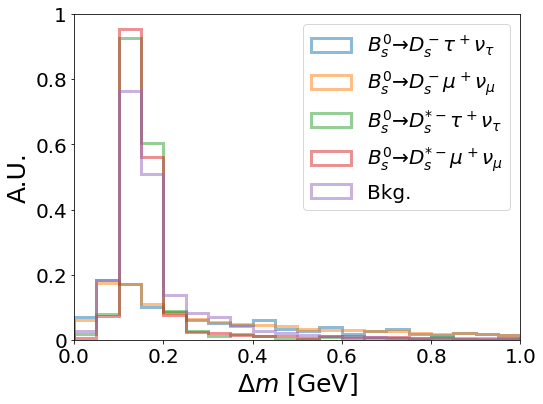}
	\caption{Normalized distributions of $\Delta m$ in the $R_{D_s^{(*)}}$ measurement. }
	\label{fig:Dsdeltam}
\end{figure}

The $D_s^-$ and $D_s^{*-}$ signal events are mutually the major backgrounds in their respective measurements (see Tab.~\ref{tab:Ds_yields}). Nevertheless, they can be distinguished by the photon from the $D_s^{*-}$ decay. 
For this purpose, we circulate all ECAL photons in the signal hemisphere to identify the one which yields a $\Delta m \equiv m(K^+K^-\pi^- \gamma) - m(K^+K^-\pi^-)$ value closest to $m_{D_s^{*-}}-m_{D_s^-}=143.8$~MeV~\cite{ParticleDataGroup:2020ssz}. The normalized $\Delta m$ distributions for the signal and background events are shown in Fig.~\ref{fig:Dsdeltam}. A clear resonant structure forms for the $D_s^{*-}$ signals but not for the $D_s^{-}$ signals. Notably, $D_s^{*-}$ mesons can be produced in the cascade and inclusive backgrounds efficiently, so a resonant structure forms in their distribution also.

\begin{figure}[h!]
\centering
\includegraphics[width=7.5 cm]{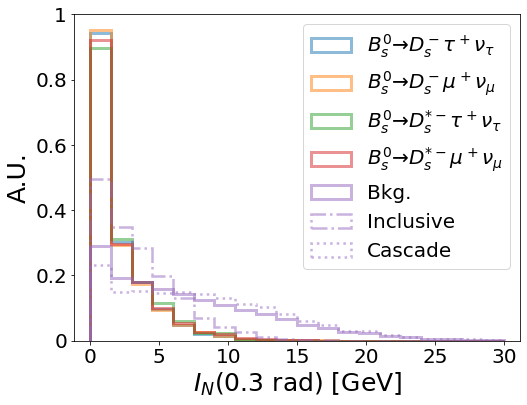}
\includegraphics[width=7.5 cm]{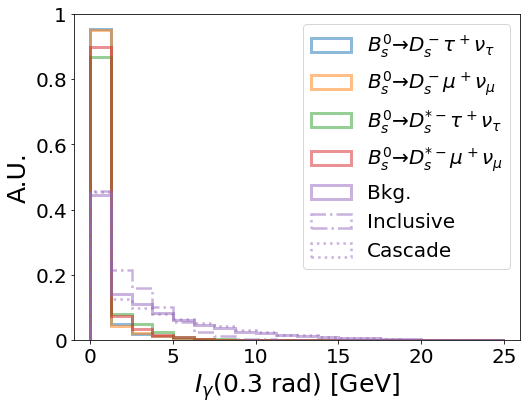}
	\caption{Distributions of $I_N(0.3$~rad$)$ and $I_\gamma(0.3$~rad$)$ in the $R_{D_s^{(*)}}$ measurement. }
\label{fig:Dsisolation}
\end{figure}

As shown in Figs.~\ref{fig:DspB} and Fig.~\ref{fig:Dsdeltam}, the observables introduced above can separate the signals of different modes from the universal backgrounds to various extents. As before, we introduce a set of isolation observables with the cone size $\Omega=0.3$ and 0.6 to further suppress these backgrounds. We show the distributions of $I_N(0.3$~rad$)$ and $I_\gamma(0.3$~rad$)$ in Fig.~\ref{fig:Dsisolation}. In both cases, the signal events tend to concentrate around zero, while the universal backgrounds are distributed more broadly.

\begin{table}[h]
\begin{center}

\fontsize{9pt}{10.8pt}\selectfont 
\begin{tabular}{ c c c c c c}
	\hline
	Channel 
	& Events at Tera-$Z$
	& $N(KK\pi\mu)$
	& $N(D_s^-)$ 
	& $N(B_s^0)$ 
	& Total eff.\\
\hline
	$B_s^0 \to D_s^- \tau^+ \nu_\tau$
	& $1.03\times 10^{6}$
	& $7.92\times 10^{5}$
	& $6.45\times 10^{5}$
	& $4.81\times 10^{5}$
	& $46.77\%$
	\\
	$B_s^0 \to D_s^- \mu^+ \nu_\mu$
	& $1.50\times 10^{7}$
	& $1.18\times 10^{7}$
	& $9.93\times 10^{6}$
	& $8.41\times 10^{6}$
	& $56.08\%$
	\\
	$B_s^0 \to D_s^{*-} \tau^+ \nu_\tau$
	& $1.72\times 10^{6}$
	& $1.30\times 10^{6}$
	& $1.05\times 10^{6}$
	& $7.65\times 10^{5}$
	& $44.61\%$
	\\
	$B_s^0 \to D_s^{*-} \mu^+ \nu_\mu$
	& $3.35\times 10^{7}$
	& $2.56\times 10^{7}$
	& $2.11\times 10^{7}$
	& $1.78\times 10^{7}$
	& $53.11\%$
	\\
	Inclusive bkg.
	& $5.78\times 10^{6}$
	& $4.28\times 10^{6}$
	& $3.28\times 10^{6}$
	& $2.72\times 10^{6}$
	& $47.03\%$
	\\
	Cascade bkg.
	& $8.44\times 10^{7}$
	& $6.20\times 10^{7}$
	& $2.33\times 10^{7}$
	& $8.71\times 10^{6}$
	& $10.33\%$
	\\
	Combinatoric bkg.
	& $1.36\times 10^{8}$
	& $1.16\times 10^{8}$
	& $2.24\times 10^{7}$
	& $2.17\times 10^{4}$
	& $0.02\%$
	\\
	Mis-ID bkg.
	& $\epsilon_{\mu\pi}\times 1.05\times 10^{10}$
	& $\epsilon_{\mu\pi}\times 4.33\times 10^{9}$
	& $\epsilon_{\mu\pi}\times 8.41\times 10^{8}$
	& $\epsilon_{\mu\pi}\times 8.50\times 10^{7}$
	& $0.81\%$
	\\
\hline
\end{tabular}
\caption{Tera-$Z$ yields for the preselected signals and the backgrounds in the $R_{D_s^*}$ and $R_{D_s}$ measurements. The preselection criteria are defined in the text.  
 \label{tab:Ds_yields}}
\end{center}
\end{table}

In this analysis, we train the BDT classifier in the five-class mode to address its four signal patterns ($D_s\mu$, $D_s\tau$, $D_s^*\mu$, $D_s^*\tau$). The full list of the discriminators is summarized below:
\begin{itemize}
\itemsep-0.5em 
\item Kinematics of the $K^+ K^- \pi^- \mu^+$ system:
	\begin{itemize}
	\itemsep-0.25em 
	\item Invariant mass: $m_{KK\pi\mu}$
	\item Energy and momentum of the reconstructed $D_s^-$ and muon: $E_{D_s^-}$, $|\vec{p}_{D_s^-}|$, $E_{\mu}$, $|\vec{p}_{\mu}|$
	\item Mass difference: $\Delta m = m_{KK\pi \gamma} - m_{KK\pi}$
	\end{itemize} 

\item Observables of the reconstructed $B_s^0$:
	\begin{itemize}
	\itemsep-0.25em 
	\item Energy and momentum of the reconstructed $B_s^0$: $E_{B_s^0}$, $|\vec{p}_{B_s^0}|$
	\item Lorentz-invariant observables: $m_{\rm miss}^2$, $q^2$
	\end{itemize}

\item Vertex information:
	\begin{itemize}
	\itemsep-0.25em 
	\item Minimal distance between the $D_s^-$ decay vertex and the muon track 
	\item Minimal distance between the deduced $B_s^0$ decay vertex and the muon track ($S_{\rm SV}$)
	\item Minimal distance between the muon track and its closest track
	\item Minimal distance between the reconstructed $D_s^-$ trajectory and its closest track
	\item Distance between the $D_s^-$ decay vertex and the PV
	\end{itemize}

\item Isolation observables: 
	\begin{itemize}
	\itemsep-0.25em 
	\item Neutral particles: $I_N(0.3 \ {\rm rad})$, $I_N(0.6 \ {\rm rad})$
	\item Neutral hadrons: $I_{H}(0.3 \ {\rm rad})$, $I_{H}(0.6 \ {\rm rad})$
	\item Photons: $I_\gamma(0.3 \ {\rm rad})$, $I_\gamma(0.6 \ {\rm rad})$
	\item Charged particles: $I_T(0.3 \ {\rm rad})$, $I_T(0.6 \ {\rm rad})$
	\item Tracks from the PV: $I_{T,\rm PV}(0.3 \ {\rm rad})$, $I_{T,\rm PV}(0.6 \ {\rm rad})$ 
	\item Tracks not from the PV: $I_{T,\rm dis}(0.3 \ {\rm rad})$, $I_{T, \rm dis}(0.6 \ {\rm rad})$ 
	\end{itemize} 
	
\item Impact parameter of the tracks in the signal hemisphere: 
	\begin{itemize}
	\itemsep-0.25em 
	\item Maximum and sum of transverse impact parameters
	\item Maximum and sum of longitudinal impact parameters
	\end{itemize} 
	
\item Some other discriminators~\cite{LHCb:2020cyw}:
	\begin{itemize}
	\itemsep-0.25em 
	\item $D_s^-\mu^+$ momentum transverse to the $B_s^0$ moving direction: $p_{\perp}(D_s^-\mu^+)$	
	\item Corrected mass: $m_{\rm corr} = \sqrt{m^2(D_s^-\mu^+)+p_\perp^2(D_s^-\mu^+)} + p_\perp(D_s^-\mu^+)$
	\end{itemize} 
\end{itemize}

\subsection{Results}

\begin{figure}[h]
	\centering
	\includegraphics[width=7.5cm]{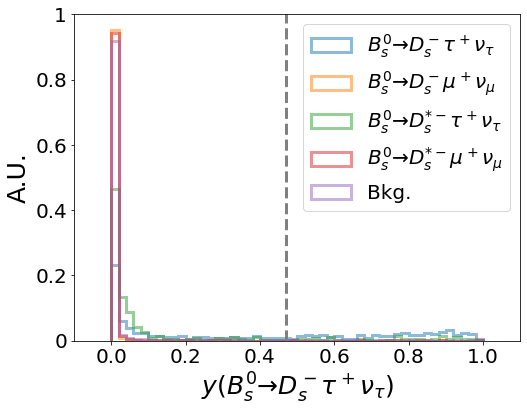}
		\includegraphics[width=7.5cm]{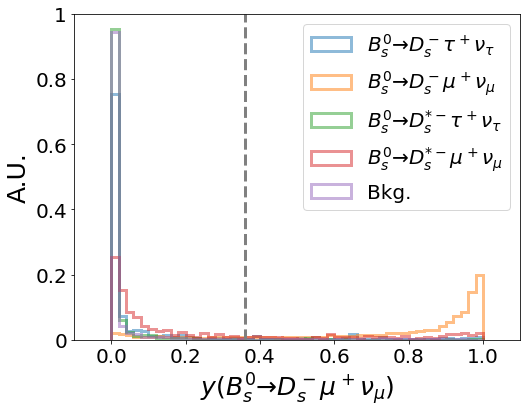}
	\includegraphics[width=7.5cm]{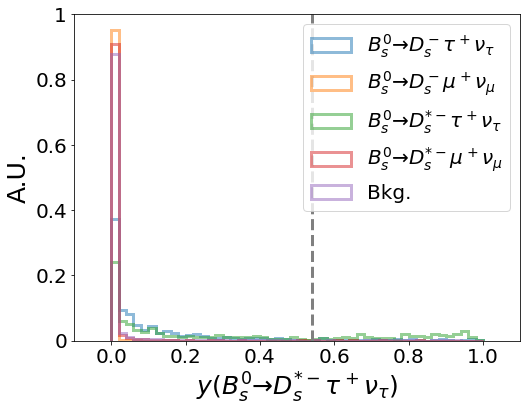}
		\includegraphics[width=7.5cm]{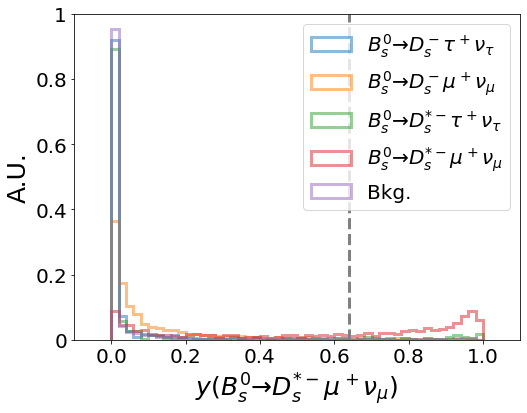}
\caption{Distributions of BDT response  in favor of $B_s^0 \to D_s^- \tau^+ \nu_\tau$ ($y_{D_s}^\tau$), $B_s^0 \to D_s^- \mu^+ \nu_\mu$ ($y_{D_s}^\mu$), $B_s^0 \to D_s^{*-} \tau^+ \nu_\tau$ ($y_{D_s^*}^\tau$) and $B_s^0 \to D_s^{*-} \mu^+ \nu_\mu$ ($y_{D_s^*}^\mu$) in the $R_{D_s^{(*)}}$ measurement. The vertical dashed lines represent optimal thresholds for sensitivity analysis.}	
	\label{fig:DsBDT}
\end{figure}

In Fig.~\ref{fig:DsBDT}, we show the distributions of BDT response in favor of $B_s^0 \to D_s^- \tau^+ \nu_\tau$, $B_s^0 \to D_s^- \mu^+ \nu_\mu$, $B_s^0 \to D_s^{*-} \tau^+ \nu_\tau$ and $B_s^0 \to D_s^{*-} \mu^+ \nu_\mu$. We summarize the event counts in the four signal regions in Tab.~\ref{tab:Ds_Final_Yield} and the expected precisions of $R_{D_s}$ and $R_{D_s^*}$ measurements at Tera-$Z$ ($10\times$Tera-$Z$) in Tab.~\ref{tab:Ds_BDT_yields} and Tab.~\ref{tab:Dss_BDT_yields}.  As before, the precisions of measuring $R_{D_s^{(*)}}$ are limited by the relatively low counts of the $\tau$-mode signal events. The two tables also show that in the high $q^2$ region where a larger $R_{D_s^{(*)}}$ has been predicted, a better precision can be achieved compared to the low $q^2$ region. Meanwhile, the relatively high $S/B$ ratios in all scenarios ensure the robustness of the sensitivity analysis of measuring  $R_{D_s^{(*)}}$ against the potential systematic uncertainties. At last, we point out that the imperfect discrimination between the $D_s$ and $D_s^\ast$ modes induces negative correlations between the $R_{D_s}$ and $R_{D_s^\ast}$ measurements (see Tab.~\ref{tab:Ds_BDT_yields} and Tab.~\ref{tab:Dss_BDT_yields}). We will discuss this feature in more details in Subsec.~\ref{ssec:photon}.

\begin{table}[h!]
\begin{center}
\fontsize{9pt}{10.8pt}\selectfont 
\begin{tabular}{c|c|c|c|c}	
\hline
	& $y_{D_s}^\tau \geq 0.47$ & $y_{D_s}^\tau <0.47$ & $y_{D_s}^\tau <0.47$ & $y_{D_s}^\tau <0.47$\\
	& $\cap ~  y_{D_s}^\mu <0.36$ & $\cap ~  y_{D_s}^\mu \geq0.36$ & $\cap ~  y_{D_s}^\mu <0.36$ & $\cap ~  y_{D_s}^\mu <0.36$\\
	& $\cap ~  y_{D_s^*}^\tau <0.54$ & $\cap ~  y_{D_s^*}^\tau <0.54$ & $\cap ~  y_{D_s^*}^\tau \geq0.54$ & $\cap ~  y_{D_s^*}^\tau <0.54$\\
	& $\cap ~  y_{D_s^*}^\mu <0.64$ & $\cap ~  y_{D_s^*}^\mu <0.64$ & $\cap ~  y_{D_s^*}^\mu <0.64$ & $\cap ~  y_{D_s^*}^\mu \geq0.64$\\
	\hline
	
	$B_s^0 \to D_s^- \tau^+ \nu_\tau$ &
	$2.05\times 10^5$ &
	$5.58\times 10^4$&
	$2.76\times 10^4$&
	$1.13\times 10^4$
	\\
	$B_s^0 \to D_s^- \mu^+ \nu_\mu$ &	
	$2.43\times 10^4$  &
	$7.11\times 10^6$ &
	$\lesssim 8.70\times 10^2$&
	$5.14\times 10^5$
	\\
	$B_s^0 \to D_s^{*-} \tau^+ \nu_\tau$ &
	$9.38\times 10^4$ &
	$2.53\times 10^4$ &
	$2.22\times 10^5$&
	$6.00\times 10^4$
	\\
	$B_s^0 \to D_s^{*-} \mu^+ \nu_\mu$ &
	$1.10\times 10^5$ & 
	$4.88\times 10^6$ &
	$1.22\times 10^5$&
	$9.03\times 10^6$
	\\
	Inclusive bkg. & 
	$4.12\times 10^4$ &
	$3.99\times 10^5$ &
	$4.35\times 10^4$&
	$2.61\times 10^5$
	\\
	Cascade bkg. &
	$6.63\times 10^4$ &
	$1.35\times 10^5$ &
	$3.66\times 10^4$&
	$4.80\times 10^4$
	\\
	Combinatoric bkg. &
	$\lesssim 3.43\times 10^3$ &
	$\lesssim 3.43\times 10^3$ &
	$\lesssim 3.43\times 10^3$ &
	$\lesssim 3.43\times 10^3$ 
	\\
	Mis-ID bkg. &
	$\epsilon_{\mu\pi}\times 4.21\times 10^5$ &
	$\epsilon_{\mu\pi}\times 6.22\times 10^6$ &
	$\epsilon_{\mu\pi}\times 3.82\times 10^5$ &
	$\epsilon_{\mu\pi}\times 1.41\times 10^6$\\
	\hline
\end{tabular}
\caption{Event counts in the signal regions of $B_s^0 \to D_s^- \tau^+ \nu_\tau$, $B_s^0 \to D_s^- \mu^+ \nu_\mu$,  $B_s^0 \to D_s^{*-} \tau^+ \nu_\tau$ and $B_s^0 \to D_s^{*-} \mu^+ \nu_\mu$ for the $R_{D_s^{(*)}}$ measurement at Tera-$Z$.}
\label{tab:Ds_Final_Yield}
\end{center}
\end{table}

\begin{table}[h]
\begin{center}
\begin{tabular}{c|cc|cc|cc}	
	\hline
	\hline
	\multirow{2}{*}{$q^2$ range}&
	\multicolumn{2}{c|}{$B_s^0 \to D_s^- \tau^+ \nu_\tau$} &
	\multicolumn{2}{c|}{$B_s^0 \to D_s^- \mu^+ \nu_\mu$} &
	$R_{D_s}$ & Correlation \\ & 
	Rel. precision & $S/B$ & 
	Rel. precision & $S/B$ & 
	Rel. precision & $\rho$  w/ $R_{D_s^*}$\\ 
	\hline
	\hline	
	\multirow{2}{*}{$q^2<7.15$~GeV$^2$}
	& $8.17\times 10^{-3}$ & \multirow{2}{*}{$0.49$}
	& $5.83\times 10^{-4}$ & \multirow{2}{*}{$1.57$} 
	& $9.37\times 10^{-3}$ & \multirow{2}{*}{$-0.56$}\\
	& $(2.58\times 10^{-3})$ & & $(1.84\times 10^{-4})$ & & $(2.96\times 10^{-3})$ &
	\\
	\hline
	\multirow{2}{*}{$q^2 \geq 7.15$~GeV$^2$}
	& $4.43\times 10^{-3}$ & \multirow{2}{*}{$0.62$} 
	& $1.39\times 10^{-3}$ & \multirow{2}{*}{$0.74$} 
	& $4.72\times 10^{-3}$ & \multirow{2}{*}{$-0.48$}\\
	& $(1.40\times 10^{-3})$ & & $(4.38\times 10^{-4})$ & & $(1.49\times 10^{-3})$ 
	\\
	\hline	
		\multirow{2}{*}{Full $q^2$} 
	& $3.81\times 10^{-3}$ & \multirow{2}{*}{$0.60$} 
	& $5.42\times 10^{-4}$ & \multirow{2}{*}{$1.28$}
	& $4.09\times 10^{-3}$ & \multirow{2}{*}{$-0.49$}\\
	& $(1.21\times 10^{-3})$ & & $(1.72\times 10^{-4})$ & & $(1.30\times 10^{-3})$ &  
	\\
	\hline
\end{tabular}
\caption{Expected BDT (relative) precisions of measuring $R_{D_s}$ at Tera-$Z$ ($10\times$Tera-$Z$). \label{tab:Ds_BDT_yields}}
\end{center}
\end{table}

\begin{table}[h]
\begin{center}
\begin{tabular}{c|cc|cc|cc}	
	\hline
	\hline
	\multirow{2}{*}{$q^2$ range} &
	\multicolumn{2}{c|}{$B_s^0 \to D_s^{*-} \tau^+ \nu_\tau$} &
	\multicolumn{2}{c|}{$B_s^0 \to D_s^{*-} \mu^+ \nu_\mu$} &
	$R_{D_s^*}$ & Correlation\\ & 
	Rel. precision & $S/B$ & 
	Rel. precision & $S/B$ & 
	Rel. precision & $\rho$ w/ $R_{D_s}$\\ 
	\hline
	\hline
	\multirow{2}{*}{$q^2<7.15$~GeV$^2$}
	& $9.93\times 10^{-3}$ & \multirow{2}{*}{$0.53$} 
	& $5.24\times 10^{-4}$ & \multirow{2}{*}{$7.90$} 
	& $9.93\times 10^{-3}$ & \multirow{2}{*}{$-0.56$}\\
	& $(3.14\times 10^{-3})$ & & $(1.66\times 10^{-4})$ & & $(3.14\times 10^{-3})$ &  
	\\
	\hline
	\multirow{2}{*}{$q^2 \geq 7.15$~GeV$^2$}
	& $3.50\times 10^{-3}$ & \multirow{2}{*}{$1.04$} 
	& $5.94\times 10^{-4}$ & \multirow{2}{*}{$15.25$} 
	& $3.49\times 10^{-3}$ & \multirow{2}{*}{$-0.48$}\\
	& $(1.11\times 10^{-3})$ & & $(1.88\times 10^{-4})$ & & $(1.10\times 10^{-3})$ &  
	\\
	\hline
		\multirow{2}{*}{Full $q^2$} 
	& $3.27\times 10^{-3}$ & \multirow{2}{*}{$0.95$} 
	& $3.94\times 10^{-4}$ & \multirow{2}{*}{$9.93$} 
	& $3.26\times 10^{-3}$ & \multirow{2}{*}{$-0.49$}\\
	& $(1.03\times 10^{-3})$ & & $(1.24\times 10^{-4})$ & & $(1.03\times 10^{-3})$ & 
	\\
	\hline
\end{tabular}
\caption{Expected BDT (relative) precisions of measuring $R_{D_s^*}$ at Tera-$Z$ ($10\times$Tera-$Z$). \label{tab:Dss_BDT_yields}}
\end{center}
\end{table}

\section{Measurement of $R_{\Lambda_c}$}
\label{sec:RLambdac}

\subsection{Method}

\begin{figure}[]
	\centering
	\includegraphics[width=8cm]{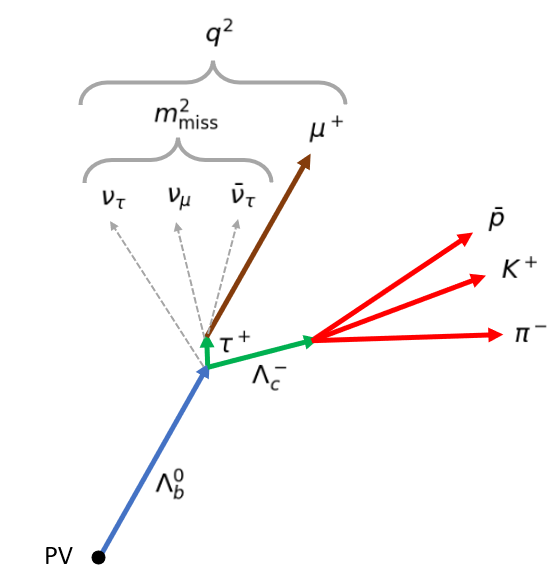}
	\caption{Schematic of the $\Lambda_b^0\to \Lambda_c^{-}\tau^+\nu$ process. Similar to $R_{D_s^{(*)}}$ that the lifetime of $\Lambda_c^-$ here is longer compared to that of $J/\psi$ in the $R_{J/\psi}$ measurement. }
	\label{fig:cartoons4}
\end{figure}

To measure $R_{\Lambda_c}$, we consider the exclusive $\Lambda_b^0$ decays, $i.e.$, $\Lambda_b^0\to \Lambda_c^- \tau^+\nu_\tau$ and $\Lambda_b^0\to \Lambda_c^- \mu^+\nu_\mu$ with  $\Lambda_c^- \to \bar{p}K^+\pi^-$, as the signals. Both signal modes contain $\bar{p}K^+\pi^- \mu^+$ in their final states. The schematic of the $\Lambda_b^0\to \Lambda_c^{-}\tau^+\nu_\tau$ process is shown in Fig.~\ref{fig:cartoons4}. Below are a set of cuts applied to preselect such events. 
\begin{itemize}

\item {\bf The $\bar{p}K^+\pi^- \mu^+$ selection.}  Candidates events that have $\bar{p}$, $K^+$ and $\pi^-$ tracks ($p_T>0.1$~GeV) sharing the same displaced decay vertex are selected. We also require exactly one muon track ($p_T>0.1$~GeV) with the same charge as the identified charged Kaon. 

\item {\bf The $\Lambda_c^-$ selection.}
The $\bar{p}K^+\pi^-$ vertex's distance from the PV must be greater than $0.5$~mm, with its invariant mass $|m_{\bar{p}K^+\pi^-}-m_{\Lambda_c}|<14$ MeV. The $\Lambda_c$ trajectory is reconstructed based on $p_{\bar{p}K^+\pi^-}$ and its decay vertex. The closest distance between the reconstructed $\Lambda_c$ system and any other track beside the identified muon must be $>0.02$~mm. 

\item {\bf The $\Lambda_b^0$ selection.}
Once the $\Lambda_c^-$ candidate is identified, two hemispheres are divided by the plane perpendicular to the displacement of $\Lambda_c^-$ decay vertex, with the signal hemisphere containing the $\Lambda_c^-$ decay vertex. The muon candidate must be found in the signal hemisphere. Similar to the requirement in Sec.~\ref{sec:RDs}, its minimal distance from other tracks, except the tagged $\bar{p}K^+\pi^-$ tracks, needs to be greater than 0.02~mm. Also, its $p_T$ has to be larger than 1.2~GeV. Finally, the invariant mass of $\bar{p}K^+\pi^-\mu^+$ has to be smaller than $m_{\Lambda_b^0}$.

\end{itemize}
The expected Tera-$Z$ yields after the preliminary cuts are shown in Tab.~\ref{tab:Lambdac_yields}.

\begin{table}[h!]
\begin{center}
\fontsize{9pt}{10.8pt}\selectfont 
\begin{tabular}{cccccc}	
	\hline
	Channel
	& Events at Tera-$Z$ 
	& $N(pK\pi\mu)$
	& $N(\Lambda_c^+)$
	& $N(\Lambda_b^0)$
	& Total eff.\\
	\hline
	
	$\Lambda_b^0\to \Lambda_c^- \tau^+\nu_\tau$
	& $ 4.46\times 10^{6}$
	& $ 3.52\times 10^{6}$
	& $ 2.96\times 10^{6}$
	& $ 2.22\times 10^{6}$
	& $49.89\%$\\
	
	$\Lambda_b^0 \to \Lambda_c^- \mu^+\nu_\mu$
	& $ 7.58\times 10^{7}$
	& $ 6.23\times 10^{7}$
	& $ 5.26\times 10^{7}$
	& $ 4.48\times 10^{7}$
	& $59.11\%$\\
	
	Inclusive bkg.
	& $ 2.75\times 10^{6}$
	& $ 2.17\times 10^{6}$
	& $ 6.75\times 10^{5}$
	& $ 5.79\times 10^{5}$
	& $21.05\%$\\
	
	Cascade bkg.
	& $ 1.03\times 10^{6}$
	& $ 8.05\times 10^{5}$
	& $ 4.05\times 10^{5}$
	& $ 2.18\times 10^{5}$
	& $ 21.19\%$\\
	
	Combinatoric bkg.
	& $ 1.57\times 10^{7}$
	& $ 1.33\times 10^{7}$
	& $ 4.93\times 10^{5}$
	& $ 7.91\times 10^{2}$
	& $0.01\%$\\

	Mis-ID bkg.
	& $\epsilon_{\mu\pi} \times 1.36\times 10^9$
	& $\epsilon_{\mu\pi} \times 5.43\times 10^8$
	& $\epsilon_{\mu\pi} \times 4.05\times 10^7$
	& $\epsilon_{\mu\pi} \times 1.52\times 10^7$
	& $1.12\%$\\
	
	\hline
\end{tabular}
\caption{Tera-$Z$ yields for the preselected signals and the backgrounds in the $R_{\Lambda_c}$ measurement. The preselection criteria are defined in the text. 
 \label{tab:Lambdac_yields}}
\end{center}
\end{table}

\begin{figure}[h!]
	\centering
	\includegraphics[width=7.65 cm]{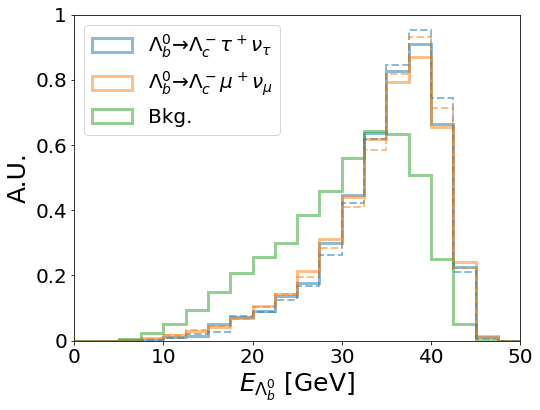}
	\includegraphics[width=7.4 cm]{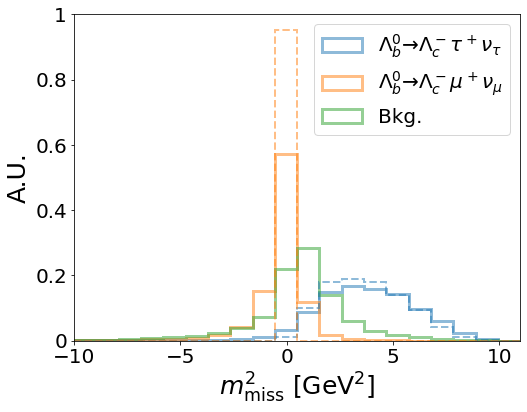}
	\includegraphics[width=7.5 cm]{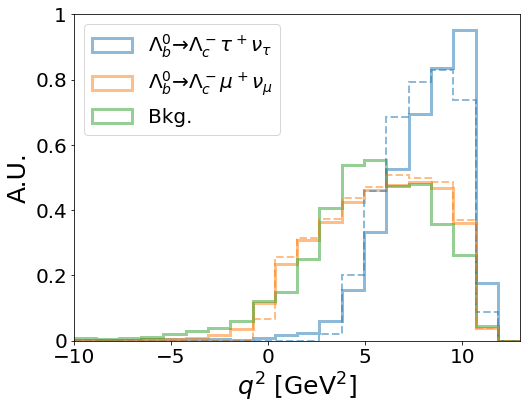}
	\includegraphics[width=7.5 cm]{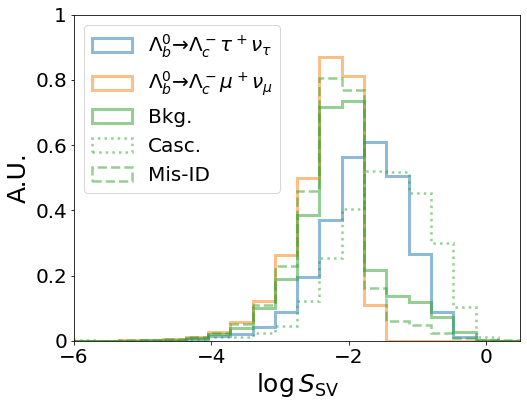} 
	\caption{Distributions of the reconstructed $E_{\Lambda_b^0}$, $q^2$, $m_{\text{miss}}^2$ and $\log{S_{\rm SV}}$ in the $R_{\Lambda_c}$ measurement. The solid and dashed lines represent the simulated and truth-level messages respectively. }	
	\label{fig:LambdacpB}
\end{figure}

\begin{figure}[h]
	\centering
	\includegraphics[width=7.5 cm]{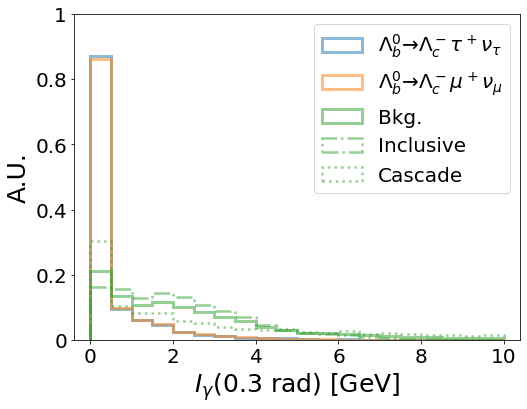}
	\caption{Distribution of $I_\gamma(0.3$~rad$)$ in the $R_{\Lambda_c}$ measurement. }	
	\label{fig:Lambdacisolation}
\end{figure} 

As has done for other signal $b$-hadrons, we can reconstruct the $\Lambda_b^0$ four-momentum using its decay vertex (or the $p_{\Lambda_b^0}$ direction) inferred from the $\Lambda_c^-$ and $\mu$ lepton kinematics, total energy
\begin{equation}
E_{\Lambda_b^0}=E_{\rm sig}-\sum_{i\in \text{sig-hem}} E_{i}+E_{\Lambda_c^-}+E_{\mu}~,
\end{equation}
and on-shell condition. 
Then we can introduce the Lorentz-invariant observables $q^2$ and $m_{\rm miss}^2$ and the minimal distance between the $\mu$ track and the secondary vertex $S_{\rm SV}$ to separate the signals of the $\tau$ and $\mu$ modes, and the set of isolation observables of $\Lambda_b^0$ to suppress the universal backgrounds. We show the distributions of these observables in Fig.~\ref{fig:LambdacpB} and Fig.~\ref{fig:Lambdacisolation}. The reconstruction errors of $q^2$ and $m_{\rm miss}^2$ are given by 1.37(1.23)~GeV$^2$ and 1.33(1.18)~GeV$^2$, respectively. 

In this analysis, we train the BDT classifier in the three-class mode to address its two signal patterns. The full list of the discriminators is summarized below:
\begin{itemize}
\itemsep-0.5em 
\item Kinematics of the $\bar{p}K^+\pi^-\mu^+$ system:
	\begin{itemize}
	\itemsep-0.25em 
	\item Invariant mass: $m_{pK\pi\mu}$
	\item Energy and momentum of the reconstructed $\Lambda_c$ and muon: $E_{\Lambda_c^-}$, $|\vec{p}_{\Lambda_c^-}|$, $E_{\mu}$, $|\vec{p}_{\mu}|$
	\end{itemize} 

\item Observables of the reconstructed $\Lambda_b^0$:
	\begin{itemize}
	\itemsep-0.25em 
	\item Energy and momentum of the reconstructed $\Lambda_b^0$: $E_{\Lambda_b^0}$, $|\vec{p}_{\Lambda_b^0}|$
	\item Lorentz-invariant observables: $m_{\rm miss}^2$, $q^2$
	\end{itemize}

\item Vertex information:
	\begin{itemize}
	\itemsep-0.25em 
	\item Minimal distance between the $\Lambda_c^-$ decay vertex and the muon track
	\item Minimal distance between the deduced $\Lambda_b^0$ decay vertex and the muon track ($S_{\rm SV}$)
	\item Minimal distance between the muon track and its closest track
	\item Minimal distance between the reconstructed $\Lambda_c^-$ trajectory and its closest track
	\item Distance between the $\Lambda_c^-$ decay vertex and the PV
	\end{itemize}

\item Isolation observables: 
	\begin{itemize}
	\itemsep-0.25em 
	\item Neutral particles: $I_N(0.3 \ {\rm rad})$, $I_N(0.6 \ {\rm rad})$
	\item Neutral hadrons: $I_{H}(0.3 \ {\rm rad})$, $I_{H}(0.6 \ {\rm rad})$
	\item Photons: $I_\gamma(0.3 \ {\rm rad})$, $I_\gamma(0.6 \ {\rm rad})$
	\item Charged particles: $I_T(0.3 \ {\rm rad})$, $I_T(0.6 \ {\rm rad})$
	\item Tracks from the PV: $I_{T,\rm PV}(0.3 \ {\rm rad})$, $I_{T,\rm PV}(0.6 \ {\rm rad})$ 
	\item Tracks not from the PV: $I_{T,\rm dis}(0.3 \ {\rm rad})$, $I_{T, \rm dis}(0.6 \ {\rm rad})$ 
	\end{itemize} 
	
\item Impact parameter of the tracks in the signal hemisphere:
	\begin{itemize}
	\itemsep-0.25em 
	\item Maximum and sum of transverse impact parameters
	\item Maximum and sum of longitudinal impact parameters
	\end{itemize} 
	
\item Some other discriminators~\cite{LHCb:2020cyw}:
	\begin{itemize}
	\itemsep-0.25em 
	\item $\Lambda_c^-\mu^+$ momentum transverse to the $\Lambda_b^0$ moving direction: $p_{\perp}(\Lambda_c^-\mu^+)$	
	\item Corrected mass: $m_{\rm corr} = \sqrt{m^2(\Lambda_c^-\mu^+)+p_\perp^2(\Lambda_c^-\mu^+)} + p_\perp(\Lambda_c^-\mu^+)$
	\end{itemize} 
\end{itemize}

\subsection{Results}

\begin{figure}[h]
	\centering
	\includegraphics[width=7.5 cm]{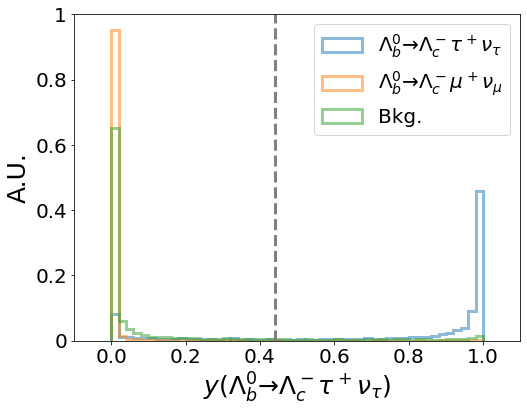}
		\includegraphics[width=7.5 cm]{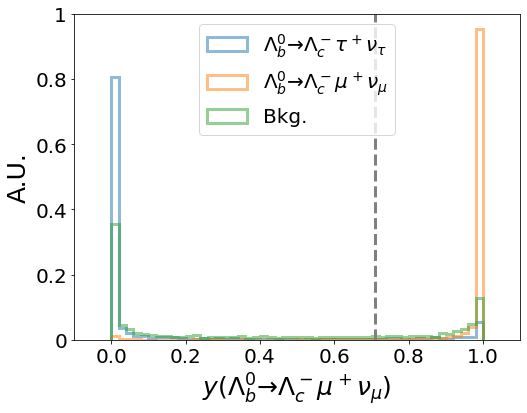}
\caption{Distributions of BDT response in favor of $\Lambda_b^0\to \Lambda_c^- \tau^+\nu_\tau$ ($y_{\Lambda_c}^\tau$) and $\Lambda_b^0\to \Lambda_c^- \mu^+\nu_\mu$ ($y_{\Lambda_c}^\mu$) in the $R_{J/\psi}$ measurement. The vertical dashed lines represent optimal thresholds for sensitivity analysis.}
	\label{fig:LBDT}
\end{figure}

\begin{table}[h]
\begin{center}
\begin{tabular}{c|c|c}	
	\hline
	& $y_{\Lambda_c}^\tau \geq0.44 ~ \cap ~ y_{\Lambda_c}^\mu<0.71$ & $y_{\Lambda_c}^\tau<0.44 ~ \cap ~ y_{\Lambda_c}^\mu\geq 0.71$\\
	\hline
	$\Lambda_b^0\to \Lambda_c^- \tau^+\nu_\tau$ &
	$1.79\times 10^6$ &
	$2.51\times 10^5$
	\\
	$\Lambda_b^0 \to \Lambda_c^- \mu^+\nu_\mu$ &	
	$5.34\times 10^5$  &
	$4.26\times 10^7$
	\\
	Inclusive bkg. & 
	$4.84\times 10^4$ &
	$2.57\times 10^5$
	\\
	Cascade bkg. &
	$4.53\times 10^4$ &
	$2.63\times 10^4$
	\\
	Combinatoric bkg. &
	$\lesssim 4.76\times 10^2$ &
	$\lesssim 4.76\times 10^2$
	\\
	Mis-ID bkg. &
	$\epsilon_{\mu\pi}\times 4.87\times 10^5$ &
	$\epsilon_{\mu\pi}\times 2.72\times 10^6$\\
	\hline		
\end{tabular}
\caption{Event counts in the signal regions of $\Lambda_b^0\to \Lambda_c^- \tau^+\nu_\tau$ and $\Lambda_b^0 \to \Lambda_c^- \mu^+\nu_\mu$ for the $R_{\Lambda_c}$ measurement at Tera-$Z$. }
\label{tab:Lambdac_final_yields}
\end{center}
\end{table}

\begin{table}[h]
\begin{center}
\begin{tabular}{c|cc|cc|c}	
\hline
	\hline
	\multirow{2}{*}{$q^2$ range} 
	& \multicolumn{2}{c|}{$\Lambda_b^0\to \Lambda_c^- \tau^+\nu_\tau$} &
	\multicolumn{2}{c|}{$\Lambda_b^0\to \Lambda_c^- \mu^+\nu_\mu$}
	& $R_{\Lambda_c}$\\
	 & 
	Rel. precision & $S/B$ & 
	Rel. precision & $S/B$ & 
	Rel. precision \\ 
	\hline
	\hline
	\multirow{2}{*}{$q^2<7.15$~GeV$^2$}
	& $2.01\times 10^{-3}$ & \multirow{2}{*}{$1.63$} 
	& $2.22\times 10^{-4}$ & \multirow{2}{*}{$71.81$} & $2.02\times 10^{-3}$\\
	& $(6.34\times 10^{-4})$ & & $(7.01\times 10^{-5})$ & & $(6.38\times 10^{-4})$\\
	\hline
	\multirow{2}{*}{$q^2 \geq 7.15$~GeV$^2$}
	& $1.10\times 10^{-3}$ & \multirow{2}{*}{$3.74$} 
	& $2.86\times 10^{-4}$ & \multirow{2}{*}{$77.94$} & $1.14\times 10^{-3}$\\
	& $(3.49\times 10^{-4})$ & & $(9.04\times 10^{-5})$ & & $(3.60\times 10^{-4})$\\
	\hline
		\multirow{2}{*}{Full $q^2$} 
	& $9.61\times 10^{-4}$ & \multirow{2}{*}{$2.83$} 
	& $1.75\times 10^{-4}$ & \multirow{2}{*}{$75.98$} & $9.77\times 10^{-4}$\\
	& $(3.04\times 10^{-4})$ & & $(5.54\times 10^{-5})$ & & $(3.09\times 10^{-4})$\\
	\hline
\end{tabular}	
\caption{Expected BDT (relative) precisions of measuring $R_{\Lambda_c}$ at Tera-$Z$ ($10\times$Tera-$Z$). \label{tab:Lambdac_BDT_yields}}
\end{center}
\end{table}

In Fig.~\ref{fig:LBDT}, we show the distributions of BDT response in favor of $\Lambda_b^0\to \Lambda_c^- \tau^+\nu_\tau$ and $\Lambda_b^0\to \Lambda_c^- \mu^+\nu_\mu$. We summarize the event counts in the two signal regions in Tab.~\ref{tab:Lambdac_final_yields} and the expected precisions of measuring $R_{\Lambda_c}$ at Tera-$Z$ ($10\times$Tera-$Z$) in Tab.~\ref{tab:Lambdac_BDT_yields}. The $S/B$ ratios are high to avoid large background systematics similar to previous $R_{H_c}$ measurements.

\section{Impacts of Detector Performance and Event Shape}
\label{sec:Results}

\subsection{Detector Tracking Resolution}
\label{ssec:trackingnoise}

In the analysis scheme developed above for measuring $R_{H_c}$, the $H_b$ reconstruction significantly relies on the determination of the $H_c$ decay vertex and the measurement of the muon track originating from the $H_b$ or $\tau$ decay. The precision of measuring $R_{H_c}$ thus could be sensitive to the tracker resolution of impact parameters. To explore the potential improvement with a better tracker resolution and test the robustness of the presented results against a worse situation, one then needs to draw a picture of the variation of the precision of measuring $R_{H_c}$ with the tracker resolution. In our previous analyses, we have simulated the tracker effects via the vertex noise and modeled it as a random vector with a reference magnitude of 10~$\mu$m. The noise is then injected to the $H_c$ decay vertex and the muon track vertex independently, following a normal distribution $\mathcal{N}(0, 100/3$)~$\mu$m in each direction such that the overall noise respects the normal distribution $\mathcal{N}(0, 100$)~$\mu$m. To generate a global picture mentioned above, below we will perform a series of studies, with the noise level varying from a perfect tracker case to more conservative resolution scenarios.

\begin{figure}[h!]
	\begin{center}
	
	\includegraphics[width=0.325\textwidth]{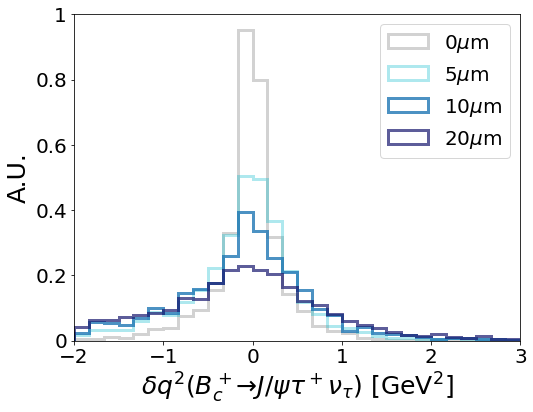}
	\includegraphics[width=0.325\textwidth]{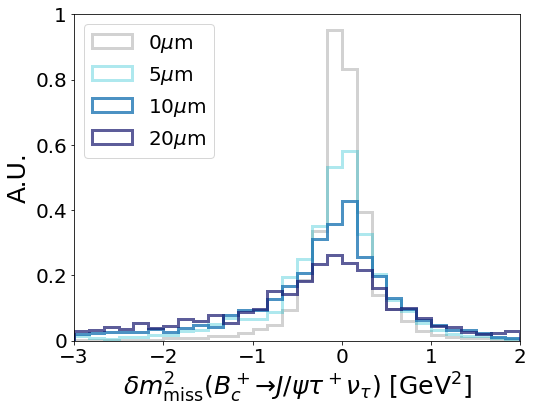}
	\includegraphics[width=0.33\textwidth]{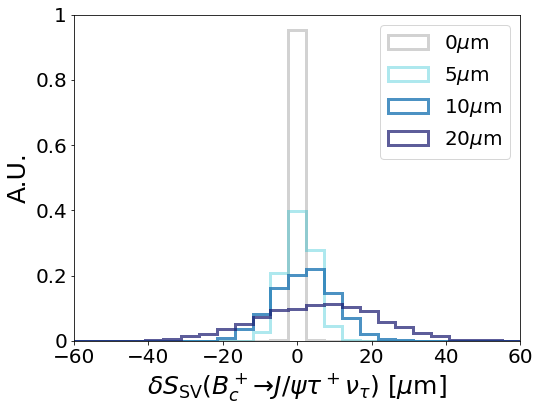}
	\end{center}
	\includegraphics[width=0.325\textwidth]{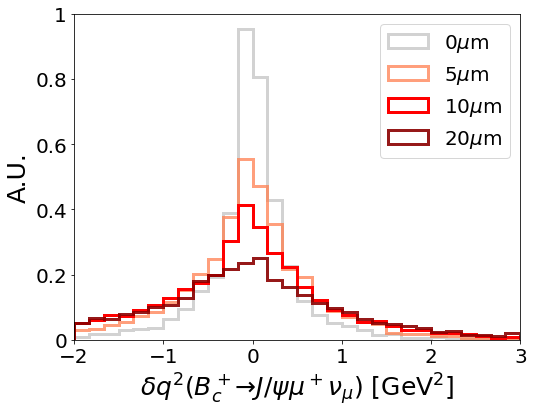}
	\includegraphics[width=0.325\textwidth]{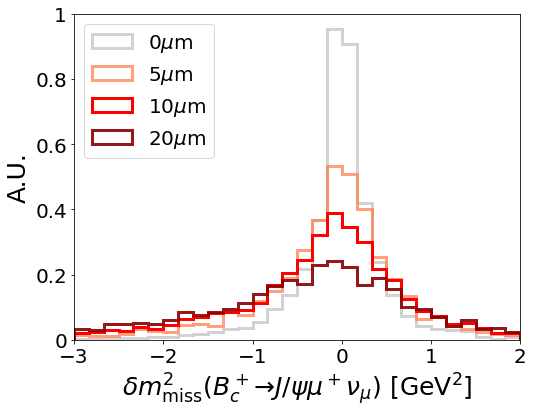}
	\includegraphics[width=0.33\textwidth]{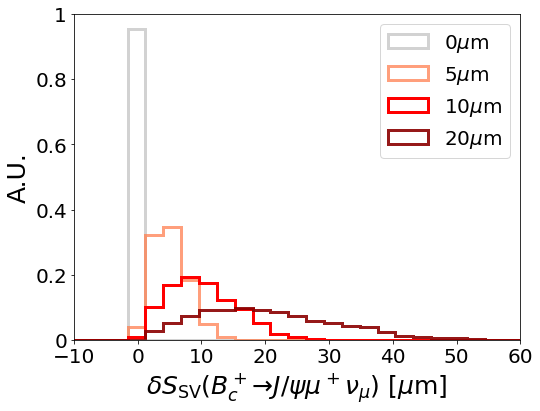}
	\caption{Distributions of $\delta q^2$, $\delta m_{\text{miss}}^2$ and $\delta S_{\rm SV}$ for $B_c^+\to J/\psi \tau^+ \nu_\tau$ and $B_c^+\to J/\psi \mu^+ \nu_\mu$, respectively. }
	\label{fig:Jpsisq2miss2_20noise}
\end{figure}

\begin{figure}[h!]
	\centering
	\includegraphics[width=0.325\textwidth]{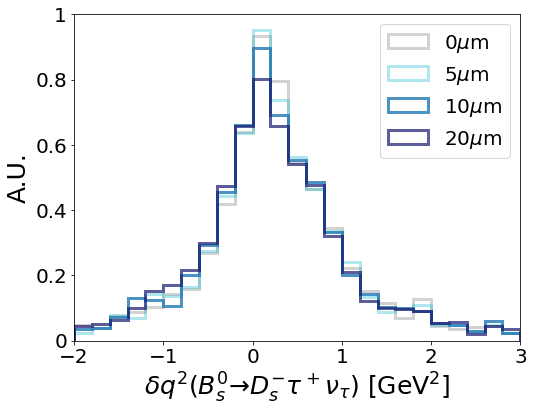}
	\includegraphics[width=0.325\textwidth]{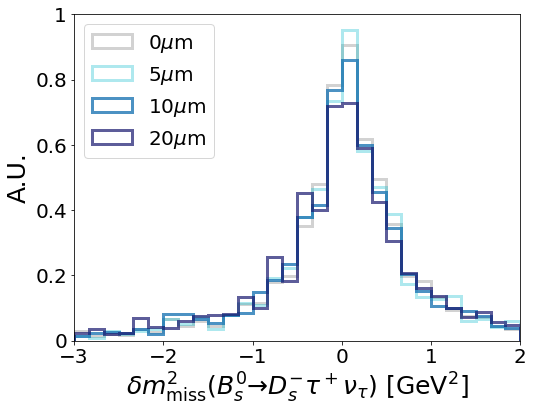}
	\includegraphics[width=0.323\textwidth]{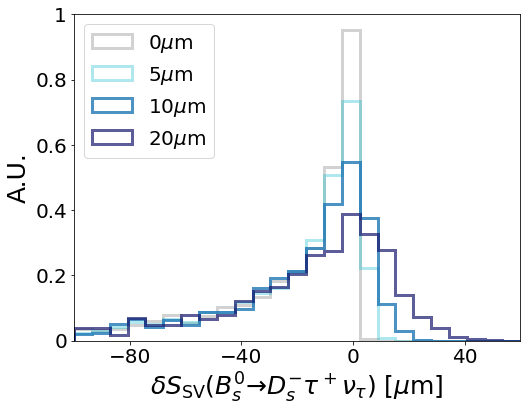}
	\includegraphics[width=0.325\textwidth]{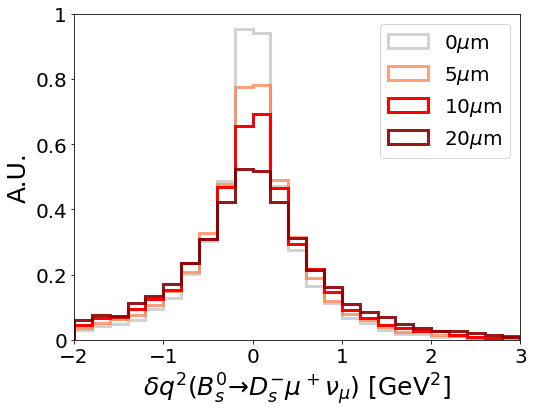}
	\includegraphics[width=0.325\textwidth]{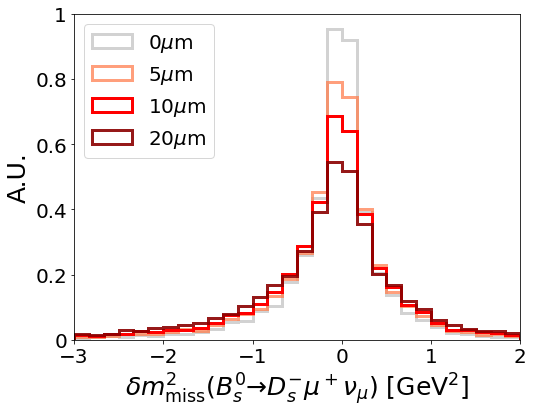}
	\includegraphics[width=0.327\textwidth]{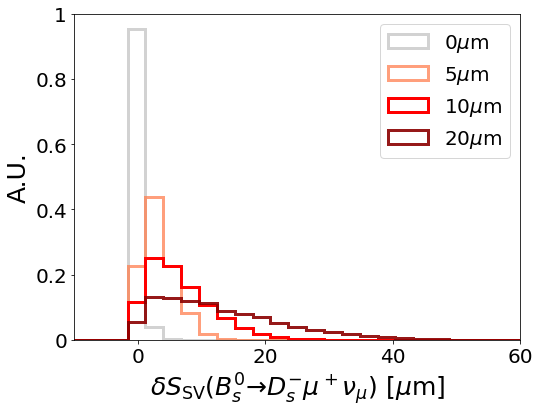}
	\caption{Distributions of $\delta q^2$, $\delta m_{\text{miss}}^2$ and $\delta S_{\rm SV}$ for $B_s^0 \to D_s^{-} \tau^+\nu_\tau$ and $B_s^0 \to D_s^{-} \mu^+\nu_\mu$, respectively. }		
	\label{fig:Dsq2miss2ssvdelatm_20noise}
\end{figure}

\begin{figure}[h!]
	\centering
	\includegraphics[width=0.325\textwidth]{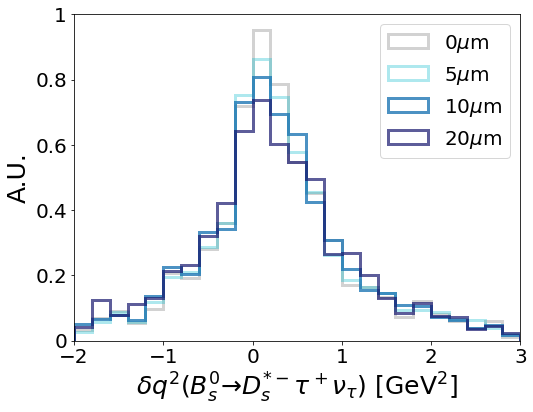}
	\includegraphics[width=0.325\textwidth]{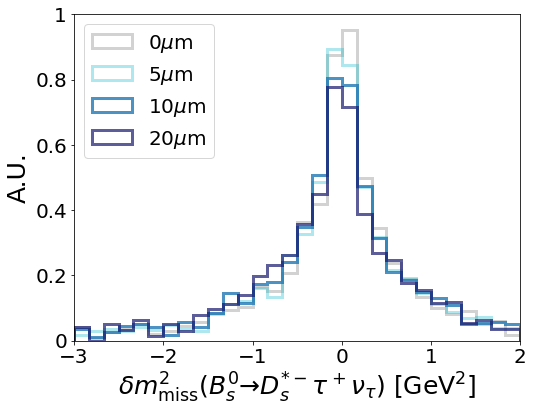}
	\includegraphics[width=0.323\textwidth]{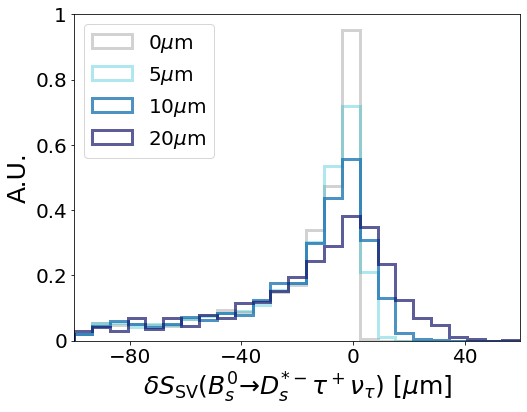}
	\includegraphics[width=0.325\textwidth]{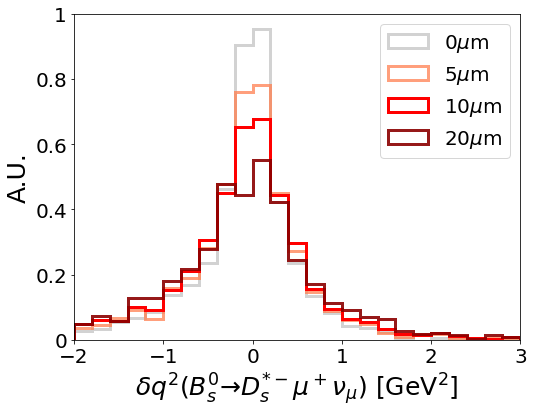}
	\includegraphics[width=0.325\textwidth]{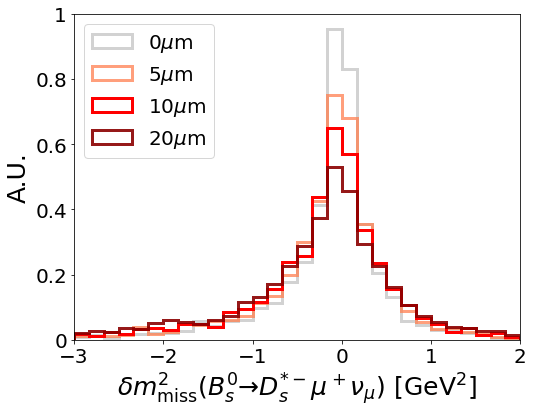}
	\includegraphics[width=0.327\textwidth]{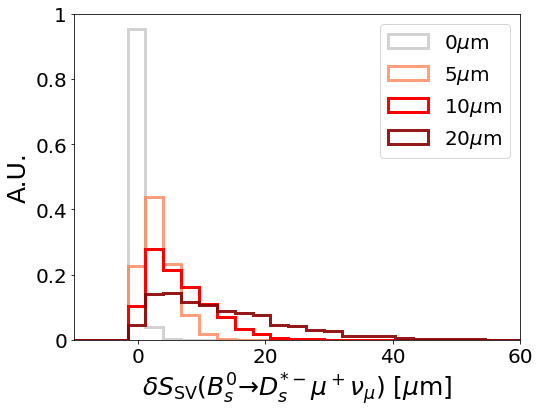}
	\caption{Distributions of $\delta q^2$, $\delta m_{\text{miss}}^2$ and $\delta S_{\rm SV}$ for $B_s^0 \to D_s^{*-} \tau^+\nu_\tau$ and $B_s^0 \to D_s^{*-} \mu^+\nu_\mu$, respectively. }		

	\label{fig:Dsstarq2miss2ssvdelatm_20noise}
\end{figure}

\begin{figure}[h!]
	\centering
	\includegraphics[width=0.325\textwidth]{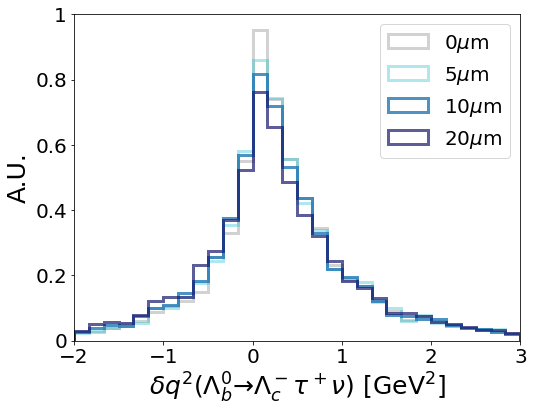}
	\includegraphics[width=0.325\textwidth]{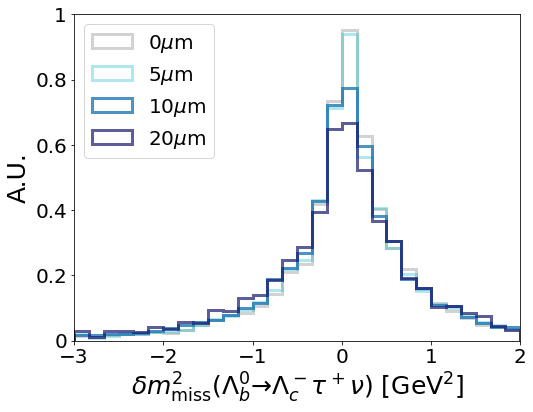}
	\includegraphics[width=0.323\textwidth]{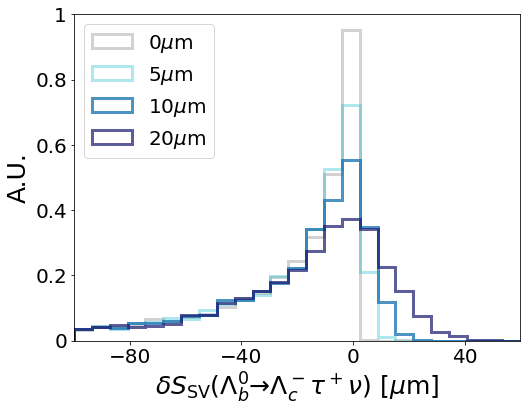}
	\includegraphics[width=0.325\textwidth]{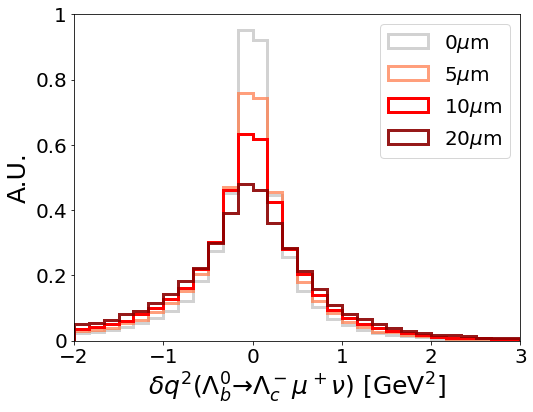}
	\includegraphics[width=0.325\textwidth]{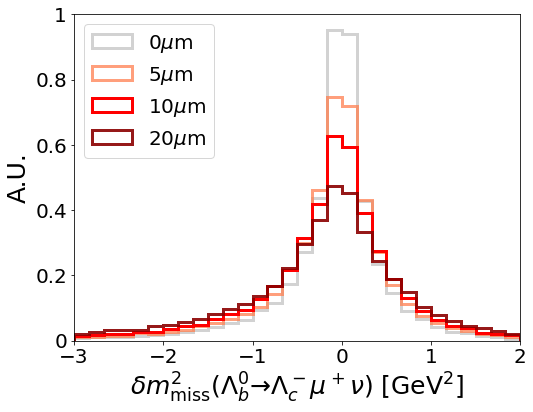}
	\includegraphics[width=0.327\textwidth]{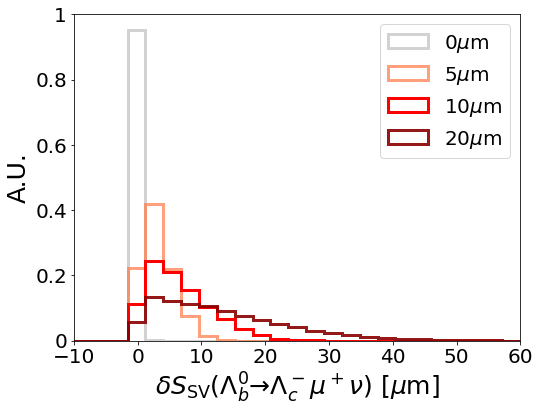}
	\caption{Distributions of $\delta q^2$, $\delta m_{\text{miss}}^2$ and $\delta S_{\rm SV}$ for $\Lambda_b^0 \to \Lambda_c^{-} \tau^+\nu_\tau$ and $\Lambda_b^0 \to \Lambda_c^{-} \mu^+\nu_\mu$, respectively. }		
	\label{fig:Lambdacq2miss2ssvdelatm_20noise}
\end{figure}

\begin{figure}[ht]
	\centering
	\includegraphics[width=0.45\textwidth]{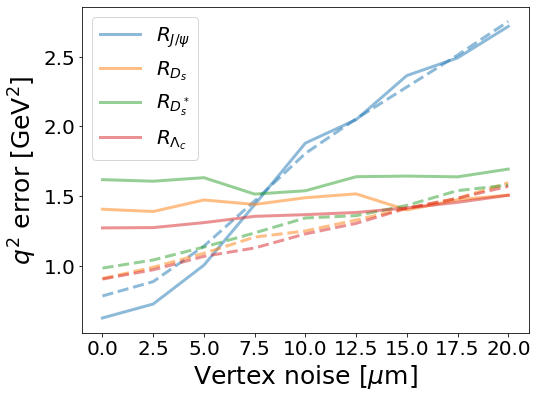}
	\includegraphics[width=0.465\textwidth]{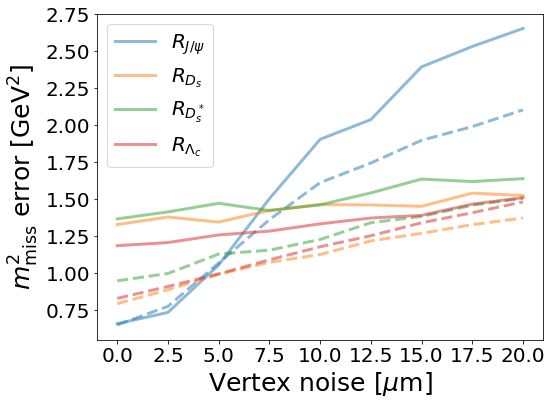} \\
	\includegraphics[width=0.445\textwidth]{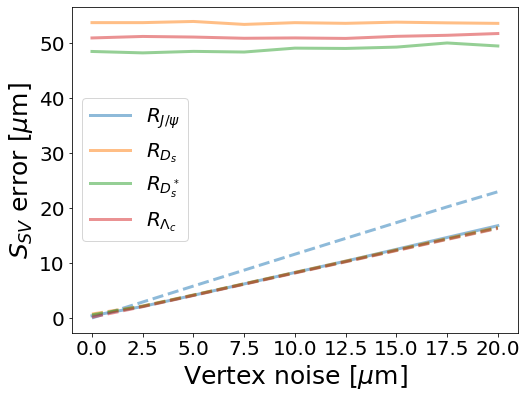}
	\includegraphics[width=0.455\textwidth]{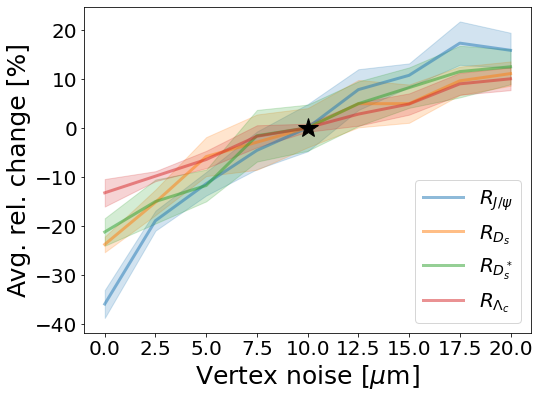}
	\caption{Reconstruction error of $q^2$ (upper-left), $m_{\text{miss}}^2$  (upper-right), $S_{\rm SV}$ (bottom-left) and averaged relative changes to the reference precision of measuring $R_{H_c}$ (bottom-right), with varied vertex noise. In the first three panels, the reconstruction error is defined to be the root mean square of $\delta X$ over the signal sample, with $X = q^2$, $m_{\text{miss}}^2$ and $S_{\rm SV}$. The solid and dashed lines correspond to the $\tau$ and $\mu$ modes, respectively. In the bottom-right panel, we have trained ten BDT classifiers, with ten random separations of the training (50\%) and testing (50\%) datasets respectively in the full simulation data samples. The averaged relative changes to the measurement precisions and their variances are denoted as solid lines and shaded bands, respectively. The reference precisions are simulated with a vertex noise of 10~$\mu$m, denoted as a black star. }
	\label{fig:error_trends}
\end{figure}

Let us consider $q^2$, $m_{\text{miss}}^2$ and $S_{\rm SV}$. As the tracker resolution correlates with the quality of $H_b$ reconstruction, these event-level observables measure the impacts on event reconstruction and $R_{H_c}$ sensitivities. We present the distributions of $\delta q^2$, $\delta m_{\text{miss}}^2$ for the four $R_{H_c}$ measurements in Fig.~\ref{fig:Jpsisq2miss2_20noise}-\ref{fig:Lambdacq2miss2ssvdelatm_20noise}, with four benchmark vertex noise levels: 0, 5, 10, and 20~$\mu$m. The dependence of their root mean square on the vertex noise level is also shown in Fig.~\ref{fig:error_trends}, where more benchmark noise levels are simulated. We have the following observations based on these figures: 
\begin{itemize}

\item For the reconstruction of $q^2$ and $m_{\text{miss}}^2$, $B_c^+\to J/\psi \tau^+ \nu_\tau$ and $B_c^+\to J/\psi \mu^+ \nu_\mu$ tend to be more sensitive to the variation of vertex noise level, compared to the other signal channels. As $J/\psi$ decays promptly, in these cases we have used the $J/\psi$ decay vertex to approximate the $b$-hadron decay vertex. So, the $b$-hadron vertex reconstruction has a higher quality in an ideal detector but is less robust against the vertex noise.

\item For the reconstruction of $S_{\rm SV}$, the muon signal modes tend to be more sensitive to the variation of vertex noise level compared to the tau signal modes. At the truth-level, we have $S_{\rm SV} \equiv 0 \mu$m for all four muon signal channels. Especially, for $B_c^+\to J/\psi \mu^+ \nu_\mu$, its $S_{\rm SV}$ can be ``perfectly'' measured in an ``ideal'' detector, due to the high-quality reconstruction of the $B_c^+$ decay vertex. But, this also implies that the reconstruction of $S_{\rm SV}$ in this case is less robust than the other three muon channels. As for the tau signal modes, we have $S_{\rm SV} \neq 0 \mu$m at the truth-level as the muon track in these cases is generated from tau decay and hence displaced from the $b$-hadron vertex. Due to the extra complexity caused by tau decay, the error of reconstructing $S_{\rm SV}$ in these cases is generally big. However, as the $B_c^+$ vertex can be well-reconstructed for the tau mode also, for $B_c^+\to J/\psi \tau^+ \nu_\tau$ the measurement of $S_{\rm SV}$ is as sensitive to the vertex noise as it is for the muon channels.

\end{itemize}
At last, we demonstrate the averaged relative precisions of measuring $R_{H_c}$ in the bottom-right panel of Fig.~\ref{fig:error_trends}, with varied vertex noise. 
Consisting with the observations above, the precision of measuring $R_{J/\psi}$ gets improved more with the reduced vertex noise, while the measurement of $R_{\Lambda_c}$ tends to be more robust against the variation of vertex noise.

\subsection{ECAL Energy Threshold}
\label{ssec:photon}

As shown in the $R_{D_s}$ and $R_{D_s^*}$ analyses in Sec.~\ref{sec:RDs}, the $B_s^0 \to D_s^{-} \tau^+ \nu_\tau (B_s^0 \to D_s^{-} \mu^+ \nu_\mu)$ and $B_s^0 \to D_s^{*-} \tau^+ \nu_\tau (B_s^0 \to D_s^{*-} \mu^+  \nu_\mu)$ events contribute mutually as one of the major backgrounds in their respective measurements. A natural discriminator between them could be the photon from the $D_s^{*-} \to D_s^{-} \gamma$ decay. So we have introduced a measure $\Delta m \equiv m(K^+K^-\pi^- \gamma) - m(K^+K^-\pi^-)$ in our analyses and reconstructed this photon as the one yielding a $\Delta m$ value closest to $m_{D_s^{*-}}-m_{D_s^-}=143.8$~MeV, among all ECAL photons in the signal hemisphere. The $\Delta m$ defined for the reconstructed $D_s^{*-}$ photon is then applied in the relevant BDT analyses.  

\begin{figure}[th!]
	\centering
	\includegraphics[width=7.5cm]{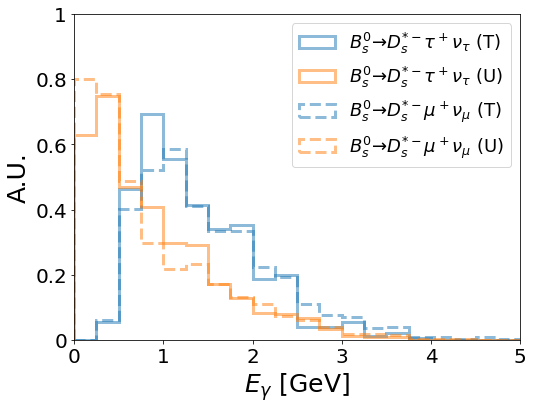} 
	\includegraphics[width=7.5cm]{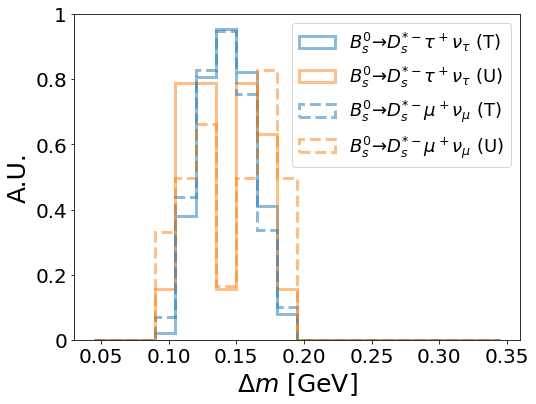} 
	\caption{Distributions of $E_\gamma$ (left) for the truth-level $D_s^{*-}$ photons and $\Delta m$ (right) for the reconstructed photons satisfying the consistency condition in footnote~\ref{foot}, in $B_s^0 \to D_s^{*-} \tau^+ \nu_\tau$ and $B_s^0 \to D_s^{*-} \mu^+ \nu_\mu$. Photons tagged (T) by the detector simulation are shown as blue curves, while untagged (U) photons are shown in orange.}
	\label{fig:Ds_pho_tag}
\end{figure}

However, the $D_s^{*-}$ photon tends to be soft, with energy typically $\lesssim \mathcal O(1)$ GeV. The performance of ECAL in detecting soft photons thus becomes highly crucial. The ECAL responds weakly to soft photons. Below some energy threshold ($E_{\rm th}$), the photons may not cause a response in the ECAL at all. We demonstrate this effect in the left panel of Fig.~\ref{fig:Ds_pho_tag}. We classify the $B_s^0 \to D_s^{*-} \tau^+ \nu_\tau$ and $B_s^0 \to D_s^{*-} \mu^+ \nu_\mu$ events into the ``tagged'' and ``untagged'' ones, with $E_{\rm th} = 0.5$~GeV, a default value in the Delphes model. In the former case, a $D_s^{*-}$ photon which is consistent with the truth in kinematics~\footnote{\label{foot} The consistency here requires the $\eta$ and $\phi$ separation between the reconstructed and truth-level $D_s^{*-}$ photons to be less than 0.01 and the energy difference to be smaller than 30\%.} and additionally yields a $\Delta m$ value closest to 143.8~MeV can be reconstructed, while in the latter case such a reconstruction fails. Following this criterion, we find that only $\sim40\%$ $D_s^{*-}$ photons are reconstructed successfully. Most of them have a truth-level energy above $E_{\rm th}$ (despite a failure of reconstruction for some ``energetic'' $D_s^{*-}$ photons due to, e.g., a collimation with other particles in the ECAL). In contrast, almost all $B_s^0 \to D_s^{*-} \tau^+ \nu_\tau$ and $B_s^0 \to D_s^{*-} \mu^+ \nu_\mu$ events containing a $D_s^{*-}$ photon with its energy below $E_{\rm th}$ leave an empty entry in the ECAL and hence are ``untagged''.  In addition to $E_{\rm th}$, the reconstruction efficiency of $D_s^{*-}$ photons can be impacted by the momentum resolution of the ECAL. This feature is shown in the right panel of Fig.~\ref{fig:Ds_pho_tag}, with a distribution of the tagged (T) and untagged (U) $D_s^{*-}$ photons w.r.t. $\Delta m$. Clearly, the reconstruction quality of $\Delta m$ tends to be lower for the untagged $D_s^{*-}$ photons. However, as the fraction of such untagged $D_s^{*-}$ photons is small in the pool, at a level of several percent only, we will focus on the effect of $E_{\rm th}$ below.

\begin{figure}[th!]
	\centering
	\includegraphics[width=7.cm]{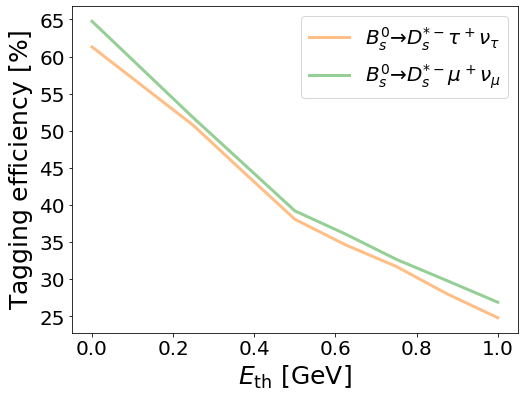}
	\includegraphics[width=8.3cm]{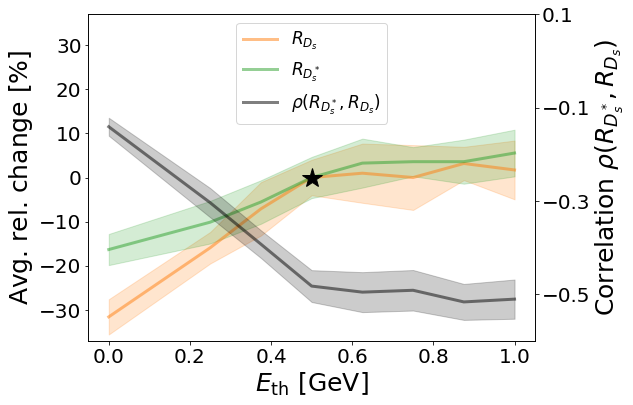}
	\caption{Tagging efficiency of $D_s^{*-}$ photons (left) and averaged relative change to the reference precisions of measuring $R_{D_s}$ and $R_{D_s^*}$ (and the correlation between these precisions) (right), with a varied value of $E_{\rm th}$. In the right panel, the averaged values (solid lines) and their variances (shaded bands) are calculated based on ten BDT analyses with their training and testing datasets defined in the caption of Fig.~\ref{fig:error_trends}. The reference precisions are simulated with $E_{\rm th}=0.5$~GeV, denoted as a black star. }
	\label{fig:Ds_variance}
\end{figure}

In Fig.~\ref{fig:Ds_variance}, we demonstrate the impacts of $E_{\rm th}$ on the tagging efficiency of $D_s^{*-}$ photons and the precisions of measuring $R_{D_s}$ and $R_{D_s^*}$ (and the correlation between these precisions). Clearly, reducing the  $E_{\rm th}$ value will improve both analyses.  It yields a positive change up to tens of percent to the tagging efficiency, relative to its reference value simulated at $E_{\rm th}=0.5$~GeV. Consistently, the expected BDT precisions of measuring $R_{D_s}$ and $R_{D_s^{*}}$ also get improved. To end this subsection, we point out that reducing $E_{\rm th}$ from its reference value will weaken the correlation between the $R_{D_s}$ and $R_{D_s^{*}}$ measurements significantly. This may further strengthen the constraints on the relevant SMEFT, a study to be performed in Sec.~\ref{sec:EFT}.

\subsection{Event Shape}
\label{ssec:eventshape}

\begin{figure}[h]
	\begin{center}
	\includegraphics[width=7.5 cm]{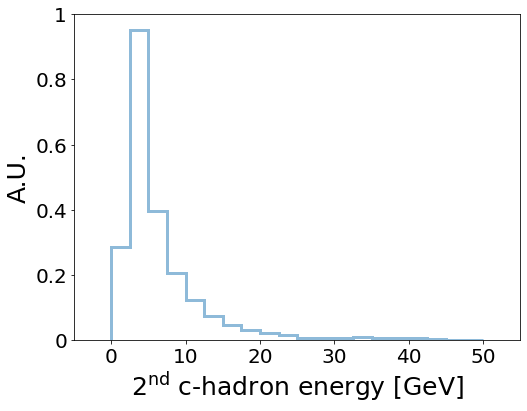} 
	\includegraphics[width=7.55 cm]{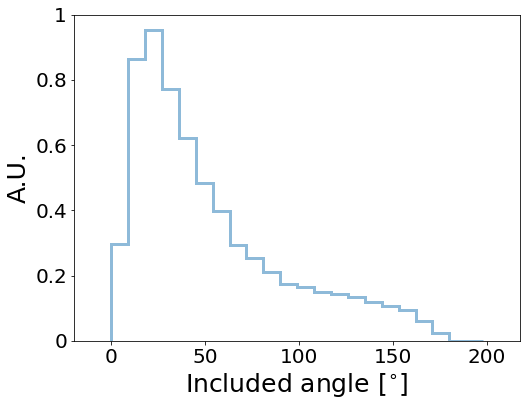} 
\end{center}
	\caption{Energy distribution of the second $c$-hadron in a $B_c^+$ event (left), and distribution of its included angle with the $B_c^+$ meson (right). 
	}
	\label{fig:2c}
\end{figure}

In the analyses above, we have focused on the features of $b$-hadron decay products. However, the kinematics of particles at event level, namely event shape~\cite{Bosch:2004cb}, may carry extra information to distinguish the signals from their backgrounds. The $B_c^+$ production in the $R_{J/\psi}$ measurement is such an example. In this process, two bottom and two charm quarks are produced~\cite{Chang:1992bb}. One charm quark and one bottom quark are then confined into a $B_c^+$ meson, while the second charm quark forms an extra $c$-hadron, as illustrated in Fig.~\ref{fig:cartoons2}. We show the energy distribution of the second $c$-hadron, and the distribution of its included angle with the $B_c^+$ meson in Fig.~\ref{fig:2c}. For many of these events, their second $c$-hadron has energy more than five or even ten GeVs, and its included angle with the $B_c^+$ meson can be quite big also. 
Such events have three hard or relatively hard heavy-flavored hadrons, {\it i.e.},  $B_c^+$ and extra $b$- and $c$- hadrons, yielding a shape different from those of the back-to-back $2b$ events and the multi-parton $4b$ events, where the heavy quarks stem from $Z$ decay or QCD radiation rather than weak decays. The $2b$ events have been known to significantly contribute to the combinatoric and muon mis-ID backgrounds. The observables of event shape thus could be applied to further improve the sensitivity of measuring $R_{J/\psi}$ by suppressing its backgrounds with the information beyond the $B_c^+$ decay~\footnote{Alternatively, one can require a successful reconstruction of extra $D$ meson via the decays such as $D^0\to K^-2\pi^+\pi^-$, $D^0\to K^+ \pi^-$ and $D^+\to K^-2\pi^+$, to improve the quality of reconstructing the $R_{J/\psi}$ signal events. The clean environment of a $Z$ factory will benefit this goal. However, the observables of event shape provide a more systematic and efficient way to look into the information beyond the $B_c^+$ decay. So we will focus on their performance in this paper.}.

The event-level observables are highly suitable for the analyses at $e^-e^+$ colliders, given no generic contaminations in hadron collisions applied such as pileups and underlying events. 
Many event-level observables have been originally proposed for the $e^-e^+$ and $e^-h$ events~\cite{Dasgupta:2003iq} rather than the $hh$ ones~\cite{Banfi:2010xy}. Especially, the definiteness of the center of mass frame for the $e^-e^+$ collision events have motivated two of the authors in this paper to build up a dictionary between the Mollweide projection of individual $e^-e^+$ collision events and the all-sky CMB map (see Tab.~2 in~\cite{Li:2020vav}), where the event-level kinematics corresponds to the anisotropy of CMB,  and accordingly a CMB-like observable scheme for collider events. In this observable scheme, the Fox-Wolfram (FW) moments~\cite{Fox:1978vu} of individual events play a leading role, just like the CMB power spectrum. 
For simplicity, we only consider the FW moments of visible energy of particles which are defined as 
\begin{equation}
H_{EE;l}= \sum_{m=-l}^l H_{EE; l,m} = \frac{4\pi}{2l +1}  \sum\limits_{i,j}   \frac{E_iE_j}{s}  \sum_{m=-l}^l \left (Y_l^m (\Omega_i)^*  Y_l^m (\Omega_j) \right) = \sum\limits_{i,j}\frac{E_iE_j}{s}P_l(\cos \Omega_{ij}) \ .  
\label{eq:FWdefinition}
\end{equation}
Here $Y_l^m (\Omega_i)$ is spherical harmonics of degree $l$ and order $m$,  $P_l(\cos \Omega_{ij})$ is Legendre polynomials, 
\begin{eqnarray}
\cos \Omega_{ij} = \cos \theta_i \cos \theta_j +  \sin \theta_i \sin \theta_j \cos (\phi_i -\phi_j)
\end{eqnarray}
is the cosine of the included angle between two visible particle $i$ and $j$. In this summation, $i$ and $j$ run over all visible particles in each event.

\begin{figure}[h]
	\begin{center}
	\includegraphics[width=\textwidth]{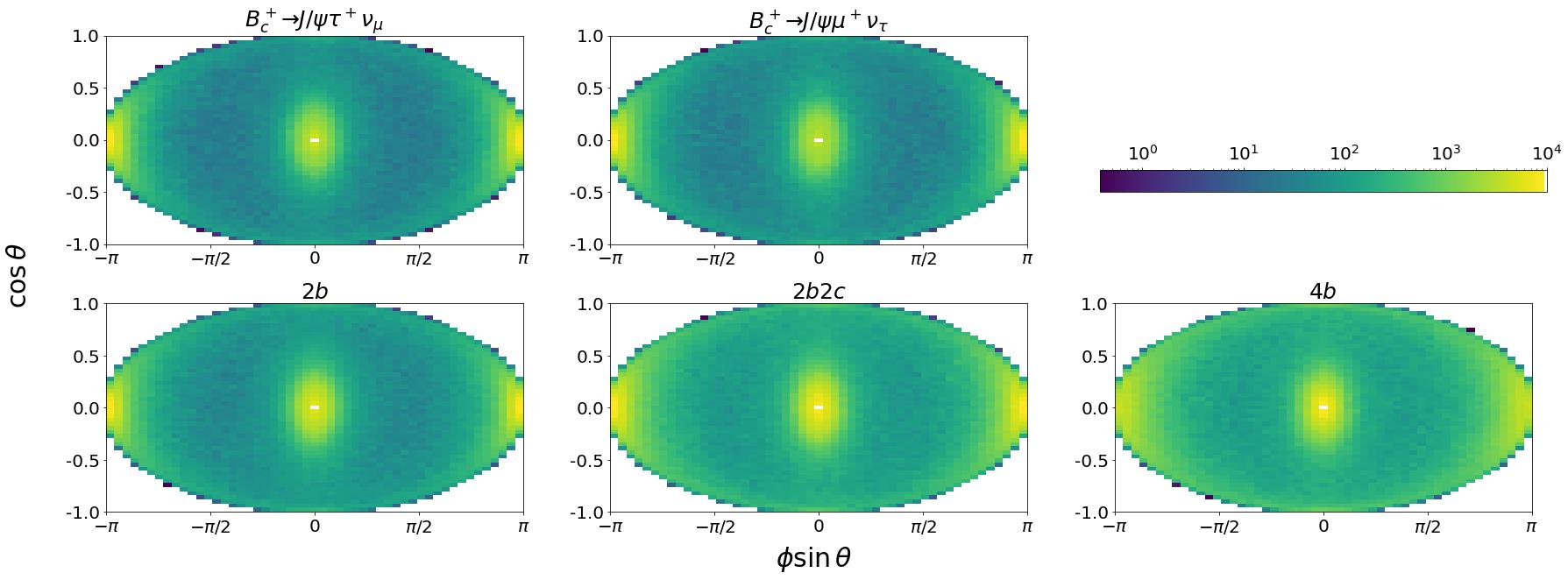} 
\end{center}
	\caption{Cumulative Mollweide projections (for details on such a projection, see Subsec. 2.1 in~\cite{Li:2020vav}) for the $B_c^+\to J/\psi\tau^+\nu_\tau$ and 
	$B_c^+\to J/\psi\mu^+\nu_\mu$ signal events (upper), and the $2b$, $4b$ and $2b2c$ combinatoric background events (bottom). In each panel, totally 10000 events have been projected. The brightness of each cell is scaled with the total energy (GeV) of the particle hits received. 
	}
	\label{fig:projection}
\end{figure}

\begin{figure}[h]
	\centering
	\includegraphics[width=7.5 cm]{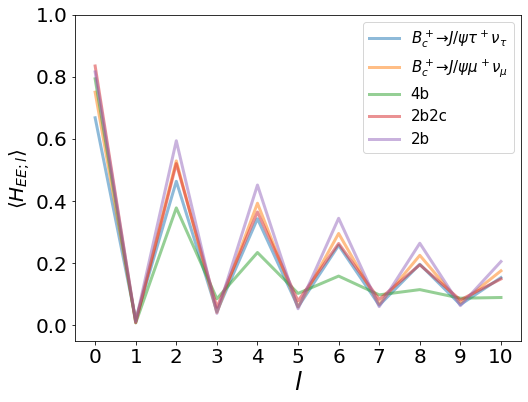}
		\includegraphics[width=7.4 cm]{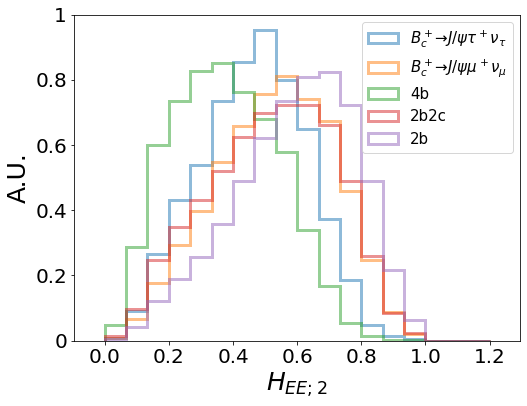}
	\caption{Averaged FW moments $\langle H_{EE;l} \rangle$ with $l=1, ... ..., 10$ (left), and event distribution w.r.t. $H_{EE;2}$ (right), for the $B_c^+\to J/\psi\tau^+\nu_\tau$ and 
	$B_c^+\to J/\psi\mu^+\nu_\mu$ signal events (upper), and the $2b$, $4b$ and $2b2c$ combinatoric background events. 
	}\label{fig:FWM}
\end{figure}

We show the cumulative Mollweide projections for the $B_c^+\to J/\psi\tau^+\nu_\tau$ and $B_c^+\to J/\psi\mu^+\nu_\mu$ events and the $2b$ and $4b$ (and also $2b2c$) background events in Fig.~\ref{fig:projection}. As the combinatoric background  receives the contributions from multiple $Z$ decay topologies, we require here all $2b$ (50.4\%), $4b$ (17.4\%) and $2b2c$ (32.3\%) events to be from this type of background. As a comparison, the inclusive background has similar event shape as that of the signals as both of them stem from the $Z\to B_c^++X$ production, while the mis-ID background is mainly from the $B\to J/\psi+\pi^++X$ decays and hence has a $2b$-like event shape. From this figure, one can see that the two bright spots in the projections are smeared more for the $4b$ and $2b2c$ events than the signal events and $2b$ events. This is consistent with our expectation. Based on such projections, we demonstrate the averaged FW moments $\langle H_{EE;l} \rangle$ (as a counterpart of the CMB power spectrum at the detector sphere~\cite{Li:2020vav}) with $l=1, ... ..., 10$ and the event distribution w.r.t. $H_{EE;2}$, for these signal and background events in Fig.~\ref{fig:FWM}. Note, the range of $l$ matches well with the angular resolution needed to look into the structure of signal events which is indicated by the right panel of Fig.~\ref{fig:2c}. Below are the main observations (for detailed discussions on the underlying physics of $\langle H_{EE;l} \rangle$  spectrum, see~\cite{Li:2020vav}). 
\begin{itemize}
\item Because of $P_l(-x) = (-1)^l P_l(x)$, the moments with odd $l$ are zero for the parity-even events such as the back-to-back $2b$ ones, which yield a zigzag structure for the spectra.

\item The tail for the $4b$ spectrum is damped more, compared to the other ones. This is because parton shower yields more particles in the final state of this class of events. The democracy of allocating visible energy among these particles tends to reduce their self-correlation contribution ({\it i.e., $H^{\rm self}_{EE;l} = \sum_i\frac{E_i^2}{s}$}) to the FW moments (which is universal to all $l$) and hence damp the spectrum tail.

\item  Note that $H_{EE;0}= \frac{(\sum_i E_i)^2}{s}$ denotes the squared share of the visible energy among the total in each event. The sorting of $\langle H_{EE;0} \rangle$ tells us that more missing energy tends to be produced for the $B_c^+\to J/\psi\tau^+\nu_\tau$ events. The event distribution w.r.t. $H_{EE;2}$ in the right panel reminds us that, unlike the CMB power spectrum, the $\langle H_{EE;l} \rangle$ spectrum is free from the ``cosmic variance'' problem, because the collider data is ample.

\item The FW moments of $B_c^+\to J/\psi\tau^+\nu_\tau$ and $B_c^+\to J/\psi\mu^+\nu_\mu$ are close to those of the $2b2c$ events. This can be understood since these signal events are  essentially the $2b2c$ events, except that they are produced with three heavy hadrons while the combinatoric background events typically contain four ones.  

\end{itemize}

\begin{table}[h!]
\begin{center}
\scalebox{0.78}{
\begin{tabular}{c|c|c|c|c|c|c}
\hline
	& \multicolumn{3}{c|}{Original} & \multicolumn{3}{c}{Original + FW moments} \\
\hline
	& Preselection & $y_{J/\psi}^\tau \geq0.03  $ & $y_{J/\psi}^\tau<0.03  $ & FW selection & $y_{J/\psi}^\tau \geq0.03  $ & $y_{J/\psi}^\tau<0.03  $ \\
	&  & $  \cap ~ y_{J/\psi}^\mu<0.97$ & $  \cap ~ y_{J/\psi}^\mu \geq 0.97$ & ($  y_{B_c^+} > 0.05$) & $ \cap ~  y_{J/\psi}^\mu<0.97$ & $ \cap ~ y_{J/\psi}^\mu \geq 0.97$ \\
	\hline
	\hline
	$B_c^+\to J/\psi\tau^+\nu_\tau$ &
	$3.08\times 10^3$ &
	$2.77\times 10^3$ &
	$2.06\times 10^2$ &
	
	$2.88\times 10^3$ &
	$2.60\times 10^3$ &
	$1.81\times 10^2$ 
	\\
	$B_c^+\to J/\psi\mu^+\nu_\mu$ &	
	$8.40\times 10^4$  &
	$4.33\times 10^3$  &
	$7.64\times 10^4$  &
	
	$6.56\times 10^4$  &
	$3.83\times 10^3$  &
	$5.95\times 10^4$
	\\
	Inclusive bkg. & 
	$3.90\times 10^3$ &
	$4.44\times 10^2$ &
	$3.67\times 10^2$ &
	
	$2.31\times 10^3$ &
	$3.76\times 10^2$ &
	$2.54\times 10^2$
	\\
	Cascade bkg. &
	$1.84\times 10^3$ &
	$1.15\times 10^2$ &
	$1.77\times 10^1$ &
	
	$1.03\times 10^3$ &
	$8.87\times 10^1$ &
	$8.87\times 10^0$
	\\
	Combinatoric bkg. &
	$7.78\times 10^4$ &
	$1.98\times 10^3$ &
	$1.60\times 10^2$ &
	
	$3.93\times 10^4$ &
	$1.61\times 10^3$ &
	$1.51\times 10^2$
	\\
	Mis-ID bkg. ($\times \epsilon_{\mu\pi}$) &
	 $1.10\times 10^8$ &
	 $2.38\times 10^5$ &
	 $7.59\times 10^4$ &
	 $2.79\times 10^7$ &
	 $1.68\times 10^5$ &
	 $2.64\times 10^4$\\
	\hline
	\hline 
	$S/B$ & - & $0.31$ & $55.62$ & - & $0.35$ & $72.41$\\
	\hline
	$R_{J/\psi} $ Rel. Precision & \multicolumn{3}{c|}{$4.12\times 10^{-2}$} & \multicolumn{3}{c}{$4.06\times 10^{-2}$} \\
	\hline
\end{tabular}
}
\caption{Sensitivities of measuring $R_{J/\psi}$. All numbers in this table are generated by averaging the results of ten BDT analyses with their training and testing datasets defined in the caption of Fig.~\ref{fig:error_trends}. In the ``Original + FW moments'' case, two BDT classifiers have been trained in each analysis: one is based on the FW moments only and another one uses the original set of observables as the inputs. Then the events are selected by the first BDT classifier before they are subject to the selection of the second BDT classifier. 
\label{tab:Jpsi_BDT_FWM}}
\end{center}
\end{table}

Finally let us consider the potential impacts of FW moments on the $R_{J/\psi}$ measurement. We perform an extra event selection with a BDT classifier developed with the $H_{EE;1-10}$ only before the BDT classifier based on the original set of observables is applied. The relevant analysis results are summarized in Tab.~\ref{tab:Jpsi_BDT_FWM}. From this table, one can see that the inclusion of FW moments for event selection yields a suppression to the backgrounds universally faster than the reduction of signal events. Thereinto, the mis-ID backgrounds are suppressed most efficiently, by a factor of nearly four.  As a result, the $S/B$ ratio for the tau and muon signal modes are  enhanced by more than $10\%$ and $30\%$ respectively, while the relative precision for measuring $R_{J/\psi}$ gets slightly improved. These outcomes suggest that the FW moments have worked as an independent discriminator beyond the  kinematics of $b$-hadron decay, making this measurement more robust. Searching for other multi-heavy-flavor processes such as exotic states~\cite{Ali:2017jda,Qin:2020zlg} may also benefit from such event-level observables. Notably, despite the gains from the FW moments, Pythia may not be accurate is simulating the event-level message especially for the multi-heavy-flavor productions~\cite{Chang:1992bb,Zheng:2017xgj}. To be conservative, we have not included the FW moments or other event-shape observables in the analyses yielding the conclusions of this paper. We hope that an improved simulation tool for such an analysis will be available~\cite{Chang:2015qea} in the near future.

\section{SMEFT Interpretation}
\label{sec:EFT}

In this section we will interpret in the SMEFT the projected sensitivities of measuring $R_{H_c}$, together with the observables involving the $b\to s\tau^+\tau^-$~\cite{Li:2020bvr} and $b\to s\nu\bar{\nu}$~\cite{Li:2022tov} transitions, at the future $Z$ factories. These measurements are performed at $\mu_b\sim m_b=4.8$~GeV, an energy scale well below the SMEFT cutoff. So we need to include the effects of the renormalization group (RG) running in this analysis. Concretely, we will take RG running for the Wilson coefficients of SMEFT from the hypothesized NP cutoff to the EW scale and match them with those of the low-energy EFT (LEFT) at this scale, and then run down the LEFT Wilson coefficients to $\mu_b$ such that they can interplay with the relevant measurements directly. Due to the generic symmetry requirement, the SMEFT Wilson coefficients are not fully independent. Their correlation is inherited by the LEFT Wilson coefficients, leaving an imprint in these measurements. Finally, the posterior distributions for the SMEFT Wilson coefficients will be analyzed by taking a Markov-Chain Monte-Carlo (MCMC) global fit.

\begin{figure}[h!]
\centering
\includegraphics[width=1\textwidth]{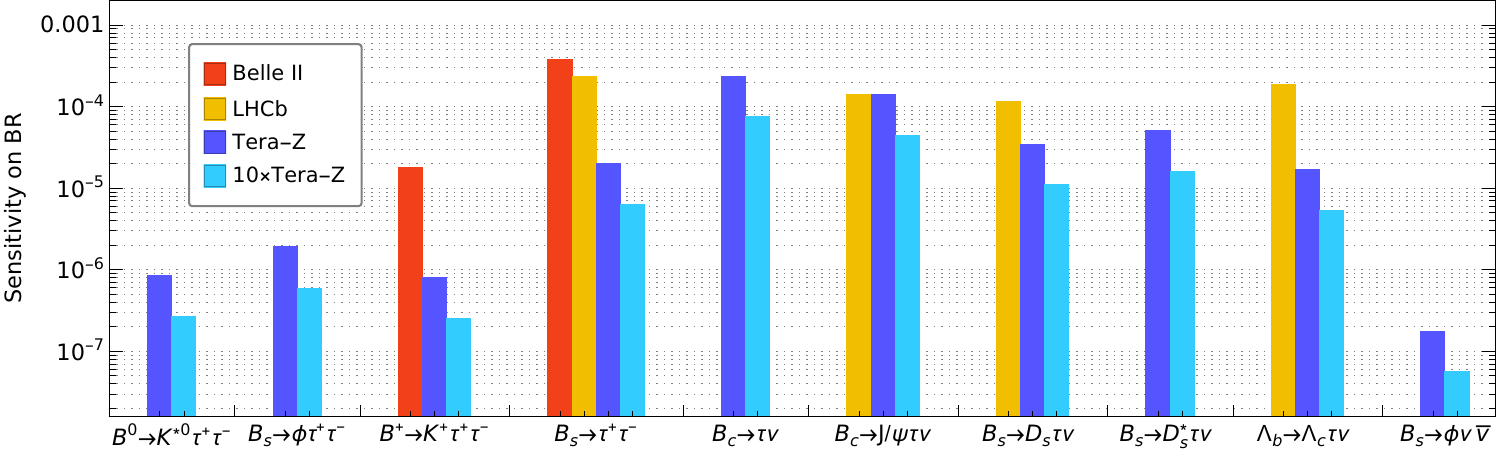}
\caption{Projected sensitivities of measuring the $b\to s\tau^+\tau^-$~\cite{Li:2020bvr}, $b\to c\tau\nu (B_c \to \tau \nu)$~\cite{Zheng:2020ult}, $b\to c\tau\nu$ (this work) and $b\to s\nu\bar{\nu}$~\cite{Li:2022tov} transitions at Tera-$Z$ and $10\times$Tera-$Z$. The sensitivities at Belle II @ 50 $\unit{ab^{-1}}$~\cite{Kou:2018nap} and LHCb Upgrade II~\cite{Bediaga:2018lhg,Bifani:2018zmi} have also been provided as a reference. Note that the sensitivities for each category might be based on different $\tau$ decay modes. For example, the LHCb sensitivities are generated by a combined analysis of $\tau^+ \to \pi^+\pi^-\pi^-(\pi^0)\nu$ and $\tau \to \mu \nu \bar \nu$. }
\label{fig:BR_overall}
\end{figure}

Here we have several comments. Firstly, for the convenience of discussions, we assume that the LFU violation is possible for the third generation only, whereas the physics of other generations have been constrained to be highly consistent with the SM by the ongoing measurements or the measurements at the future $Z$ factories. Secondly, we assume that the measured values for the relevant observables are centered at their SM predictions~\cite{Buras:2014fpa,Angelescu:2018tyl,Feruglio:2018fxo,Hu:2018veh,Alasfar:2020mne,Fajfer:2021cxa,Cornella:2021sby}. The expected measurement precisions are then summarized in Tab.~\ref{tab:SM_Values}.  We also present the expected precisions of measuring the $b\to s\tau^+\tau^-$~\cite{Li:2020bvr}, $b\to c\tau\nu$ and $b\to s\nu\bar{\nu}$~\cite{Li:2022tov} transitions in Fig.~\ref{fig:BR_overall}, as a specific demonstration of the $Z$-factory performance in exploring the FCNC and FCCC physics with the third-generation leptons. Thirdly, we ignore the systematics of measuring $R_{H_c}$ and the errors of calculating $R_{H_c}$. The former is expected to be canceled to some extent since $R_{H_c}$ denotes a ratio of two parallel measurements (this is also the reason that we apply the measurements of $R_{H_c}$ instead of Br$(H_b \to H_c \tau \nu_\tau)$ to constrain the SMEFT here). But the latter, which mainly arises from the uncertainty of the hadron decay form factors, is typically $\sim \mathcal{O}(10\%)$. This could be bigger than the statistical errors of the  $R_{H_c}$ measurements at $Z$ pole, and hence downgrade their capability to probe the SMEFT. We hope that the theoretical and experimental developments later will bring these uncertainties down to a level comparable to or even below the statistical errors of these measurements by the time of operating the future $Z$ factories.

\begin{table}[h!]
\centering
\fontsize{10pt}{12pt}\selectfont
\begin{tabular}{ccccccc}
\hline
& Physical Quantity & SM Value & Tera-$Z$ & $10\times$Tera-$Z$  & Belle II  & LHCb \\
\hline
& $R_{J/\psi}$ &  0.289 & $4.25\times 10^{-2}$ & $1.35\times 10^{-2}$ & - & -\\
\hline
& $R_{D_s}$ & 0.393 & $4.09\times 10^{-3}$ & $1.30\times 10^{-3}$ & - & -\\
\hline
& $R_{D_s^*}$ & 0.303 & $3.26\times 10^{-3}$ & $1.03\times 10^{-3}$ & - & - \\
\hline
& $R_{\Lambda_c}$ & 0.334 & $9.77\times 10^{-4}$ & $3.09\times 10^{-4}$ & - & -\\
\hline
& $\text{BR}(B_c\to \tau\nu)$ & $2.36\times 10^{-2}$~\cite{Zheng:2020emi} & 0.01~\cite{Zheng:2020emi} & $3.16\times 10^{-3}$ & -& - \\
\hline
& $\text{BR}(B^+\to K^+\tau^+\tau^-)$ &~ $1.01\times 10^{-7}$ & 7.92~\cite{Li:2020bvr} & 2.48~\cite{Li:2020bvr}  & 198~\cite{Kou:2018nap}& - \\
\hline
& $\text{BR}(B^0\to K^{\ast 0}\tau^+\tau^-)$ &~$0.825\times 10^{-7}$ & 10.3~\cite{Li:2020bvr} & 3.27~\cite{Li:2020bvr} & - & -\\
\hline
& $\text{BR}(B_s\to \phi\tau^+\tau^-)$ & $0.777\times 10^{-7}$ & 24.5~\cite{Li:2020bvr}& 7.59~\cite{Li:2020bvr} & -&- \\
\hline
& $\text{BR}(B_s\to \tau^+\tau^-)$ & $7.12\times 10^{-7}$ & 28.1~\cite{Li:2020bvr}& 8.85~\cite{Li:2020bvr} & - & 702~\cite{Bediaga:2018lhg} \\
\hline
& $\text{BR}(B^+\to K^+\bar\nu \nu)$ & $4.6\times 10^{-6}$~\cite{Kou:2018nap} & - & - & 0.11~\cite{Kou:2018nap}&- \\
\hline
& $\text{BR}(B^0\to K^{\ast 0}\bar\nu \nu)$ & $9.6\times 10^{-6}$~\cite{Kou:2018nap} & - & - & 0.096~\cite{Kou:2018nap} & -\\
\hline
& $\text{BR}(B_s\to \phi\bar\nu\nu)$  & $9.93\times 10^{-6}$~\cite{Li:2022tov} & $1.78\times 10^{-2}$~\cite{Li:2022tov} & $5.63\times 10^{-3}$ & - & -\\
\hline
\end{tabular}
\caption{SM predictions for the relevant observables and relative precisions for their measurements at Belle II @ 50 $\unit{ab^{-1}}$, LHCb Upgrade II, Tera-$Z$ and $10\times$Tera-$Z$. }
\label{tab:SM_Values}
\end{table}

\subsection{Low-Energy EFT}
\subsubsection{$b\to c\tau\nu$}
\label{sssec:bctaunuEFT}

In the 6D LEFT, the $b\to c\tau\nu$ transitions are described by
\begin{equation}
\mathcal{L}^{\rm LE}_{b\to c\tau\nu} =  -\frac{4 G_F V_{cb}}{\sqrt{2}} [(C_{V_L}^{\tau}|_{\rm SM}+\delta C_{V_L}^{\tau}) O_{V_L}^\tau+ C_{V_R}^{{\tau}} O_{V_R}^\tau+C_{S_L}^\tau O_{S_L}^\tau+C_{S_R}^\tau O_{S_R}^\tau+C_T^\tau O_T^\tau]+\rm{h.c.},
\label{eq:CCOV}
\end{equation}
where
\begin{eqnarray}
&&O^\tau_{V_L}=[\bar{c}\gamma^\mu P_L  b ][\bar{\tau}\gamma_\mu P_{L}\nu]~,  \quad O^\tau_{V_R}=[\bar{c}\gamma^\mu P_R b ][\bar{\tau}\gamma_\mu P_{L}\nu]~, 
 \nonumber \\
&&O^\tau_{S_L}=[\bar{c} P_L b ][\bar{\tau} P_{L}\nu]~,  \quad  \quad  \quad
 O^\tau_{S_R}=[\bar{c}P_R b ][\bar{\tau} P_{L}\nu]~,  \nonumber  \\ && O^\tau_{T}=[\bar{c}\sigma^{\mu\nu} b ][\bar{\tau}\sigma_{\mu\nu} P_{L}\nu]~.  
\end{eqnarray}
The subscripts ``$V_L$'', ``$V_R$'', ``$S_L$'', ``$S_R$'', and ``$T$'' denote the left- and right-handed vector currents, left- and right-handed scalar currents and the tensor current, respectively. Note, the SM contribution to the left-handed vector current is non-trivial due to $W$ boson emission, leaving $C_{V_L}^{\tau}|_{\rm SM}=1$ at $\mu_b=4.8$ GeV. The superscript ``$\tau$'' implies that any deviations of these Wilson coefficients from their SM predictions will violate the LFU explicitly.

Now we are able to calculate the LEFT predictions for $R_{H_c}$, which are given by (the details for these calculations are summarized in Appendix~\ref{app:calc}): 
\begin{equation}
\begin{aligned}
\frac{R_{J/\psi}}{R_{J/\psi}^{\rm SM}} =& ~1.0 + \text{Re} ( 0.12 C_{S_L}^{\tau} + 0.034 |C_{S_L}^{\tau}|^2 - 0.12 C_{S_R}^{\tau} - 0.068 C_{S_L}^{\tau} C_{S_R}^{\tau\ast} + 0.034 |C_{S_R}^{\tau}|^2 \\
& - 5.3 C_{T}^{\tau} + 13 |C_{T}^{\tau}|^2 - 1.9 C_{V_R}^{\tau} - 0.12 C_{S_L}^{\tau} C_{V_R}^{\tau\ast} + 0.12 C_{S_R}^{\tau} C_{V_R}^{\tau\ast} \\
& + 5.8 C_{T}^{\tau} C_{V_R}^{\tau\ast} + 1.0 |C_{V_R}^{\tau}|^2 + 2.0 \delta C_{V_L}^{\tau} + 0.12 C_{S_L}^{\tau} \delta C_{V_L}^{\tau\ast} \\
&- 0.12 C_{S_R}^{\tau} \delta C_{V_L}^{\tau\ast}- 5.3 C_{T}^{\tau} \delta C_{V_L}^{\tau\ast} - 1.9 C_{V_R}^{\tau} \delta C_{V_L}^{\tau\ast} + 1.0 |\delta C_{V_L}^{\tau}|^2  )   \ ,
\end{aligned}
\label{eq:RJpsinumerical}
\end{equation}
\begin{equation}
\begin{aligned}
\frac{R_{D_s}}{R_{D_s}^{\rm SM}}=& ~1.0 + \text{Re}  ( 1.6 C_{S_L}^{\tau} + 1.2 |C_{S_L}^{\tau}|^2 + 1.6 C_{S_R}^{\tau} + 2.4 C_{S_L}^{\tau} C_{S_R}^{\tau\ast} + 1.2 |C_{S_R}^{\tau}|^2 \\
& + 1.4 C_{T}^{\tau} + 1.4 |C_{T}^{\tau}|^2 + 2.0 C_{V_R}^{\tau} + 1.6 C_{S_L}^{\tau} C_{V_R}^{\tau\ast} + 1.6 C_{S_R}^{\tau} C_{V_R}^{\tau\ast} \\
& + 1.4 C_{T}^{\tau} C_{V_R}^{\tau\ast} + 1.0 |C_{V_R}^{\tau}|^2 + 2.0 \delta C_{V_L}^{\tau} + 1.6 C_{S_L}^{\tau} \delta C_{V_L}^{\tau\ast} \\
& + 1.6 C_{S_R}^{\tau} \delta C_{V_L}^{\tau\ast} + 1.4 C_{T}^{\tau} \delta C_{V_L}^{\tau\ast} + 2.0 C_{V_R}^{\tau} \delta C_{V_L}^{\tau\ast} + 1.0 |\delta C_{V_L}^{\tau}|^2  )   \ ,
\end{aligned}
\label{eq:RDsnumerical}
\end{equation}
\begin{equation}
\begin{aligned}
\frac{R_{D_s^*}}{R_{D_s^*}^{\rm SM}}=& ~1.0 + \text{Re}  ( 0.085 C_{S_L}^{ \tau} + 0.026 |C_{S_L}^{ \tau}|^2 - 0.085 C_{S_R}^{ \tau} - 0.052 C_{S_L}^{ \tau} C_{S_R}^{ \tau\ast} \\ 
& + 0.026 |C_{S_R}^{ \tau}|^2 - 4.6 C_{T}^{ \tau} + 15 |C_{T}^{ \tau}|^2 - 1.8 C_{V_R}^{ \tau} - 0.085 C_{S_L}^{ \tau} C_{V_R}^{ \tau\ast} \\
& + 0.085 C_{S_R}^{ \tau} C_{V_R}^{ \tau\ast} + 6.4 C_{T}^{ \tau} C_{V_R}^{ \tau\ast} + 1.0 |C_{V_R}^{ \tau}|^2 + 2.0 \delta C_{V_L}^{ \tau} + 0.085 C_{S_L}^{ \tau} \delta C_{V_L}^{ \tau\ast} \\
& - 0.085 C_{S_R}^{ \tau} \delta C_{V_L}^{ \tau\ast} - 4.6 C_{T}^{ \tau} \delta C_{V_L}^{ \tau\ast} - 1.8 C_{V_R}^{ \tau} \delta C_{V_L}^{ \tau\ast} + 1.0 |\delta C_{V_L}^{ \tau }|^2  )   \ ,
\end{aligned}
\label{eq:RDsstarnumerical}
\end{equation}
\begin{equation}
\begin{aligned}
\frac{R_{\Lambda_c}}{R_{\Lambda_c}^{\rm SM}} =& ~1.0 + \text{Re} ( 0.39 C_{S_L}^{ \tau} + 0.34 |C_{S_L}^{ \tau}|^2 + 0.49 C_{S_R}^{ \tau} + 0.61 C_{S_L}^{ \tau} C_{S_R}^{ \tau\ast} + 0.34 |C_{S_R}^{ \tau }|^2 \\
& + 1.1 C_{T}^{ \tau} + 12 |C_{T}^{ \tau }|^2 - 0.71 C_{V_R}^{ \tau} + 0.49 C_{S_L}^{ \tau} C_{V_R}^{ \tau\ast} + 0.39 C_{S_R}^{ \tau} C_{V_R}^{ \tau\ast} \\
& - 1.7 C_{T}^{ \tau} C_{V_R}^{ \tau\ast} + 1.0 |C_{V_R}^{ \tau}|^2 + 2.0 \delta C_{V_L}^{ \tau} + 0.39 C_{S_L}^{ \tau} \delta C_{V_L}^{ \tau\ast} \\
&+ 0.49 C_{S_R}^{ \tau} \delta C_{V_L}^{ \tau\ast} + 1.1 C_{T}^{ \tau} \delta C_{V_L}^{ \tau\ast} - 0.71 C_{V_R}^{ \tau} \delta C_{V_L}^{ \tau\ast} + 1.0 |\delta C_{V_L}^{ \tau }|^2  )   \ .
\end{aligned}
\label{eq:RLambdacnumerical}
\end{equation}
We also include the measurement of BR$(B_c\to \tau\nu)$ in this analysis. This channel is sensitive to the axial vector ($C_{V_L}^\tau-C_{V_R}^\tau$) and pseudoscalar ($C_{S_L}^\tau-C_{S_R}^\tau$) combinations only. Here we take the results reported in~\cite{Zheng:2020emi}:
\begin{equation}
\begin{aligned}
\frac{\text{BR}(B_c\to \tau\nu)}{\text{BR}(B_c\to \tau\nu)^{\rm SM}}= & ~1.0 + \text{Re} ( 7.1 C_{S_L}^{\tau} + 13 |C_{S_L}^{\tau}|^2 - 7.1 C_{S_R}^{\tau} - 26 C_{S_L}^{\tau} C_{S_R}^{\tau\ast} + 13 |C_{S_R}^{\tau}|^2\\
& - 2.0 C_{V_R}^{\tau} - 7.1 C_{S_L}^{\tau} C_{V_R}^{\tau\ast} + 7.1 C_{S_R}^{\tau} C_{V_R}^{\tau\ast} +1.0 |C_{V_R}^{\tau}|^2 + 2.0 \delta C_{V_L}^{\tau} \\
&+ 7.1 C_{S_L}^{\tau} \delta C_{V_L}^{\tau\ast} - 7.1 C_{S_R}^{\tau} \delta C_{V_L}^{\tau\ast} - 2.0 C_{V_R}^{\tau} \delta C_{V_L}^{\tau\ast} +1.0 |\delta C_{V_L}^{\tau}|^2   )   \ .
\end{aligned}
\label{eq:Bctaununumerical}
\end{equation}

Notably, these channels represent four types of the $b\to c\tau\nu$ transitions: the vector type ($R_{J/\psi}$ and $R_{D_s^\ast}$), the pseudoscalar type ($R_{D_s}$), the baryon type ($R_{\Lambda_c}$) and the annihilation type (BR$(B_c\to \tau\nu)$). The responses to the NP tend to be aligned for the channels of the same types, as indicated by Eq.~(\ref{eq:RJpsinumerical}) and Eq.~(\ref{eq:RDsstarnumerical}). The difference between them mainly arises from meson masses and decay form factors, which are usually small. So it is important to combine all four types of measurements for more accurate EFT interpretation.

\subsubsection{$b\to s\tau^+\tau^-$}
\label{sssec:bstautauEFT}
In the 6D LEFT, the $b\to s\tau^+\tau^-$ transitions are described by
\begin{equation}
\begin{aligned}
\mathcal{L}^{\rm LE}_{b\to s\tau^+\tau^-} =  \frac{4 G_F V_{tb}V^\ast_{ts}}{\sqrt{2}}[& (C^\tau_9|_{\rm SM}+\delta C^\tau_9) O_9^\tau+(C^\tau_{10}|_{\rm SM}+\delta C^\tau_{10}) O_{10}^\tau+C_{9}^{\prime\tau} O_{9}^{\prime\tau}+ C_{10}^{\prime\tau}  O_{10}^{\prime\tau}\\
& + C_S^\tau O_S^{\tau} + C_{S}^{\prime\tau}  O_{S}^{\prime\tau} + C_P^\tau O_P^{\tau} + C_{P}^{\prime\tau}  O_{P}^{\prime\tau} \\
&+ C_T^\tau O_T^{\tau} + C_{T5}^{\tau}  O_{T5}^{\tau}] +\rm{h.c.}~,
\end{aligned}
\label{eq:NCH}
\end{equation}
where
\begin{eqnarray}
&&
O_{9(10)}^{\tau}=\frac{\alpha}{4\pi}[\bar{s}\gamma^\mu P_L b ][\bar{\tau}\gamma_\mu(\gamma^5)\tau]~, \quad O_{9(10)}^{\prime\tau}=\frac{\alpha}{4\pi}[\bar{s}\gamma^\mu P_R b ][\bar{\tau}\gamma_\mu(\gamma^5)\tau]~,  \nonumber \\
&& O_{S(P)}^{\tau}=\frac{\alpha}{4\pi}[\bar{s} P_R b ][\bar{\tau}(\gamma^5)\tau]~, \quad  \quad  \quad O_{S(P)}^{\prime\tau}=\frac{\alpha}{4\pi}[\bar{s} P_L b ][\bar{\tau}(\gamma^5)\tau]~,  \nonumber \\
&& O_{T(T5)}^{\tau}=\frac{\alpha}{4\pi}[\bar{s} \sigma_{\mu\nu} b ][\bar{\tau}\sigma^{\mu\nu}(\gamma^5)\tau]~.
\label{eq:NCOT}
\end{eqnarray}
As it occurs to the left-handed vector current of $b\to c\tau\nu$, the SM contributes to $O_9^\tau$ and $O_{10}^\tau$. The contributions include the gluon penguin diagrams with  extra quark loop and the radiative $b\to s\gamma^\ast \to s\tau^+\tau^-$ processes, yielding $C^\tau_{9(10)}|_{\rm SM} \approx 4.07(-4.31)$ at $\mu_b=4.8$GeV~\cite{DescotesGenon:2011yn}. $\alpha$ is running fine-structure constant. Note, we tolerate the abuse of notation here for the $O_T^\tau$ operator and its Wilson coefficient. This notation has been used for the tensor-current operator in the $b\to c\tau\nu$ LEFT defined in Eq.~(\ref{eq:CCOV}). We will see later that both $O_{T}^{\tau}$ and $O_{T5}^{\tau}$ in this Lagrangian are irrelevant to the SMEFT interpretation.

The LEFT predictions for Br$(B^+\to K^+\tau^+\tau^-)$, Br$(B^0\to K^{\ast 0}\tau^+\tau^-)$, Br$(B_s\to \phi\tau^+\tau^-)$ and Br$(B_s\to\tau^+\tau^-)$ are given below~\footnote{These relations are slightly different from those in~\cite{Capdevila:2017iqn}. The main reason is that we have not taken a full consideration on the uncertainties of the decay form factors for simplicity.  Moreover, unlike~\cite{Capdevila:2017iqn} where four LEFT operators are turned on, here we consider totally ten LEFT operators instead.}:
\begin{equation}
\begin{aligned}
\frac{\text{BR}(B^+\to K^+\tau^+\tau^-)}{\text{BR}(B^+\to K^+\tau^+\tau^-)^{\rm SM}}= &  ~1.0 + \text{Re} ( -0.35 C_{10}^{\prime \tau} + 0.041 |C_{10}^{\prime \tau}|^2 + 0.14 C_{9}^{\prime \tau} + 0.019 |C_{9}^{\prime \tau}|^2 \\ 
& - 0.34 C_{P}^{\tau} + 0.079 C_{10}^{\prime \tau} C_{P}^{\tau\ast}+ 0.043 |C_{P}^{ \tau}|^2 - 0.34 C_{P}^{\prime \tau} + 0.079 C_{10}^{\prime \tau} C_{P}^{\prime \tau\ast} \\ 
& + 0.086 C_{P}^{\tau} C_{P}^{\prime \tau\ast} + 0.043 |C_{P}^{\prime \tau}|^2 + 0.014 |C_{S}^{ \tau}|^2 + 0.027 C_{S}^{\tau} C_{S}^{\prime \tau\ast} \\
& + 0.014 |C_{S}^{\prime \tau}|^2 + 0.018 |C_{T5}^{ \tau}|^2 + 0.37 C_{T}^{\tau} + 0.10 C_{9}^{\prime \tau} C_{T}^{\tau\ast} \\
& + 0.15 |C_{T}^{ \tau }|^2 - 0.35 \delta C_{10}^{\tau} + 0.082 C_{10}^{\prime \tau} \delta C_{10}^{\tau\ast} + 0.079 C_{P}^{\tau} \delta C_{10}^{\tau\ast} \\
& + 0.079 C_{P}^{\prime \tau} \delta C_{10}^{\tau\ast} + 0.041 |\delta C_{10}^{ \tau}|^2 + 0.14 \delta C_{9}^{\tau} + 0.038 C_{9}^{\prime \tau} \delta C_{9}^{\tau\ast}\\
& + 0.10 C_{T}^{\tau} \delta C_{9}^{\tau\ast} + 0.019 |\delta C_{9}^{ \tau}|^2 ) \ ,
\end{aligned}
\label{eq:BKtautaunumerical}
\end{equation}
\begin{equation}
\begin{aligned}
\frac{\text{BR}(B^0\to K^{\ast 0}\tau^+\tau^-)}{\text{BR}(B^0\to K^{\ast 0}\tau^+\tau^-)^{\rm SM}}=&  ~1.0 + \text{Re}( 0.13 C_{10}^{\prime \tau} + 0.018 |C_{10}^{\prime \tau}|^2 - 0.31 C_{9}^{\prime \tau} + 0.059 |C_{9}^{\prime \tau }|^2 \\
& - 0.057 C_{P}^{\tau}  - 0.013 C_{10}^{\prime \tau} C_{P}^{\tau\ast} + 0.0062 |C_{P}^{ \tau}|^2 + 0.057 C_{P}^{\prime \tau} \\
& + 0.013 C_{10}^{\prime \tau} C_{P}^{\prime \tau\ast} - 0.012 C_{P}^{\tau} C_{P}^{\prime \tau\ast}  + 0.0062 |C_{P}^{\prime \tau }|^2 + 0.0014 |C_{S}^{ \tau}|^2 \\
& - 0.0029 C_{S}^{\tau} C_{S}^{\prime \tau\ast} + 0.0014 |C_{S}^{\prime \tau}|^2 + 1.3 C_{T5}^{\tau} - 0.39 C_{9}^{\prime \tau} C_{T5}^{\tau\ast} \\
& + 0.77 |C_{T5}^{ \tau}|^2 + 0.22 C_{T}^{\tau} + 0.068 C_{9}^{\prime \tau} C_{T}^{\tau\ast} + 0.24 |C_{T}^{ \tau }|^2 \\
& - 0.15 \delta C_{10}^{\tau} - 0.030 C_{10}^{\prime \tau} \delta C_{10}^{\tau\ast} + 0.013 C_{P}^{\tau} \delta C_{10}^{\tau\ast} - 0.013 C_{P}^{\prime \tau} \delta C_{10}^{\tau\ast}\\
& + 0.018 |\delta C_{10}^{ \tau }|^2 + 0.40 \delta C_{9}^{\tau} - 0.090 C_{9}^{\prime \tau} \delta C_{9}^{\tau\ast} + 0.39 C_{T5}^{\tau} \delta C_{9}^{\tau\ast}\\
& + 0.068 C_{T}^{\tau} \delta C_{9}^{\tau\ast} + 0.059 |\delta C_{9}^{ \tau}|^2 ) \ ,
\end{aligned}
\label{eq:BKstartautaunumerical}
\end{equation}
\begin{equation}
\begin{aligned}
\frac{\text{BR}(B_s\to \phi\tau^+\tau^-)}{\text{BR}(B_s\to \phi\tau^+\tau^-)^{\rm SM}}= & ~1.0 + \text{Re}( 0.14 C_{10}^{\prime \tau} + 0.017 |C_{10}^{\prime \tau}|^2 - 0.33 C_{9}^{\prime \tau} + 0.060 |C_{9}^{\prime \tau}|^2 \\
& - 0.057 C_{P}^{\tau} - 0.013 C_{10}^{\prime \tau} C_{P}^{\tau\ast} + 0.0062 |C_{P}^{ \tau}|^2 + 0.057 C_{P}^{\prime \tau} \\
& + 0.013 C_{10}^{\prime \tau} C_{P}^{\prime \tau\ast} - 0.012 C_{P}^{\tau} C_{P}^{\prime \tau\ast}  + 0.0062 |C_{P}^{\prime \tau}|^2 + 0.0014 |C_{S}^{ \tau}|^2 \\
& - 0.0028 C_{S}^{\tau} C_{S}^{\prime \tau\ast} + 0.0014 |C_{S}^{\prime \tau}|^2 + 1.4 C_{T5}^{\tau} - 0.41 C_{9}^{\prime \tau} C_{T5}^{\tau\ast} \\
& + 0.80 |C_{T5}^{ \tau}|^2 + 0.17 C_{T}^{\tau} + 0.054 C_{9}^{\prime \tau} C_{T}^{\tau\ast} + 0.21 |C_{T}^{ \tau}|^2 \\
& - 0.15 \delta C_{10}^{\tau} - 0.031 C_{10}^{\prime \tau} \delta C_{10}^{\tau\ast} + 0.013 C_{P}^{\tau} \delta C_{10}^{\tau\ast} - 0.013 C_{P}^{\prime \tau} \delta C_{10}^{\tau\ast} \\
& + 0.018 |\delta C_{10}^{ \tau}|^2 + 0.40 \delta C_{9}^{\tau} - 0.097 C_{9}^{\prime \tau} \delta C_{9}^{\tau\ast} + 0.41 C_{T5}^{\tau} \delta C_{9}^{\tau\ast} \\
& + 0.054 C_{T}^{\tau} \delta C_{9}^{\tau\ast} + 0.060 |\delta C_{9}^{ \tau}|^2 ) \ ,
\end{aligned}
\label{eq:Bsphitautaunumerical}
\end{equation}
\begin{equation}
\begin{aligned}
\frac{\text{BR}(B_s\to \tau^+\tau^-)}{\text{BR}(B_s\to \tau^+\tau^-)^{\rm SM}}= & ~1.0 + \text{Re} ( 0.46 C_{10}^{\prime \tau} + 0.054 |C_{10}^{\prime \tau}|^2 - 0.78 C_{P}^{\tau} - 0.18 C_{10}^{\prime \tau} C_{P}^{\tau\ast} \\
& + 0.15 |C_{P}^{ \tau}|^2 + 0.78 C_{P}^{\prime \tau} + 0.18 C_{10}^{\prime \tau} C_{P}^{\prime \tau\ast} - 0.31 C_{P}^{\tau} C_{P}^{\prime \tau\ast} \\
& + 0.15 |C_{P}^{\prime \tau}|^2 + 0.086 |C_{S}^{ \tau}|^2 - 0.17 C_{S}^{\tau} C_{S}^{\prime \tau\ast}\\
& + 0.086 |C_{S}^{\prime \tau}|^2 - 0.46 \delta C_{10}^{\tau} - 0.11 C_{10}^{\prime \tau} \delta C_{10}^{\tau\ast} \\
& + 0.18 C_{P}^{\tau} \delta C_{10}^{\tau\ast} - 0.18 C_{P}^{\prime \tau} \delta C_{10}^{\tau\ast} + 0.054 |\delta C_{10}^{ \tau}|^2 ) \ .
\end{aligned}
\label{eq:Bstautaunumerical}
\end{equation}

The collider phenomenology on the $b\to s\tau^+\tau^-$ transitions have been studied in various contexts~\cite{Kamenik:2017ghi,Capdevila:2017iqn,Li:2020bvr}. Currently, the upper limits set  by BaBar and LHCb for their branching ratios are $\sim \mathcal{O}(10^{-3})$~\cite{TheBaBar:2016xwe,Aaij:2017xqt}. They are much higher than the SM predictions which are typically $\sim \mathcal{O}(10^{-7})$. Recently, a systematic study performed in~\cite{Li:2020bvr} indicates that these limits (except Br$(B_s \to \tau^+\tau^-)$) can be improved to $\sim \mathcal{O}(10^{-7})$ at Tera-$Z$ and even more for $10\times$Tera-$Z$. The relevant sensitivity inputs on the $b\to s\tau^+\tau^-$ measurements at the $Z$ pole will be mainly based on this paper.

\subsubsection{$b\to s\bar\nu\nu$}
\label{sssec:bsnunuEFT}
In the 6D LEFT, the $b\to s\bar\nu\nu$ transitions are described by 
\begin{equation}
\begin{aligned}
\mathcal{L}^{\rm LE}_{b\to s\bar\nu\nu} = + \frac{4 G_F V_{tb}V^\ast_{ts}}{\sqrt{2}}[(C_L^{\nu}|_{\rm SM}+\delta C_L^{\nu})O_L^{\nu}+C_R^{\nu}O_R^{\nu}] +\rm{h.c.}~,
\end{aligned}
\label{eq:NCnuH}
\end{equation}
where
\begin{equation}
O_{L(R)}^{\nu}=\frac{\alpha}{4\pi}[\bar{s}\gamma^\mu P_{L(R)} b ][\bar{\nu}\gamma_\mu(1-\gamma^5)\nu]~.
\label{eq:NCnuOLR}
\end{equation}
$O_L^{\nu}$ receives contributions from the SM at loop level. Combining the EW contributions and the NLO QCD corrections yields $C^\nu_{L}|_{\rm SM} \approx -6.47$~\cite{Buras:2014fpa}. Notably, the three flavors of neutrinos all contribute at colliders and are mutually indistinguishable. Here we assume a deviation from the SM prediction to be possible for the third generation only, as discussed above.

The LEFT predictions for $\text{BR}(B^+\to K^+\bar\nu \nu)$, $\text{BR}(B^0\to K^{\ast 0}\bar\nu \nu)$ and $\text{BR}(B_s\to \phi\bar\nu \nu)$ are given below~\cite{Buras:2014fpa}:
\begin{equation}
\begin{aligned}
\frac{\text{BR}(B^+\to K^+\bar\nu \nu)}{\text{BR}(B^+\to K^+\bar\nu \nu)^{\rm SM}}= & \frac{1}{3}\left [2+ (1-2\eta)\epsilon^2 \right ] \ , \\
\frac{\text{BR}(B^0\to K^{\ast 0}\bar\nu \nu)}{\text{BR}(B^0\to K^{\ast 0}\bar\nu \nu)^{\rm SM}}= & \frac{1}{3}\left [2+ (1+\kappa_{K^{\ast 0}}\eta)\epsilon^2 \right ] \ ,\\
\frac{\text{BR}(B_s\to \phi\bar\nu \nu)}{\text{BR}(B_s\to \phi\bar\nu \nu)^{\rm SM}}= & \frac{1}{3}\left [2+ (1+\kappa_\phi\eta)\epsilon^2 \right ] \ , 
\end{aligned}
\label{eq:dinudecay}
\end{equation}
where  $\kappa_{K^{\ast 0}}=1.34$~\cite{Buras:2014fpa}, $\kappa_\phi=1.56$~\cite{Li:2022tov} and 
\begin{equation}
\epsilon=\frac{\sqrt{|C_L^\nu|^2+|C_R^\nu|^2}}{C^\nu_{L}|_{\rm SM}}~,~ \ \ \eta=\frac{-\text{Re}(C_L^\nu C_R^{\nu\ast})}{|C_L^\nu|^2+|C_R^\nu|^2} \ .
\label{eq:epsiloneta}
\end{equation}
At Belle II with 50~ab$^{-1}$, relative sensitivities up to $11\%$ and $9.6\%$ could be achieved for Br$(B^+\to K^+\bar\nu \nu)$ and Br$(B^0\to K^{\ast 0}\bar\nu \nu)$, respectively~\cite{Kou:2018nap}. The CEPC may constrain Br$(B_s\to \phi\bar\nu \nu)$ with a relative sensitivity $\sim 1.78\%$, as reported in~\cite{Li:2022tov}.

\subsection{SMEFT and Matching}

The SMEFT respects the SM gauge symmetries, namely $SU(3)_c\times SU(2)_L\times U(1)_Y$. Its 6D operators contributing to the $b\to c\tau\nu$, $b\to s\tau^+\tau^-$ and $b\to s\bar\nu\nu$ transitions are given by~\cite{Grzadkowski:2010es,Azatov:2018knx} 
\begin{align}
\mathcal{L}^{\rm SM} \supset \frac{1}{\Lambda^2} \sum_{i,j,k,l} ( & [C^{(1)}_{\ell q}]_{ijkl} [O^{(1)}_{\ell q}]_{ijkl} + [C^{(3)}_{\ell q}]_{ijkl} [O^{(3)}_{\ell q}]_{ijkl}+ [C_{ed}]_{ijkl} [O_{ed}]_{ijkl} \\\nonumber
& + [C_{\ell d}]_{ijkl} [O_{\ell d}]_{ijkl} + [C_{qe}]_{ijkl} [O_{qe}]_{ijkl} + [C_{\ell edq}]_{ijkl} [O_{\ell edq}]_{ijkl}  \\
\nonumber
& + [C^{(1)}_{\ell equ}]_{ijkl} [O^{(1)}_{\ell equ}]_{ijkl} + [C^{(3)}_{\ell equ}]_{ijkl} [O^{(3)}_{\ell equ}]_{ijkl}  ) +{\rm h.c.}~,
\end{align}
where 
\begin{eqnarray}
&&[O^{(1)}_{\ell q}]_{ijkl}=[\bar{L}_i\gamma_\mu L_j][\bar{Q}_k\gamma^\mu Q_l]~, \quad [O^{(3)}_{\ell q}]_{ijkl}=[\bar{L}_i\gamma_\mu \sigma^a L_j][\bar{Q}_k\gamma^\mu \sigma^a Q_l]~,  \nonumber \\
&&[O_{ed}]_{ijkl}=[\bar{\ell}_i\gamma_\mu  \ell_j][\bar{d}_k\gamma^\mu  d_l]~, \quad \quad  \ [O_{\ell d}]_{ijkl}=[\bar{L}_i\gamma_\mu  L_j][\bar{d}_k\gamma^\mu  d_l]~, \nonumber \\
&&[O_{qe}]_{ijkl}=[\bar{\ell}_i\gamma_\mu \ell_j][\bar{Q}_k\gamma^\mu Q_l]~, \quad \ \  [O_{\ell edq}]_{ijkl}=[\bar{L}_i^I \ell_j][\bar{d}_k Q_l^I]~, \nonumber \\
&&[O_{\ell equ}^{(1)}]_{ijkl}=[\bar{L}_i^I \ell_j]\epsilon_{IJ}[\bar{Q}_k^J u_l]~, \quad \ \ [O_{\ell equ}^{(3)}]_{ijkl}=[\bar{L}_i^I \sigma_{\mu\nu}\ell_j]\epsilon_{IJ}[\bar{Q}_k^J\sigma^{\mu\nu} u_l]~, 
\end{eqnarray}
with $i,j,k,$ and $l$ denoting the quark/lepton flavor and $I$ and $J$ representing the $SU(2)_L$ symmetry index. The scale $\Lambda \gtrsim $ the electroweak scale is the cutoff of EFT, corresponding to the scale of new physics. For concreteness, we focus on the operators unsuppressed by the CKM elements, $i.e.$, the ones containing exactly one bottom quark and one strange or one charm quark. As only the third-generation leptons are allowed to deviate their physics from the SM,  there are nine 6D operators of SMEFT to consider in total. These operators are summarized in Tab.~\ref{tab:SMEFT_LEFT}.

\begin{table}[h]
\begin{center}
\begin{tabular}{c|c}
\hline
SMEFT operators
& SMEFT operators (down basis) \\
\hline
$[O_{lq}^{(1)}]_{3332}$
& $[\bar{\nu}\gamma^\mu P_L \nu + \bar{\tau}\gamma^\mu P_L \tau][\bar{b}\gamma_\mu P_L s] $ \\
\hline
$[O_{lq}^{(3)}]_{3332}$
& $2V_{cs}^\ast[\bar{\nu}\gamma^\mu P_L \tau ][\bar{b}\gamma_\mu P_L c] -  [\bar{\nu}\gamma^\mu P_L \nu - \bar{\tau}\gamma^\mu P_L \tau][\bar{b}\gamma_\mu P_L s]$ \\
\hline
$[O_{ed}]_{3332}$
& $[\bar{\tau}\gamma^\mu P_R \tau][\bar{b}\gamma_\mu P_R s]$ \\
\hline
$[O_{ld}]_{3332}$
& $[\bar{\nu}\gamma^\mu P_L \nu + \bar{\tau}\gamma^\mu P_L \tau][\bar{b}\gamma_\mu P_R s] $ \\
\hline
$[O_{qe}]_{3332}$
& $[\bar{\tau}\gamma^\mu P_R \tau][\bar{b}\gamma_\mu P_L s]$ \\
\hline
$[O_{ledq}]_{3332}$
& $V_{cs}^\ast[\bar{\nu} P_R \tau][\bar{b} P_L c] + [\bar{\tau} P_R \tau][\bar{b} P_L s]$ \\
\hline
$[O_{ledq}]_{3323}$
& $[\bar{\tau} P_R \tau][\bar{s} P_L b]$ \\
\hline
$[O_{lequ}^{(1)}]_{3332}$
& $V_{cs}^\ast[\bar{\nu} P_R \tau][\bar{b} P_R c]$\\
\hline
$[O_{lequ}^{(3)}]_{3332}$
& $V_{cs}^\ast [\bar{\nu} \sigma^{\mu\nu}P_R \tau][\bar{b} \sigma_{\mu\nu}P_R c]$\\
\hline

\end{tabular}
\caption{SMEFT Operators which are relevant to this study. In the second column, the operators are shown in the down basis, where $Q_i=\{V_{ji}^\ast u_j,d_i\}$ and $L_i= \{ \nu_i, \ell_i \}$.}\label{tab:SMEFT_LEFT}
\end{center}
\end{table}

We calculate the SMEFT and LEFT Wilson coefficients with RG running using the Wilson package~\cite{Aebischer:2018bkb}.  These two theories are then matched at the scale of $m_Z$ by demanding 
\begin{eqnarray}
\mathcal{L}^{\rm SM} (m_Z)   =  \mathcal{L}^{\rm LE}  (m_Z)  \ .
\end{eqnarray}
The LEFT operators $O_{V_R}^\tau$ for $b\to c \tau \nu$ and $O_T^\tau$ and $O_{T5}^\tau$ for $b\to s \tau^+\tau^-$ are irrelevant to matching and hence are turned off. 
As for the left fourteen LEFT operators, only nine are independent due to the relations inherited from the SM gauge symmetries.  
We take $C_{S_L}^\tau$ and $C_T^\tau$ from Eq.~(\ref{eq:CCOV}), $\delta C_9^\tau$, $C_9^{\prime \tau}$, $\delta C_{10}^\tau$, $C_{10}^{\prime \tau}$, $C_S^\tau$, $C_S^{\prime \tau}$ from Eq.~(\ref{eq:NCH}) and $\delta C_L^\nu$ from Eq.~(\ref{eq:NCnuH}) to define the basis of the constrained LEFT Wilson coefficients without losing any generality. Then we have 
\begin{align}
[C_{\ell q}^{(i)\ast}]_{3332} &= \frac{G_F \alpha V_{tb}V_{ts}^\ast \Lambda^2}{2\sqrt{2}\pi} (\delta C_9^\tau-\delta C_{10}^\tau\pm 2\delta C_L^{\nu} )~, ~i=\{1,3\}~, \nonumber \\
[C_{ed(\ell d)}^\ast]_{3332} &= \frac{G_F \alpha V_{tb}V_{ts}^\ast \Lambda^2}{\sqrt{2}\pi} ( C_9^{\prime\tau} \pm C_{10}^{\prime\tau} )~,  \nonumber \\
[C_{qe}^\ast]_{3332} &= \frac{G_F \alpha V_{tb}V_{ts}^\ast \Lambda^2}{\sqrt{2}\pi} (\delta C_9^{\tau} + \delta C_{10}^{\tau} ),  \nonumber \\
[C_{\ell edq}^\ast]_{3332} &= \frac{\sqrt{2} G_F \alpha V_{tb}V_{ts}^\ast \Lambda^2}{\pi} C_S^\tau ~,  \nonumber \\
[C_{\ell edq}]_{3323} &= \frac{\sqrt{2} G_F \alpha V_{tb}V_{ts}^\ast \Lambda^2}{\pi} C_S^{\tau\prime} ~,  \nonumber \\
[C_{\ell equ}^{(1)\ast}]_{3332} &= -\frac{4 G_F V_{cb}\Lambda^2}{\sqrt{2}V_{cs}} C_{S_L}^{\tau}~,  \nonumber \\
[C_{\ell equ}^{(3)\ast}]_{3332} &= -\frac{4 G_F V_{cb}\Lambda^2}{\sqrt{2}V_{cs}} C_{T}^{\tau}~.
\label{eqn:LEFTmatch1}
\end{align}
At the matching scale around $m_Z$, the other LEFT Wilson coefficients then satisfy the following relations:
\begin{eqnarray}
&& C_{S_R}^{\tau} = -\frac{\alpha V_{tb}V_{ts}^\ast V_{cs}}{2\pi V_{cb}} C_S^{\tau}~, \quad
\delta C_{V_L}^{\tau} = \frac{\alpha V_{tb}V_{ts}^\ast V_{cs}}{4\pi V_{cb}}(\delta C_{10}^\tau-\delta C_{9}^\tau+2\delta C_L^{\nu} )~,  \nonumber   \\
&& C_P^\tau = -C_S^\tau~, \quad C_P^{\prime \tau} = C_S^{\prime\tau}~, \quad
C_R^\nu = \frac{1}{2}(C_9^{\prime\tau}-C_{10}^{\prime\tau})~ .
\label{eqn:LEFTmatch2}
\end{eqnarray}

\subsection{SMEFT Interpretation}

To generate the posterior distributions of the SMEFT Wilson coefficients at the cutoff scale, we sample totally $10^5$ points in the space of $\mathcal{L}^{\rm LE} (\mu_b)$ with the {\it emcee} package~\cite{Foreman_Mackey_2013} to fit the data in Tab.~\ref{tab:SM_Values}. These points are then projected to the space of $\mathcal{L}^{\rm SM} (\Lambda = 10{\rm TeV})$, using the Wilson package for RG running~\cite{Aebischer:2018bkb}, where marginalization is performed with the {\it corner} package~\cite{corner}. For the convenience of analysis, we implement the matching conditions at the scale of $\mu_b$ instead~\cite{Murgui:2019czp}. The matching conditions subject to an effect of RG running. However, the relations in Eq.~(\ref{eqn:LEFTmatch2}) are preserved by the QCD effect~\cite{Aebischer:2017gaw, Jenkins:2017dyc}, as the operators involved in each relation share identical quark spinor structures. At one-loop level, they are deformed by electroweak coupling and quadratic product of Wilson coefficients only. We thus take Eq.~(\ref{eqn:LEFTmatch2}) to be an approximation of the matching conditions at $\mu_b$. Numerical work indicates that, for the data sampling in such a manner, the caused deviation from Eq.~(\ref{eqn:LEFTmatch2}) at the scale of $m_Z$ is at most of a level of several percents.

\begin{figure}[h!]
\centering
\includegraphics[width=0.9\textwidth]{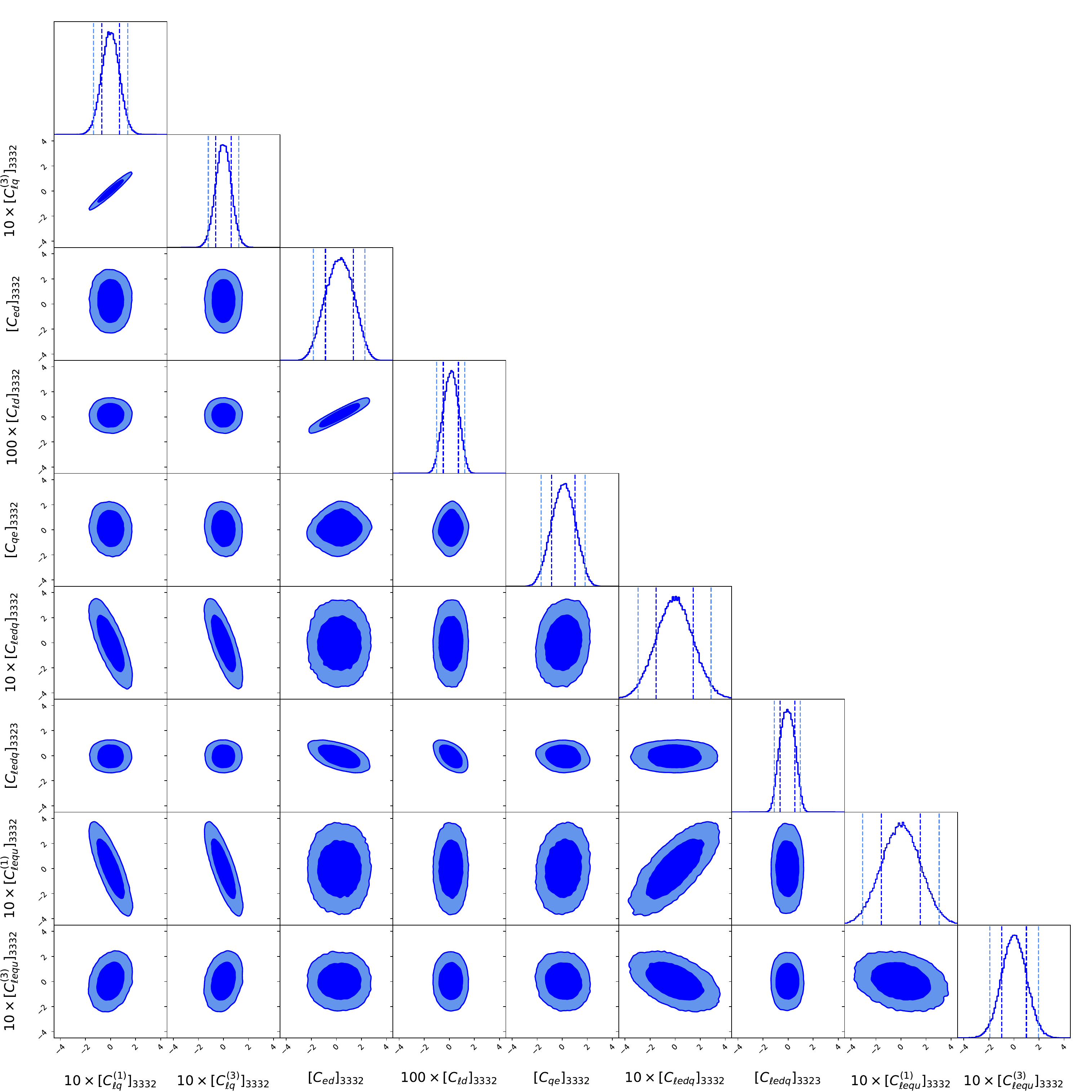}
\caption{2D posterior distributions of the SMEFT Wilson coefficients ($@\Lambda = 10$~TeV) at Tera-$Z$, with 68\% (dark blue) and 95\% (light blue) confidence levels. The fitting inputs are summarized in Tab.~\ref{tab:SM_Values}.}
\label{fig:SMEFT2Dpdf}
\end{figure}

\begin{figure}[h!]
\centering
\includegraphics[width=0.9\textwidth]{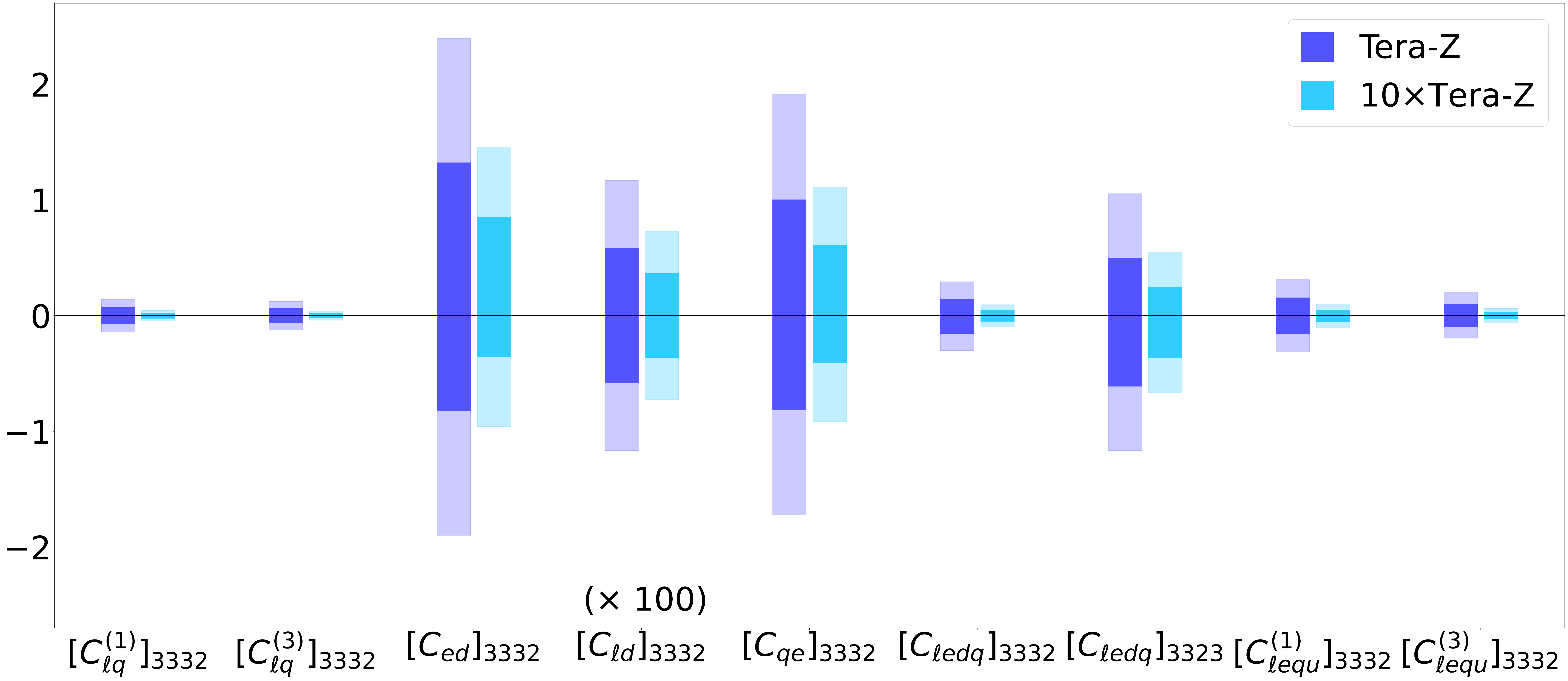}
\caption{1D posterior distributions of the SMEFT Wilson coefficients ($@\Lambda = 10$~TeV) at Tera-$Z$ and $10\times$Tera-$Z$, with $68\%$ (dark) and $95\%$ (light) confidence levels. The fitting inputs are summarized in Tab.~\ref{tab:SM_Values}.}
\label{fig:SMEFT1Dpdf}
\end{figure}

We present the 2D posterior distributions of the SMEFT Wilson coefficients ($@\Lambda = 10$~TeV) at Tera-$Z$ in Fig.~\ref{fig:SMEFT2Dpdf} and their 1D posterior distributions at Tera-$Z$ and $10\times$Tera-$Z$ in Fig.~\ref{fig:SMEFT1Dpdf}. These parameters are constrained to be $\lesssim \mathcal O(1)$ with $68\%$ confidence level by the Tera-$Z$, but not at a comparable level. As summarized in Tab.~\ref{tab:SM_Values} (also see Fig.~\ref{fig:BR_overall}), the measurements of $b\to c \tau \nu$ transitions demonstrate a universally high precision (though the precisions for $R_{J/\psi}$ and Br($B_c\to \tau \nu$) are one order of magnitude lower than those of $R_{D_s^{(*)}}$ and $R_{\Lambda_c}$ due to the relatively low production rate of $B_c$ mesons). The relative precision for measuring $\text{BR}(B_s\to \phi\bar\nu\nu)$ is also high, though the SM prediction for its absolute value is tiny. The three operators $[O_{ed}]_{3332}$, $[O_{qe}]_{3332}$, and $[O_{\ell edq}]_{3323}$ do not contribute to any of them except the $b\to s \tau\tau$ transitions. So the constraints for their Wilson coefficients are a few times weaker than those of the other ones.

\section{Summary and Conclusion}
\label{sec:conclusion}

The LFU is one of the hypothetical principles in the SM, and should be measured with a precision as high as possible such that the physics violating this principle can be fully tested. The future $Z$ factories provide a great opportunity to perform this task. At $Z$-pole, the $b$-hadrons are produced to be highly boosted, with relatively few contaminations from the environment. A higher precision of measuring particle energy and vertex and efficiency for reconstructing the signal events thus can be achieved, compared to those at Belle II and LHCb. For the $H_b\to H_c \tau (\mu) \nu$ measurements studied in this paper, we have developed an algorithm to reconstruct $H_c$ and $H_b$, where the total four-momentum of neutrinos or the missing momentum in each signal event can be inferred, by employing these advantages. Moreover, heavy $b$-hadrons such as $\Lambda_b$ can be produced at $Z$-pole with significant statistics. This opens new avenues to test the LFU. If the LFU violation is observed, such a multiplicity of signal modes may greatly benefit exploring the nature of LFU-violating new physics, e.g., parity and spin of the relevant mediators.

The study performed in this paper was mainly based on the $b\to c \tau (\mu) \nu$ transitions. Concretely, we analyzed the sensitivity of measuring $R_{H_c}$ in four representative scenarios: $B_c \to J/\phi \tau (\mu) \nu$, $B_s \to D_s^{(*)} \tau (\mu) \nu$ and $\Lambda_b \to\Lambda_c \tau (\mu) \nu$, with $\tau\to \mu\nu\bar{\nu}$. The statistics for all of them at Belle II are significantly lower than those expected to achieve at the future $Z$ factories. Because of the relatively high efficiency for event reconstruction (see, e.g., the left-upper panel of Fig.~\ref{fig:JpsipB}, Fig.~\ref{fig:DspB} and Fig.~\ref{fig:LambdacpB}), we are allowed to introduce the invariant mass of off-shell $W$ boson $q^2$ and missing momenta $m^2_{\rm miss}$, and the minimal distance between the $\mu_3$ track and the $H_b$ decay vertex $S_{\rm SV}$ to distinguish the $\mu$-mode and $\tau$-mode signal events which serve as mutual backgrounds in their respective measurements.  These observables can be also applied to discriminate the signals from the universal backgrounds. The universal backgrounds have been classified into five categories in this study.
Despite their multiplicity, these backgrounds tend to be less isolated compared to the signal $B_c$ mesons. So we also turned on a set of isolation measures to further suppress these backgrounds. Finally these observables and some others are integrated using the tool of BDT in the sensitivity analysis, yielding $S/B \gtrsim 0.3$ for various relevant scenarios.

The algorithm developed for reconstructing $H_c$ and $H_b$ relies on the messages of tracks and decay vertex significantly. So we further explored the robustness of sensitivity against the tracker performance and the potential improvement with a better tracker resolution.  We showed that the variation of tracker resolution, which is manifested as vertex noise, from a perfect case to more conservative scenarios causes a change of $\sim \mathcal O(10\%)$ to sensitivity. Specifically, the precision of measuring $R_{J/\psi}$ could be improved more with the reduced vertex noise, while the measurement of $R_{\Lambda_c}$ tends to be more robust against the variation of vertex noise level. The reason is simple. Unlike the $\Lambda_b$ one, the $B_c$ vertex can be well-approximated by the $H_c$ vertex. This leads to more accurate reconstruction for $B_c$, which in turn leaves smaller space to resolve the variation of vertex noise. In addition, we investigated the impacts of ECAL energy threshold on soft photon tagging. The latter plays a central role in the $R_{D_s^{(\ast)}}$ measurements, as the separation between $D_s$ and $D_s^\ast$ relies on the resonance reconstruction of $D_s^\ast \to D_s \gamma$. We found that, as the photon energy threshold decreases, the precisions of measuring $R_{D_s^{(\ast)}}$ could be improved by $\sim\mathcal O(10\%)$ and meanwhile the correlation between them gets weakened. Finally, for the first time we scrutinized the effect of event shape in distinguishing the signal and background events. We considered the $R_{J/\psi}$ measurement as an example, as multiple heavy-flavor quarks can be produced in the $Z\to B_c+X$ events. The message beyond the $B_c$ decays indeed yields a suppression of the background events, especially the ones from the back-to-back $Z\to b\bar{b}$ decays. By including the F-W moments as the inputs for the BDT analysis, we showed that the signal-to-background ratio is increased by several percents, and the signal significance can be slightly improved.

Finally, we have interpreted in the SMEFT the projected sensitivities of measuring $R_{H_c}$, together with the $b\to s\tau^+\tau^-$ and $b\to s\nu\bar{\nu}$ observables, at the future $Z$ factories. These measurements are performed at an energy scale well below the SMEFT cutoff. So we included the effects of the RG running in this analysis, with the EW-scale matching conditions being implemented. Due to the generic constraints of symmetries for the SMEFT, these observables are entangled with each other. For example, the operator $[O_{lq}^{(3)}]_{3332}$ correlates the measurements of all three types of relevant FCNC/FCCC transitions (see Tab.~\ref{tab:SMEFT_LEFT}) at the low energy scale. The MCMC posterior distributions for the SMEFT Wilson coefficients then indicate - for $\lesssim \mathcal{O}(1)$ Wilson coefficients, the LFU-violating physics can be probed up to a scale $\sim \mathcal{O}(10)$~TeV at Tera-$Z$.

Notably, to demonstrate the sensitivity potential of testing the LFU at the future $Z$ factories, we have taken the $b\to c \tau \nu$ measurements with $\tau\to \mu\nu\bar{\nu}$ as a benchmark. Such a scenario is representative but not complete. Actually, the accuracy of electron identification in a $Z$ factory is also high. One can thus generalize the analysis to the mode of $\tau\to e \nu\bar{\nu}$ straightforwardly, to further improve the measurement precision. Alternative decay modes for $H_c$ could be also considered for the $R_{H_c}$ measurements, if we can reconstruct the $H_c$ hadron well. By including electron modes, we will double the effective statistics for the $R_{D_s^{(\ast)}}$ and $R_{\Lambda_c}$ measurements and quadruple that for the $R_{J/\psi}$ measurement. By including hadronic modes, we will gain more. A combination of these analyses will certainly generate positive impacts on the sensitivity reach of the LFU tests at the future $Z$ factories. Moreover, the strategies developed here could be applied to other tasks at $Z$ pole also, such as the differential and $CP$ measurements in semi-leptonic $b$-hadron decays. We leave these explorations to future work.

\begin{acknowledgments}

This work is supported partly by the Area of Excellence (AoE) under the Grant No.~AoE/P-404/18-3, and partly by the General Research Fund (GRF) under Grant No.~16304321. Both of the AoE and GRF grants are issued by the Research Grants Council of Hong Kong S.A.R. LL is also supported by the DOE grant DE-SC-0010010. We would like to thank Manqi Ruan for highly valuable and constructive comments on the studies performed in Sec. 6 ``Impacts of Detector Performance and Event Shape''. We would also thank Lorenzo Calibbi, Jibo He, Fengkun Guo and Wei Wang for useful discussions.

\end{acknowledgments}

\newpage

\appendix
\section{Relevant Observables in Low-Energy EFT}
\label{app:calc}

The LEFT predictions for the $b\to c \tau \nu$, $b\to s \tau^+\tau^-$ and $b\to s \bar\nu\nu$ observables have been analytically or semi-analytically studied in literatures. However, to apply them to the SMEFT interpretation performed in Sec.~\ref{sec:EFT}, we need these predictions to be numerically calculated in terms of the LEFT Wilson coefficients first. Below is a summary of the analytical formulae that we have used for such a calculation (a summary of the parameter values used for this purpose can be found in Tab.~\ref{tab:relvt_param}), which include the ones for $R_{J/\psi}$, $R_{D_s^{(*)}}$, $R_{\Lambda_c}$ and Br$(B^+\to K^+ \tau^+\tau^-)$, Br$(B^0\to K^{\ast 0} \tau^+\tau^-)$, Br$(B_s\to \phi \tau^+\tau^-)$ and Br$(B_s\to \tau^+\tau^-)$. Note, the numerical formulae for  $\text{BR}(B^+\to K^+\bar\nu \nu)$, $\text{BR}(B^0\to K^{\ast 0}\bar\nu \nu)$ and $\text{BR}(B_s\to \phi\bar\nu \nu)$ have been presented in~\cite{Buras:2014fpa,Li:2022tov}. So we quote them directly in the main text.

\begin{table}
\begin{center}
\begin{tabular}{||ll|ll||}
\hline \hline
$\alpha(\mu_b)$ & 1/133 \hfill~\cite{Bobeth:2007dw, Kamenik:2017ghi}
&
$G_F$ & $\unit[1.166\times 10^{-5}]{GeV^{-2}}$ \hfill~\cite{Tanabashi:2018oca}
\\
$m_\mu$ & $\unit[0.1057]{GeV}$\hfill~\cite{Tanabashi:2018oca}
&
$m_{\Lambda_c}$ & $\unit[2.286]{GeV}$\hfill~\cite{Tanabashi:2018oca}
\\
$m_\tau$ & $\unit[1.777]{GeV}$\hfill~\cite{Tanabashi:2018oca}
&
$m_{K^+}$ & $\unit[0.4937]{GeV}$\hfill~\cite{Tanabashi:2018oca}
\\
$m_b$ & $\unit[4.8]{GeV}$\hfill~\cite{Tanabashi:2018oca}
&
$m_{K^\ast}$ & $\unit[0.8917]{GeV}$\hfill~\cite{Tanabashi:2018oca}
\\
$m_c$ & $\unit[1.67]{GeV}$\hfill~\cite{Tanabashi:2018oca}
&
$m_{\phi}$ & $\unit[1.019]{GeV}$\hfill~\cite{Tanabashi:2018oca}
\\
$m_s$ & $\unit[0.093]{GeV}$\hfill~\cite{Tanabashi:2018oca}
&
$\tau_{B^0}$ & $\unit[1.519]{ps}$\hfill~\cite{Tanabashi:2018oca}
\\
$m_{B^0}$ & $\unit[5.279]{GeV}$\hfill~\cite{Tanabashi:2018oca}
&
$\tau_{B_s}$ & $\unit[1.516]{ps}$\hfill~\cite{Tanabashi:2018oca}
\\
$m_{B^+}$ & $\unit[5.279]{GeV}$\hfill~\cite{Tanabashi:2018oca}
&
$\tau_{B^+}$ & $\unit[1.638]{ps}$\hfill~\cite{Tanabashi:2018oca}
\\
$m_{B_c}$ & $\unit[6.274]{GeV}$\hfill~\cite{Tanabashi:2018oca}
&
$|V_{tb}|$ & 1.013\hfill~\cite{Tanabashi:2018oca}
\\
$m_{B_s}$ & $\unit[5.367]{GeV}$\hfill~\cite{Tanabashi:2018oca}
&
$|V_{ts}|$ & 0.0388\hfill~\cite{Tanabashi:2018oca}
\\
$m_{\Lambda_b}$ & $\unit[5.620]{GeV}$\hfill~\cite{Tanabashi:2018oca}
&
$|V_{cs}|$ & 0.987\hfill~\cite{Tanabashi:2018oca}
\\
$m_{J/\psi}$ & $\unit[3.097]{GeV}$\hfill~\cite{Tanabashi:2018oca}
&
$C_7^\tau|_{\rm SM}$ & -0.292\hfill~\cite{DescotesGenon:2011yn}
\\
$m_{D_s}$ & $\unit[1.968]{GeV}$\hfill~\cite{Tanabashi:2018oca}
&
$C_9^\tau|_{\rm SM}$ & 4.07\hfill~\cite{DescotesGenon:2011yn}
\\
$m_{D_s^\ast}$ & $\unit[2.112]{GeV}$\hfill~\cite{Tanabashi:2018oca}
&
$C_{10}^\tau|_{\rm SM}$ & -4.31\hfill~\cite{DescotesGenon:2011yn}
\\
$f_{B_s}$ & $\unit[0.234]{GeV}$\hfill \cite{Becirevic:2012fy}
&
$C_L^\nu|_{\rm SM}$ & -6.47\hfill~\cite{Buras:2014fpa}
\\
\hline \hline
\end{tabular}
\end{center}
\caption{Parameter values used for numerically calculating the $b\to c \tau \nu$, $b\to s \tau\tau$ and $b\to s \bar\nu\nu$ observables in the LEFT. }
\label{tab:relvt_param}
\end{table}

\subsection{$R_{J/\psi}$ and $R_{D_s^\ast}$}

$R_{J/\psi}$ and $R_{D_s^\ast}$ involve the decay of $b$-meson with a vector meson. Their calculations in the LEFT are essentially the same. Consider $R_{D_s^\ast}$ as an example. We have (following~\cite{Sakaki:2013bfa}) 
\begin{align}
\frac{d \Gamma_{B_s\to D_s^\ast\tau\nu}}{d q^2} = &\frac{G_F^2|V_{cb}|^2}{192\pi^3 m_{B_s}^3}q^2 \sqrt{\lambda(q^2)}\bigg(1-\frac{m_\tau^2}{q^2}\bigg)^2 \times  \nonumber \\
&\bigg\{(|1+\delta C_{V_L} ^{\tau}|^2+|C_{V_R}^{\tau}|^2)\bigg[\bigg(1+\frac{m_\tau^2}{2q^2}\bigg) (H_{V_+}^2+H_{V_-}^2+H_{V_0}^2) +\frac{3 m_\tau^2}{2 q^2} H_{V_t}^2\bigg] \nonumber \\
&-2 \text{Re} [(1+\delta C_{V_L} ^\tau)C_{V_R}^{\tau\ast}] \bigg[\bigg(1+\frac{m_\tau^2}{2q^2}\bigg) (H_{V_+}^2+2H_{V_-}H_{V_0}) +\frac{3 m_\tau^2}{2 q^2} H_{V_t}^2\bigg] \nonumber \\
&+\frac{3}{2}|C_{S_L}^{\tau}-C_{S_R}^{\tau}|^2 H_S^2+ 8|C_T^\tau|^2 \bigg(1+\frac{2 m_\tau^2}{q^2}\bigg) (H_{V_+}^2+H_{V_-}^2+H_{V_0}^2) \nonumber \\
&+3\text{Re}[(1+\delta C_{V_L} ^{\tau}-C_{V_R}^{\tau})(C_{S_L}^{\tau\ast}-C_{S_R}^{\tau\ast})] \frac{m_\tau}{\sqrt{q^2}} H_S H_{V_t}\nonumber \\
&-12 \text{Re}[(1+\delta C_{V_L} ^{\tau}) C_T^{\tau\ast}] \frac{m_\tau}{\sqrt{q^2}}(H_{V_+} H_{T_+}- H_{V_-}H_{T_-}+H_{V_0}H_{T_0}) \nonumber \\
&+12 \text{Re}(C_{V_R}^{\tau} C_T^{\tau\ast}) \frac{m_\tau}{\sqrt{q^2}}(H_{V_-} H_{T_+}- H_{V_+}H_{T_-}+H_{V_0}H_{T_0}) \bigg\}~,
\end{align}
with 
\begin{equation}
\lambda(q^2) = [(m_{B_s}-m_{D_s^\ast})^2-q^2 ] [ (m_{B_s}+m_{D_s^\ast})^2-q^2  ]  \ .
\end{equation}
Here the $H$-quantities are hadronic helicity amplitudes, given by 
\begin{align}   
H_{V_{\pm}} &= (m_{B_s}+m_{D_s^\ast})A_1 \mp\frac{\lambda}{m_{B_s}+m_{D_s^\ast}}V \ , \\
H_{V_0} &= \frac{m_{B_s}+m_{D_s^\ast}}{2m_{D_s^\ast}\sqrt{q^2}}\bigg [ -(m^2_{B_s}-m^2_{D_s^\ast}-q^2)A_1 + \frac{\lambda}{(m_{B_s}+m_{D_s^\ast})^2}A_2 \bigg ] \ ,\\
H_{V_t} &= -\sqrt{\frac{\lambda}{q^2}}A_0 \ , \\
H_S &= -\frac{\sqrt{\lambda}}{m_b+m_c}A_0 \ , \\
H_{T_{\pm}} &= \frac{1}{\sqrt{q^2}} [ \pm (m^2_{B_s}-m^2_{D_s^\ast})T_2+\sqrt{\lambda}T_1  ] \ ,\\
H_{T_0} &= \frac{1}{2m_{D_s^\ast}}\bigg [ -(m^2_{B_s}+3m^2_{D_s^\ast}-q^2)T_2+\frac{\lambda}{m^2_{B_s}-m^2_{D_s^\ast}}T_3 \bigg ] \ .
\end{align}
$A_{0,1,2}(q^2)$, $V(q^2)$ and $T_{1,2,3}(q^2)$ are form factors. With the convention of $\epsilon^{0123}=+1$, they parametrize the relevant hadronic matrix elements as 
\begin{align}
\langle D_s^\ast(k, \varepsilon) | \bar c \gamma^\mu b | B_s(p) \rangle &= \frac{2iV(q^2)}{m_{B_s}+m_{D_s^\ast}}\epsilon^{\mu\nu\rho\sigma} \varepsilon^\ast_\nu p_\rho k_\sigma \ , \\
\langle D_s^\ast(k, \varepsilon) | \bar c \gamma^\mu\gamma^5 b | B_s(p) \rangle &= 2m_{D_s^\ast}A_0(q^2)\frac{\varepsilon^\ast\cdot q}{q^2}q^\mu + (m_{B_s}+m_{D_s^\ast})A_1(q^2)\bigg ( \varepsilon^{\ast\mu}-\frac{\varepsilon^\ast \cdot q}{q^2}q^\mu \bigg ) \nonumber \\
&~~~-A_2(q^2)\frac{\varepsilon^\ast\cdot q}{m_{B_s}+m_{D_s^\ast}} \bigg (p^{\mu}+k^{\mu}-\frac{m^2_{B_s}-m^2_{D_s^\ast}}{q^2}q^\mu \bigg ) \ , \\
\langle D_s^\ast(k, \varepsilon) | \bar c \sigma^{\mu\nu}q_\nu b | B_s(p) \rangle &= 2T_1(q^2)\epsilon^{\mu\nu\rho\sigma} \varepsilon^\ast_\nu p_\rho k_\sigma\ , \\
\langle D_s^\ast(k, \varepsilon) | \bar c \sigma^{\mu\nu}\gamma^5q_\nu b | B_s(p) \rangle &= -T_2(q^2)\bigg [ (m^2_{B_s}-m^2_{D_s^\ast})\varepsilon^{\ast\mu}- (\varepsilon^\ast\cdot q)(p+k)^\mu \bigg ] \nonumber \\
&~~~ -T_3(q^2)(\varepsilon^\ast\cdot q)\bigg [ q^\mu -\frac{q^2}{m^2_{B_s}-m^2_{D_s^\ast}}(p+k)^\mu \bigg ] \ , 
\end{align}
where~\cite{Sakaki:2013bfa} 
\begin{align}
T_1(q^2) &= \frac{m_b+m_c}{m_{B_s}+m_{D_s^\ast}}V(q^2) \ , \\
T_2(q^2) &= \frac{m_b-m_c}{m_{B_s}-m_{D_s^\ast}}A_1(q^2) \ , \\
T_3(q^2) &= -\frac{m_b-m_c}{q^2}\{ m_{B_s}[A_1(q^2)-A_2(q^2) ]+m_{D_s^\ast}[A_2(q^2)+A_1(q^2)-2A_0(q^2)] \} \ .
\end{align}
In our analysis, we take the formulae of $A_{0,1,2}(q^2)$ and $V(q^2)$ from~\cite{Wang:2008xt, Fan:2013kqa} (and their counterparts in the $R_{J/\psi}$ analysis from~\cite{Watanabe:2017mip, Wang:2012lrc}). 
With $\frac{d \Gamma_{B_s\to D_s^\ast\mu\nu}}{d q^2}$ being calculated by replacing $m_\tau$ with $m_\mu$ and turning off all Wilson coefficients, 
finally we have 
\begin{align}
R_{D_s^\ast}=\frac{\int_{m_\tau^2}^{q_{\text{max}}^2}dq^2 ~~d\Gamma_{B_s\to D_s^\ast\tau\nu} / dq^2}{\int_{m_\mu^2}^{q_{\text{max}}^2}dq^2~~d\Gamma_{B_s\to D_s^\ast\mu\nu}/dq^2} \ ,
\end{align}
with $q_{\text{max}}^2=(m_{B_s}-m_{D_s^\ast})^2$.

\subsection{$R_{D_s}$}

$R_{D_s}$ involves the decay of $b$-meson with a pseudoscalar meson.
Following~\cite{Sakaki:2013bfa}, we have:
\begin{align}
\frac{d \Gamma_{B_s\to D_s\tau\nu}}{d q^2} = &\frac{G_F^2|V_{cb}|^2}{192\pi^3 m_{B_s}^3}q^2 \sqrt{\lambda(q^2)}\bigg(1-\frac{m_\tau^2}{q^2}\bigg)^2 \times \\\nonumber
&\bigg\{|1+\delta C_{V_L} ^{\tau}+C_{V_R}^{\tau}|^2\bigg[\bigg(1+\frac{m_\tau^2}{2q^2}\bigg) H_{V_0}^{s~2} +\frac{3 m_\tau^2}{2 q^2} H_{V_t}^{s~2}\bigg] \\\nonumber
&+\frac{3}{2}|C_{S_L}^{\tau}+C_{S_R}^{\tau}|^2 H_S^{s~2}+ 8|C_T^\tau|^2 \bigg(1+\frac{2 m_\tau^2}{q^2}\bigg) H_T^{s~2} \\\nonumber
&+3\text{Re}[(1+\delta C_{V_L} ^{\tau}+C_{V_R}^{\tau})(C_{S_L}^{\tau\ast}+C_{S_R}^{\tau\ast})] \frac{m_\tau}{\sqrt{q^2}} H_S^s H_{V_t}^s\\\nonumber
&-12 \text{Re}[(1+\delta C_{V_L} ^{\tau}+ C_{V_R} ^{\tau}) C_T^{\tau\ast}] \frac{m_\tau}{\sqrt{q^2}}H_T^s H_{V_0}^s \bigg\}~.
\end{align}
Here the hadronic helicity amplitudes ($H_{V_0}^s$, $H_{V_t}^s$, $H_S^s$ and $H_T^s$) are given by~\cite{Tanaka:2012nw} 
\begin{align}
H_{V_0}^s &= \sqrt{\frac{\lambda}{q^2}}F_1  \ , \\
H_{V_t}^s &= \frac{m^2_{B_s}-m^2_{D_s}}{\sqrt{q^2}}F_0 \ , \\
H_S^s &= \frac{m^2_{B_s}-m^2_{D_s}}{m_b-m_c}F_0 \ , \\
H_T^s &= -\frac{\sqrt{\lambda}}{m_{B_s}+m_{D_s}}F_T \ .
\end{align}
The form factors ($F_0$, $F_1$ and $F_T$) parameterize the relevant matrix elements as~\cite{Wang:2008xt}
\begin{align}
\langle D_s(k) |\bar c \gamma^\mu b | B_s(p) \rangle &= \bigg [ (p+k)^\mu - q^\mu\frac{m^2_{B_s}-m^2_{D_s}}{q^2}\bigg ] F_1(q^2) + q^\mu\frac{m^2_{B_s}-m^2_{D_s}}{q^2} F_0(q^2) \ , \\
\langle D_s(k) |\bar c \sigma^{\mu\nu} b | B_s(p) \rangle &= -\frac{2i(p^\mu k^\nu - p^\nu k^\mu)}{m_{B_s}+m_{D_s}}F_T(q^2) \ ,  \\
\langle D_s(k) |\bar c b | B_s(p) \rangle & = \frac{m^2_{B_s}-m^2_{D_s}}{m_b-m_c}F_0(q^2) \ , 
\end{align}
where $F_1(0)=F_0(0)$ has been taken to cancel the divergence at $q^2=0$. In our analysis, we take the formulae for these form factors from~\cite{Fan:2013kqa}. 
With $\frac{d \Gamma_{B_s\to D_s\mu\nu}}{d q^2}$ being calculated by replacing $m_\tau$ with $m_\mu$ and turning off all Wilson coefficients, finally we have  
\begin{align}
R_{D_s}=\frac{\int_{m_\tau^2}^{q_{\text{max}}^2}dq^2 ~~d\Gamma_{B_s\to D_s\tau\nu} / dq^2}{\int_{m_\mu^2}^{q_{\text{max}}^2}dq^2~~d\Gamma_{B_s\to D_s\mu\nu}/dq^2}\ ,
\end{align}
with $q_{\text{max}}^2=(m_{B_s}-m_{D_s})^2$.

\subsection{$R_{\Lambda_{c}}$}

$R_{\Lambda_{c}}$ involves baryonic decay of $b$-hadron. Following~\cite{Shivashankara:2015cta,Dutta:2015ueb,Datta:2017aue}, we have
\begin{align}
\frac{d \Gamma_{\Lambda_{b}\to \Lambda_{c}\tau\nu}}{d q^2} = &\frac{G_F^2|V_{cb}|^2}{384\pi^3 m_{\Lambda_{b}}^3}q^2 \sqrt{\lambda(q^2)} \bigg(1-\frac{m_\tau^2}{q^2}\bigg)^2 \times \\ \nonumber
& \bigg [ A_1 + \frac{m_\tau^2}{2q^2}A_2+\frac{3}{2}A_3 +\frac{3m_\tau}{\sqrt{q^2}}A_4 + 2\bigg (1+\frac{2m_\tau^2}{q^2}\bigg )A_5 + \frac{6m_\tau}{\sqrt{q^2}}A_6 \bigg ]~,
\end{align}
where $A_{1,2,3,4}$ are contributed by scalar and vector operators~\cite{Shivashankara:2015cta,Dutta:2015ueb}, $A_5$ is contributed by tensor operators, and $A_6$ is contributed by both~\cite{Datta:2017aue}. Explicitly, these $A$ terms are given by
\begin{align}
A_1 =& |H_{1/2,0}|^2 + |H_{-1/2,0}|^2 + |H_{1/2,1}|^2 + |H_{-1/2,-1}|^2 \ , \\
A_2 =& A_1 + 3 |H_{1/2,t}|^2+3|H_{-1/2,t}|^2 \ , \\
A_3 =& |H_{1/2,0}^{SP}|^2+ |H_{-1/2,0}^{SP}|^2 \ , \\
A_4 =& \text{Re}  ( H_{1/2,t}H_{1/2,0}^{SP~\ast} + H_{-1/2,t}H_{-1/2,0}^{SP~\ast} )\ ,\\
A_5 =& | H_{1/2,t,0}^{(T)1/2} + H_{1/2,-1,1}^{(T)1/2} |^2  +  | H_{-1/2,t,-1}^{(T)1/2} + H_{-1/2,-1,0}^{(T)1/2} |^2  +  | H_{1/2,0,1}^{(T)-1/2} + H_{1/2,t,1}^{(T)-1/2} |^2  \nonumber \\
&+ | H_{-1/2,-1,1}^{(T)-1/2} + H_{-1/2,t,0}^{(T)-1/2} |^2  \ ,\\
A_6 =& \text{Re} [ H_{1/2,0}^\ast ( H_{1/2,-1,1}^{(T)1/2} + H_{1/2,t,0}^{(T)1/2} )  ]      +     \text{Re} [ H_{1/2,1}^\ast ( H_{1/2,0,1}^{(T)-1/2} + H_{1/2,t,1}^{(T)-1/2} )  ] \nonumber \\
&+    \text{Re} [ H_{-1/2,0}^\ast ( H_{-1/2,-1,1}^{(T)-1/2} + H_{-1/2,t,0}^{(T)-1/2} )  ]      +     \text{Re} [ H_{-1/2,-1}^\ast ( H_{-1/2,-1,0}^{(T)1/2} + H_{-1/2,t,-1}^{(T)1/2} )  ] \  ,
\end{align}
with $H_{\lambda_{\Lambda_{c}},\lambda_W} = H_{\lambda_{\Lambda_{c}},\lambda_W}^V - H_{\lambda_{\Lambda_{c}},\lambda_W}^A$ and $H_{\lambda_{\Lambda_{c}},\lambda_{\rm NP}}^{SP} = H_{\lambda_{\Lambda_{c}},\lambda_{\rm NP}}^{S} + H_{\lambda_{\Lambda_{c}},\lambda_{\rm NP}}^{P}$. Here $H_{\lambda_{\Lambda_{c}},\lambda_W}^{V(A)}$, $H_{\lambda_{\Lambda_{c}},\lambda_{\rm NP}}^{S(P)}$ and $H_{\lambda_{\Lambda_{c}},\lambda,\lambda^\prime}^{(T)\lambda_{\Lambda_{b}}}$ denote the (axial-)vector, (pseudo-)scalar and tensor helicity amplitudes, respectively. They are characterized by the helicities of $\Lambda_{b}$ ($\lambda_{\Lambda_{b}}$),  $\Lambda_c$ ($\lambda_{\Lambda_{c}}$), intermediate off-shell $W$ boson ($\lambda_W$)~\footnote{$\lambda_W=0$ is allowed for both $J_W = 0$ and 1. Here $J_W$ is the angular momentum of $W$ boson. To distinguish these two cases, we follow~\cite{Gutsche:2015mxa} and use $\lambda_W=t$ for $J_W=0$ case and $\lambda_W=0$ for $J_W=1$ case.} and new-physics particle ($\lambda_{\rm NP}$), and the possible tensor degrees of freedom ($\lambda$ and $\lambda^\prime$) together. These helicity amplitudes are then found to be 
\begin{align}
H_{1/2,0}^V &= (1+\delta C_{V_L}^\tau + C_{V_R}^\tau)\frac{\sqrt{Q_-}}{\sqrt{q^2}} [ (m_{\Lambda_{b}}+ m_{\Lambda_{c}})f_1-q^2f_2 ] \ ,\\
H_{1/2,0}^A &= (1+\delta C_{V_L}^\tau - C_{V_R}^\tau)\frac{\sqrt{Q_+}}{\sqrt{q^2}} [ (m_{\Lambda_{b}}- m_{\Lambda_{c}})g_1+q^2g_2 ] \ ,\\
H_{1/2,1}^V &= (1+\delta C_{V_L}^\tau + C_{V_R}^\tau)\sqrt{2Q_-} [ f_1-(m_{\Lambda_{b}}+ m_{\Lambda_{c}})f_2 ] \ ,\\
H_{1/2,1}^A &= (1+\delta C_{V_L}^\tau - C_{V_R}^\tau)\sqrt{2Q_+} [ g_1+(m_{\Lambda_{b}}- m_{\Lambda_{c}})g_2 ] \ ,\\
H_{1/2,t}^V &= (1+\delta C_{V_L}^\tau + C_{V_R}^\tau)\frac{\sqrt{Q_+}}{\sqrt{q^2}} [ (m_{\Lambda_{b}}- m_{\Lambda_{c}})f_1+q^2f_3 ]  \ ,\\
H_{1/2,t}^A &= (1+\delta C_{V_L}^\tau - C_{V_R}^\tau)\frac{\sqrt{Q_-}}{\sqrt{q^2}} [ (m_{\Lambda_{b}}+ m_{\Lambda_{c}})g_1-q^2g_3 ]\ , \\
H_{1/2,0}^{S} &= (C_{S_L}^\tau+C_{S_R}^\tau)\frac{\sqrt{Q_+}}{m_b-m_c} [ (m_{\Lambda_{b}}-m_{\Lambda_{c}})f_1+q^2f_3 ] \ , \\
H_{1/2,0}^{P} &= (C_{S_L}^\tau-C_{S_R}^\tau)\frac{\sqrt{Q_-}}{m_b+m_c} [ (m_{\Lambda_{b}}+m_{\Lambda_{c}})g_1-q^2g_3 ] \ , \\
H_{-1/2,t,0}^{(T)-1/2} &= C_T^\tau ( h_+\sqrt{Q_-}-\tilde{h}_+\sqrt{Q_+} ) \ ,\\
H_{1/2,t,0}^{(T)1/2} &= C_T^\tau ( h_+\sqrt{Q_-}+\tilde{h}_+\sqrt{Q_+} ) \ ,\\
H_{1/2,t,1}^{(T)-1/2} &= -C_T^\tau\frac{\sqrt{2}}{\sqrt{q^2}}[ h_\perp(m_{\Lambda_{b}}+m_{\Lambda_{c}})\sqrt{Q_-}+\tilde{h}_\perp(m_{\Lambda_{b}}-m_{\Lambda_{c}})\sqrt{Q_+} ] \ ,\\
H_{-1/2,t,-1}^{(T)1/2} &= -C_T^\tau\frac{\sqrt{2}}{\sqrt{q^2}} [ h_\perp(m_{\Lambda_{b}}+m_{\Lambda_{c}})\sqrt{Q_-}-\tilde{h}_\perp(m_{\Lambda_{b}}-m_{\Lambda_{c}})\sqrt{Q_+} ]\ ,\\
H_{1/2,0,1}^{(T)-1/2} &= H_{1/2,t,1}^{(T)-1/2}\ , \\
H_{-1/2,0,-1}^{(T)1/2} &= -H_{-1/2,t,-1}^{(T)1/2}\ , \\
H_{1/2,1,-1}^{(T)1/2} &= -H_{1/2,t,0}^{(T)1/2}\ ,  \\
H_{-1/2,1,-1}^{(T)-1/2} &= -H_{-1/2,t,0}^{(T)-1/2}~,
\end{align}
where $Q_\pm = (m_{\Lambda_{b}}\pm m_{\Lambda_{c}})^2-q^2$. In addition, some useful properties on these helicity amplitudes have also been applied, including
\begin{align}
H_{\lambda_{\Lambda_{c}},\lambda_W}^V &= H_{-\lambda_{\Lambda_{c}},-\lambda_W}^V \ ,\\
H_{\lambda_{\Lambda_{c}},\lambda_W}^A &= -H_{-\lambda_{\Lambda_{c}},-\lambda_W}^A \ , \\
H_{\lambda_{\Lambda_{c}},\lambda_{\rm NP}}^S &= H_{-\lambda_{\Lambda_{c}},-\lambda_{\rm NP}}^S \ ,\\
H_{\lambda_{\Lambda_{c}},\lambda_{\rm NP}}^P &= -H_{-\lambda_{\Lambda_{c}},-\lambda_{\rm NP}}^P \ ,  \\
H_{\lambda_{\Lambda_{c}},\lambda,\lambda^\prime}^{(T)\lambda_{\Lambda_{b}}} &= -H_{\lambda_{\Lambda_{c}},\lambda^\prime,\lambda}^{(T)\lambda_{\Lambda_{b}}}~.
\end{align}
As for the ten form factors introduced in these calculations, 
$f$s and $g$s parameterize the vector and axial vector matrix elements as~\cite{Shivashankara:2015cta}
\begin{align}
\langle \Lambda_{c} | \bar c \gamma^\mu b | \Lambda_{b} \rangle &= \bar{u}_{\Lambda_{c}} (f_1\gamma^\mu + if_2\sigma^{\mu\nu}q_\nu + f_3q^\mu) u_{\Lambda_{b}}\ , \\
\langle \Lambda_{c} | \bar c \gamma^\mu\gamma^5 b | \Lambda_{b} \rangle &= \bar{u}_{\Lambda_{c}}(g_1\gamma^\mu + ig_2\sigma^{\mu\nu}q_\nu + g_3q^\mu)\gamma^5 u_{\Lambda_{b}}~,
\end{align}
and the scalar and pseudoscalar matrix elements as
\begin{align}
\langle \Lambda_{c} | \bar c b | \Lambda_{b} \rangle &= \frac{1}{m_b-m_c}\bar{u}_{\Lambda_{c}} (f_1\slashed q + f_3q^2)u_{\Lambda_{b}} \ ,\\
\langle \Lambda_{c} | \bar c\gamma^5 b | \Lambda_{b} \rangle &= \frac{-1}{m_b+m_c}\bar{u}_{\Lambda_{c}} (g_1\slashed q + g_3q^2)\gamma^5 u_{\Lambda_{b}}~,
\end{align}
while $h_+$, $h_\perp$, $\tilde{h}_+$ and $\tilde{h}_\perp$ parametrize the tensor matrix elements as~\cite{Datta:2017aue}:
\begin{align}
\langle \Lambda_{c} | \bar ci\sigma^{\mu\nu} b | \Lambda_{b} \rangle =& ~ \bar{u}_{\Lambda_{c}} \bigg \{ 2h_+\frac{p_{\Lambda_{b}}^\mu p_{\Lambda_{c}}^\nu - p_{\Lambda_{b}}^\nu p_{\Lambda_{c}}^\mu}{Q_+}   + h_\perp \bigg [\frac{m_{\Lambda_{b}}+m_{\Lambda_{c}}}{q^2}(q^\mu \gamma^\nu-q^\nu \gamma^\mu)  \nonumber \\
& -2\bigg (\frac{1}{q^2}+\frac{1}{Q_+}\bigg )(p_{\Lambda_{b}}^\mu p_{\Lambda_{c}}^\nu - p_{\Lambda_{b}}^\nu p_{\Lambda_{c}}^\mu)\bigg ]  + \tilde{h}_+ \bigg \{ i\sigma^{\mu\nu}-\frac{2}{Q_-}\bigg [m_{\Lambda_{b}}(p_{\Lambda_{c}}^\mu \gamma^\nu-p_{\Lambda_{c}}^\nu \gamma^\mu    ) \nonumber \\
& -    m_{\Lambda_{c}}(p_{\Lambda_{b}}^\mu \gamma^\nu-p_{\Lambda_{b}}^\nu \gamma^\mu    ) + p_{\Lambda_{b}}^\mu p_{\Lambda_{c}}^\nu - p_{\Lambda_{b}}^\nu p_{\Lambda_{c}}^\mu \bigg ] \bigg \}  + \tilde{h}_\perp\frac{m_{\Lambda_{b}}-m_{\Lambda_{c}}}{q^2Q_-} \nonumber \\
&   [ (m_{\Lambda_{b}}^2-m_{\Lambda_{c}}^2-q^2)(\gamma^\mu p_{\Lambda_{b}}^\nu - \gamma^\nu p_{\Lambda_{b}}^\mu) - (m_{\Lambda_{b}}^2-m_{\Lambda_{c}}^2+q^2)(\gamma^\mu p_{\Lambda_{c}}^\nu - \gamma^\nu p_{\Lambda_{c}}^\mu) \nonumber \\
&  +2(m_{\Lambda_{b}}-m_{\Lambda_{c}})(p_{\Lambda_{b}}^\mu p_{\Lambda_{c}}^\nu - p_{\Lambda_{b}}^\nu p_{\Lambda_{c}}^\mu)  ]\bigg \} u_{\Lambda_{b}}~,
\end{align}
(the parametrization of $\langle \Lambda_{c} | \bar ci\sigma^{\mu\nu} \gamma^5 b | \Lambda_{b} \rangle$ can be found using the relation 
$2i\sigma^{\mu\nu}\gamma^5= \epsilon^{\mu\nu\alpha\beta}\sigma_{\alpha \beta}$).
In our analysis, we take the formulae for the form factors $f$s and $g$s from~\cite{Detmold:2015aaa} and the other four from~\cite{Datta:2017aue, Detmold:2016pkz}.
With $d\Gamma_{\Lambda_b\to\Lambda_c\mu\nu}/dq^2$ being calculated by 
replacing $m_\tau$ with $m_\mu$ and turning off all Wilson coefficients, finally we have
\begin{align}
R_{\Lambda_{c}}=\frac{\int_{m_\tau^2}^{q_{\text{max}}^2}dq^2 ~~d\Gamma_{\Lambda_b\to\Lambda_c\tau\nu} / dq^2}{\int_{m_\mu^2}^{q_{\text{max}}^2}dq^2~~d\Gamma_{\Lambda_b\to\Lambda_c\mu\nu}/dq^2}~,
\end{align}
with $q_{\text{max}}^2=(m_{\Lambda_b}-m_{\Lambda_c})^2$.

\subsection{Br$(B^+\to K^+ \tau^+ \tau^-)$}

Br$(B^+\to K^+ \tau^+ \tau^-)$ involves the $b$-meson decay into a pseudoscalar meson. According to~\cite{Bobeth:2007dw, Becirevic:2012fy}, we have: 
\begin{align}
\frac{d \Gamma_{B^+\to K^+ \tau^+ \tau^-}}{d q^2} = &\frac{G_F^2 \alpha^2 |V_{tb}V_{ts}^\ast|^2}{256\pi^5 m_{B^+}^3} \sqrt{\lambda(q^2)}\beta_\tau \bigg (A + \frac{1}{3}C\bigg )\ ,
\end{align}
where
\begin{align}
\beta_\tau &= \sqrt{1-4\frac{m_\tau^2}{q^2}}\ ,\\
A &= q^2 (\beta_\tau^2 |F_S|^2 + |F_P|^2) + \frac{\lambda}{4} (|F_A|^2+|F_V|^2)+ 4m_\tau^2 m_{B^+}^2 |F_A|^2 \nonumber \\
& + 2m_\tau(m_{B^+}^2-m_{K^+}^2+q^2)\text{Re}(F_PF_A^\ast) \ ,\\
C &= q^2 (\beta_\tau^2 |F_T|^2 + |F_{T5}|^2) - \frac{\lambda\beta_\tau^2}{4} (|F_A|^2+|F_V|^2) + 2m_\tau\sqrt{\lambda}\beta_\tau \text{Re}(F_TF_V^\ast)~.
\end{align}
Here $F(q^2)$s are given by:
\begin{align}
F_V &= (C_9^\tau|_{\text{SM}} + \delta C_9^\tau + C_9^{\prime\tau})f_+ + \frac{2m_b}{m_{B^+}+m_{K^+}}\bigg (C_7^\tau|_{\rm SM} 
+ \frac{4m_\tau}{m_b}C_T^\tau \bigg )f_T \ ,\\
F_A &= (C_{10}^\tau|_{\text{SM}} + \delta C_{10}^\tau + C_{10}^{\prime\tau})f_+ \ ,\\
F_S &= (C_S^\tau +C_S^{\prime\tau})\frac{m_{B^+}^2-m_{K^+}^2}{2m_b}f_0 \ ,\\
F_P &= (C_P^\tau +C_P^{\prime\tau})\frac{m_{B^+}^2-m_{K^+}^2}{2m_b}f_0 \nonumber \\
&-m_\tau(C_{10}^\tau|_{\text{SM}} + \delta C_{10}^\tau + C_{10}^{\prime\tau})\bigg [f_+ - \frac{m_{B^+}^2-m_{K^+}^2}{q^2}(f_0-f_+)\bigg ] \ ,\\
F_T &= 2C_T^\tau \frac{\beta_\tau\sqrt{\lambda}}{m_{B^+}+m_{K^+}}f_T ,\\
F_{T5} &= 2C_{T5}^\tau \frac{\beta_\tau\sqrt{\lambda}}{m_{B^+}+m_{K^+}}f_T~.
\end{align}
$f_{+,0,T}$ are form factors which parameterize the relevant matrix elements as
\begin{align}
\langle K^+(k) |\bar s \gamma^\mu b | B^+(p) \rangle &= \bigg [ (p+k)^\mu - q^\mu\frac{m^2_{B^+}-m^2_{K^+}}{q^2}\bigg ] f_+(q^2) + q^\mu\frac{m^2_{B^+}-m^2_{K^+}}{q^2} f_0(q^2)\ ,\\
\langle K^+(k) |\bar s \sigma^{\mu\nu} b | B^+(p) \rangle &= -\frac{2i(p^\mu k^\nu - p^\nu k^\mu)}{m_{B^+}+m_{K^+}}f_T(q^2)~.
\end{align}
In this analysis their lattice-QCD-based values are taken from~\cite{Bailey:2015dka}. The branching ratio is finally given by 
\begin{align}
\text{Br}(B^+\to K^+\tau^+\tau^-) = \tau_{B^+}\int_{q_{\text{min}}^2}^{q_{\text{max}}^2}dq^2 ~~d\Gamma_{B^+\to K^+ \tau^+ \tau^-} / dq^2 ,
\end{align}
where $\tau_{B^+}$ is the lifetime of $B^+$ and $q^2$ ranges from 15 $\unit{GeV^2}$ to $(m_{B^+}-m_{K^+})^2$.

\subsection{Br$(B^0\to K^{\ast 0} \tau^+\tau^-)$ and Br$(B_s\to \phi\tau^+\tau^-)$}

Br$(B^0\to K^{\ast 0} \tau^+\tau^-)$ and Br$(B_s\to \phi\tau^+\tau^-)$ invovle the decay of $b$-meson into a vector meson. Here we consider Br$(B^0\to K^{\ast 0} \tau^+\tau^-)$ and the calculation of Br$(B_s\to \phi\tau^+\tau^-)$ is similar. Following~\cite{Bobeth:2010wg, Bobeth:2012vn}, we have:
\begin{align}
\frac{d \Gamma_{B^0\to K^{\ast 0} \tau^+\tau^-}}{d q^2} = 2J_{1s}+J_{1c}-\frac{2J_{2s}+J_{2c}}{3}\ ,
\end{align}
where
\begin{align}
J_{1s} &= \frac{3(2+\beta_\tau^2)}{16} ( |A_\perp^L|^2 + |A_\parallel^L|^2 + |A_\perp^R|^2 + |A_\parallel^R|^2 ) + \frac{3m_\tau^2}{q^2}\text{Re} (A_\perp^L A_\perp^{R\ast}+A_\parallel^L A_\parallel^{R\ast} ) \nonumber \\
& + 3\beta_\tau^2(|A_{0\perp}|^2+|A_{0\parallel}|^2)+3(4-3\beta_\tau^2)(|A_{t\perp}|^2+|A_{t\parallel}|^2) \nonumber \\
& +\frac{6\sqrt{2}m_\tau}{q^2}\text{Re} [ (A_\parallel^L+A_\parallel^R)A_{t\parallel}^\ast+(A_\perp^L+A_\perp^R)A_{t\perp}^\ast ] \ ,\\
J_{1c} &= \frac{3}{4}(|A_0^L|^2+|A_0^R|^2 + \beta_\tau^2|A_S|^2)+\frac{3m_\tau^2}{q^2} [ |A_t|^2+2\text{Re}(A_0^LA_0^{R\ast}) ] +6(2-\beta_\tau^2)|A_{t0}|^2 \nonumber \\
&  + 6\beta_\tau^2 |A_{\parallel\perp}|^2 + \frac{12m_\tau}{\sqrt{q^2}}\text{Re} [(A_0^L+A_0^R)A_{t0}^\ast ] \ ,\\
J_{2s} &= \frac{3\beta_\tau^2}{16}( |A_\perp^L|^2+ |A_\parallel^L|^2+ |A_\perp^R|^2+ |A_\parallel^R|^2) - 3\beta_\tau^2 (|A_{t\perp}|^2+|A_{t\parallel}|^2+|A_{0\perp}|^2+|A_{0\parallel}|^2)\  ,\\
J_{2c} &= \frac{3\beta_\tau^2}{4} [8(|A_{t0}|^2 + |A_{\parallel\perp}|^2) -|A_0^L|^2 -|A_0^R|^2 ]\ ,
\end{align}
with $\beta_\tau = \sqrt{1-4\frac{m_\tau^2}{q^2}}$. Here all $A$-quantities are transversity amplitudes. They are given by 
\begin{align}
A_\perp^{L,R} &= N\sqrt{2\lambda}\bigg \{ [(C_9^\tau|_{\text{SM}} + \delta C_9^\tau + C_9^{\prime\tau})\mp (C_{10}^\tau|_{\text{SM}} + \delta C_{10}^\tau + C_{10}^{\prime\tau}) ]\frac{V}{m_{B^0}+m_{K^{\ast 0}}}\nonumber \\
&+\frac{2m_b}{q^2}(C_7^\tau+C_7^{\prime\tau})T_1\bigg \}  \ ,\\
A_\parallel^{L,R} &= -N\sqrt{2}(m_{B^0}^2-m_{K^{\ast 0}}^2)\bigg \{   [(C_9^\tau|_{\text{SM}} + \delta C_9^\tau - C_9^{\prime\tau}) \nonumber \\
&\mp (C_{10}^\tau|_{\text{SM}} + \delta C_{10}^\tau - C_{10}^{\prime\tau}) ] \frac{A_1}{m_{B^0}-m_{K^{\ast 0}}} +\frac{2m_b}{q^2}(C_7^\tau-C_7^{\prime\tau})T_2\bigg \} \ , \\
A_0^{L,R} &= -\frac{N}{2m_{K^{\ast 0}}\sqrt{q^2}}\bigg \{     [(C_9^\tau|_{\text{SM}} + \delta C_9^\tau - C_9^{\prime\tau})\mp (C_{10}^\tau|_{\text{SM}} + \delta C_{10}^\tau - C_{10}^{\prime\tau}) ] \times\nonumber \\
& \bigg [(m_{B^0}^2-m_{K^{\ast 0}}^2-q^2)(m_{B^0}+m_{K^{\ast 0}})A_1 -\frac{\lambda A_2}{m_{B^0}+m_{K^{\ast 0}}}\bigg ] + 2m_b (C_7^\tau - C_7^{\prime\tau})\times \nonumber \\
& \bigg [ (m_{B^0}^2+3m_{K^{\ast 0}}^2-q^2)T_2 - \frac{\lambda T_3}{m_{B^0}^2-m_{K^{\ast 0}}^2}\bigg ]\bigg \} \ , \\
A_t &= N\frac{\sqrt{\lambda}}{\sqrt{q^2}}\bigg [ 2(C_{10}^\tau|_{\text{SM}} + \delta C_{10}^\tau - C_{10}^{\prime\tau}) + \frac{q^2(C_P^\tau-C_P^{\prime\tau})}{m_\tau m_b}\bigg ] A_0 \ ,\\
A_S &= -2N\sqrt{\lambda}\frac{C_S^\tau-C_S^{\prime\tau}}{m_b}A_0 \ ,\\
A_{\parallel\perp (t0)} &= \pm N\frac{C_{T(5)}^\tau}{m_{K^{\ast 0}}}\bigg [ (m_{B^0}^2+3m_{K^{\ast 0}}^2-q^2)T_2- \frac{\lambda T_3}{m_{B^0}^2-m_{K^{\ast 0}}^2}\bigg ] \ ,\\
A_{t\perp (0\perp)} &= \pm 2N \frac{\sqrt{\lambda}}{\sqrt{q^2}}C_{T(5)}^\tau T_1 \ ,\\
A_{0\parallel (t\parallel)} &= \pm 2N \frac{m_{B^0}^2-m_{K^{\ast 0}}^2}{\sqrt{q^2}}C_{T(5)}^\tau T_2 \ ,
\end{align}
with
\begin{align}
N &= G_F\alpha V_{tb}V_{ts}^\ast \sqrt{\frac{q^2\beta_\tau\sqrt{\lambda}}{3072\pi^5m_{B^0}^3}} \ ,\\
\lambda(q^2) &=  [(m_{B^0}-m_{K^{\ast 0}})^2-q^2  ]  [ (m_{B^0}+m_{K^{\ast 0}})^2-q^2  ]  \ .
\end{align}
$A_{0,1,2}(q^2)$, $V(q^2)$ and $T_{1,2,3}(q^2)$ are form factors. 
They parameterize the relevant matrix elements as~\cite{Horgan:2013hoa} 
\begin{align}
\langle K^{\ast 0}(k, \varepsilon) | \bar c \gamma^\mu b | B^0(p) \rangle &= \frac{2iV}{m_{B^0}+m_{K^{\ast 0}}}\epsilon^{\mu\nu\rho\sigma} \varepsilon^\ast_\nu p_\rho k_\sigma \ , \\
\langle K^{\ast 0}(k, \varepsilon) | \bar c \gamma^\mu\gamma^5 b | B^0(p) \rangle &= 2m_{K^{\ast 0}}A_0\frac{\varepsilon^\ast\cdot q}{q^2}q^\mu + (m_{B^0}+m_{K^{\ast 0}})A_1\bigg ( \varepsilon^{\ast\mu}-\frac{\varepsilon^\ast \cdot q}{q^2}q^\mu \bigg ) \nonumber \\
&~~~-A_2\frac{\varepsilon^\ast\cdot q}{m_{B^0}+m_{K^{\ast 0}}} \bigg (p^{\mu}+k^{\mu}-\frac{m^2_{B^0}-m^2_{K^{\ast 0}}}{q^2}q^\mu \bigg )\ , \\
\langle K^{\ast 0}(k, \varepsilon) | \bar c \sigma^{\mu\nu}q_\nu b | B^0(p) \rangle &= 2T_1\epsilon^{\mu\nu\rho\sigma} \varepsilon^\ast_\nu p_\rho k_\sigma \ , \\
\langle K^{\ast 0}(k, \varepsilon) | \bar c \sigma^{\mu\nu}\gamma^5q_\nu b | B^0(p) \rangle &= -T_2\bigg [ (m^2_{B^0}-m^2_{K^{\ast 0}})\varepsilon^{\ast\mu}- (\varepsilon^\ast\cdot q)(p+k)^\mu \bigg ] \nonumber \\
&~~~ -T_3(\varepsilon^\ast\cdot q)\bigg [ q^\mu -\frac{q^2}{m^2_{B^0}-m^2_{K^{\ast 0}}}(p+k)^\mu \bigg ]~.
\end{align}
In this analysis, their lattice-QCD-based values are taken from~\cite{Horgan:2015vla}. 
We then have 
\begin{eqnarray}
\text{Br}(B^0\to K^{\ast 0}\tau^+\tau^-) = \tau_{B^0}\int_{q_{\text{min}}^2}^{q_{\text{max}}^2}  dq^2 ~~  \frac{d\Gamma_{B^0\to K^{\ast 0}\tau^+\tau^-} }{ dq^2} ~.
\end{eqnarray}
Here $\tau_{B^0}$ is the lifetime of $B^0$ and $q^2$ ranges from 15 $\unit{GeV^2}$ to $(m_{B^0}-m_{K^{\ast 0}})^2$. 

\subsection{Br$(B_s\to \tau^+\tau^-)$}

As studied in~\cite{Becirevic:2012fy}, Br$(B_s\to \tau^+\tau^-)$ is given by
\begin{align}
\text{Br}(B_s\to \tau^+\tau^-) &=  \tau_{B_s} f_{B_s}^2 m_{B_s}^3 \frac{G_F^2\alpha^2}{64\pi^3} |V_{tb}V_{ts}^*| \beta_\tau (m_{B_s}^2) \bigg [ \frac{m_{B_s}^2}{m_b^2}\big |C_S^\tau - C_S^{\prime\tau} \big |^2\bigg (1-\frac{4m_\tau^2}{m_{B_s}^2}\bigg ) \nonumber \\
& + \bigg | \frac{m_{B_s}}{m_b}(C_P^\tau - C_P^{\prime\tau})+\frac{2m_\tau}{m_{B_s}}(C_{10}^\tau - C_{10}^{\prime\tau})\bigg |^2 \bigg ] \ .
\end{align}
Here $f_{B_s}$ is a form factor parametrizing the hadronic matrix element 
\begin{align}
\langle 0| \bar s\gamma^\mu P_L b |B_s(p)\rangle = \frac{i}{2}f_{B_s}p^\mu \ , 
\end{align}
$\tau_{B_s}$ is the lifetime of $B_s$, and $\beta_\tau(q^2)$ is a function of $q^2$, defined as 
\begin{align}
\beta_\tau (q^2) = \sqrt{1-\frac{4m_\tau^2}{q^2}}\ .
\end{align}

\newpage

\bibliography{CEPCb}

\end{document}